%% file: MAIN.tex
\documentclass[sigconf]{acmart}

\AtBeginDocument{%
  }

\copyrightyear{2026}
\acmYear{2026}
\setcopyright{cc}
\setcctype{by}
\acmConference[CHI '26]{Proceedings of the 2026 CHI Conference on Human Factors in Computing Systems}{April 13--17, 2026}{Barcelona, Spain}
\acmBooktitle{Proceedings of the 2026 CHI Conference on Human Factors in Computing Systems (CHI '26), April 13--17, 2026, Barcelona, Spain}
\acmPrice{}
\acmDOI{10.1145/3772318.3790879}
\acmISBN{979-8-4007-2278-3/2026/04}

\usepackage{subcaption}
\usepackage{booktabs}
\usepackage{listings}
\usepackage{xcolor}
\usepackage{pifont}
\usepackage{tabularx}
\usepackage{tabularray}
\usepackage{graphicx}

\newcommand{\uparrowgreen}{\textcolor{green}{$\uparrow$}}
\newcommand{\downarrowred}{\textcolor{red}{$\downarrow$}}

\newcommand{\re}[1]{\textcolor{black}{#1}}

\begin{document}

\title{Through the Lens of Human-Human Collaboration: A Configurable Research Platform for Exploring Human-Agent Collaboration}

\author{Bingsheng Yao}
\authornotemark[1]
\email{b.yao@northeastern.edu}
\affiliation{
  \institution{Northeastern University}
  \city{Boston}
  \state{Massachusetts}
  \country{USA}
  }
\orcid{0009-0004-8329-4610}

\author{Jiaju Chen}
\authornote{Both authors contributed equally to this research.}
\email{jiaju.chen@rice.edu}
\affiliation{
  \institution{Rice University}
  \city{Houston}
  \state{Texas}
  \country{USA}
  }
\orcid{0009-0005-9598-5838}

\author{Chaoran Chen}
\email{cchen25@nd.edu}
\affiliation{
\department{Department of Computer Science and
Engineering}
  \institution{University of Notre Dame}
  \city{Notre Dame}
  \state{Indiana}
  \country{USA}
  }
\orcid{0000-0002-9161-4088}

\author{April Wang}
\email{april.wang@inf.ethz.ch}
\affiliation{
  \institution{ETH Zurich}
  \city{Zurich}
  \country{Switzerland}
  }
\orcid{0000-0001-8724-4662}

\author{Toby Jia-jun Li}
\email{toby.j.li@nd.edu}
\affiliation{
\department{Department of Computer Science and
Engineering}
  \institution{University of Notre Dame}
  \city{Notre Dame}
  \state{Indiana}
  \country{USA}
  }
\orcid{0000-0001-7902-7625}

\author{Dakuo Wang}
\authornote{Corresponding author d.wang@northeastern.edu}
\email{d.wang@northeastern.edu}
\affiliation{
  \institution{Northeastern University}
  \city{Boston}
  \state{Massachusetts}
  \country{USA}
  }
\orcid{0000-0001-9371-9441}

\renewcommand{\shortauthors}{Yao and Chen et al.}
\renewcommand{\shorttitle}{A Configurable Research Platform for Exploring Human-Agent Collaboration}

\input{sections/0abstract}



\input{sections/teaserfigure}


\maketitle

\input{sections/1introduction-v7}

\input{sections/2relatedwork-v3}

\input{sections/3system-v3}
\input{sections/4-evaluations}

\input{sections/5-discussion}

\input{sections/6-futureworkandlimitation}

\input{sections/7-conclusion}


\input{sections/acknowledge}

\bibliographystyle{ACM-Reference-Format}
\bibliography{reference}

\include{sections/appendix}

\input{sections/Questionnaires}

\input{sections/configlanguage}

\input{sections/hiddenprofileresults}

\input{sections/adaptation}

\include{sections/prompts}

\include{sections/interfaces}

\end{document}

%% file: sections/0abstract.tex
\begin{abstract}
Intelligent systems have traditionally been designed as tools rather than collaborators, often lacking critical characteristics that collaboration partnerships require.
Recent advances in large language model (LLM) agents open new opportunities for \textbf{human-LLM-agent collaboration} 
by enabling natural communication and various social and cognitive behaviors.
Yet it remains unclear whether principles of computer-mediated collaboration established in HCI and CSCW persist, change, or fail when humans collaborate with LLM agents.
To support systematic investigations of these questions, we introduce an open and configurable research platform for HCI researchers\footnote{Our platform is available at \url{https://github.com/neuhai/human-agent-collab}.}.
The platform's modular design allows seamless adaptation of classic CSCW experiments and manipulation of theory-grounded interaction controls.
We demonstrate the platform's \re{research efficacy} and usability through \re{three} case studies: \re{(1) two \textit{Shape Factory} experiments for resource negotiation with 16 participants, (2) one \textit{Hidden Profile} experiment for information pooling with 16 participants,} and (3) a participatory cognitive walkthrough with five HCI researchers to refine workflows of researcher interface for experiment setup and analysis.

\end{abstract}

%% file: sections/teaserfigure.tex
\begin{teaserfigure}
\centering
  \includegraphics[width=.9\linewidth]{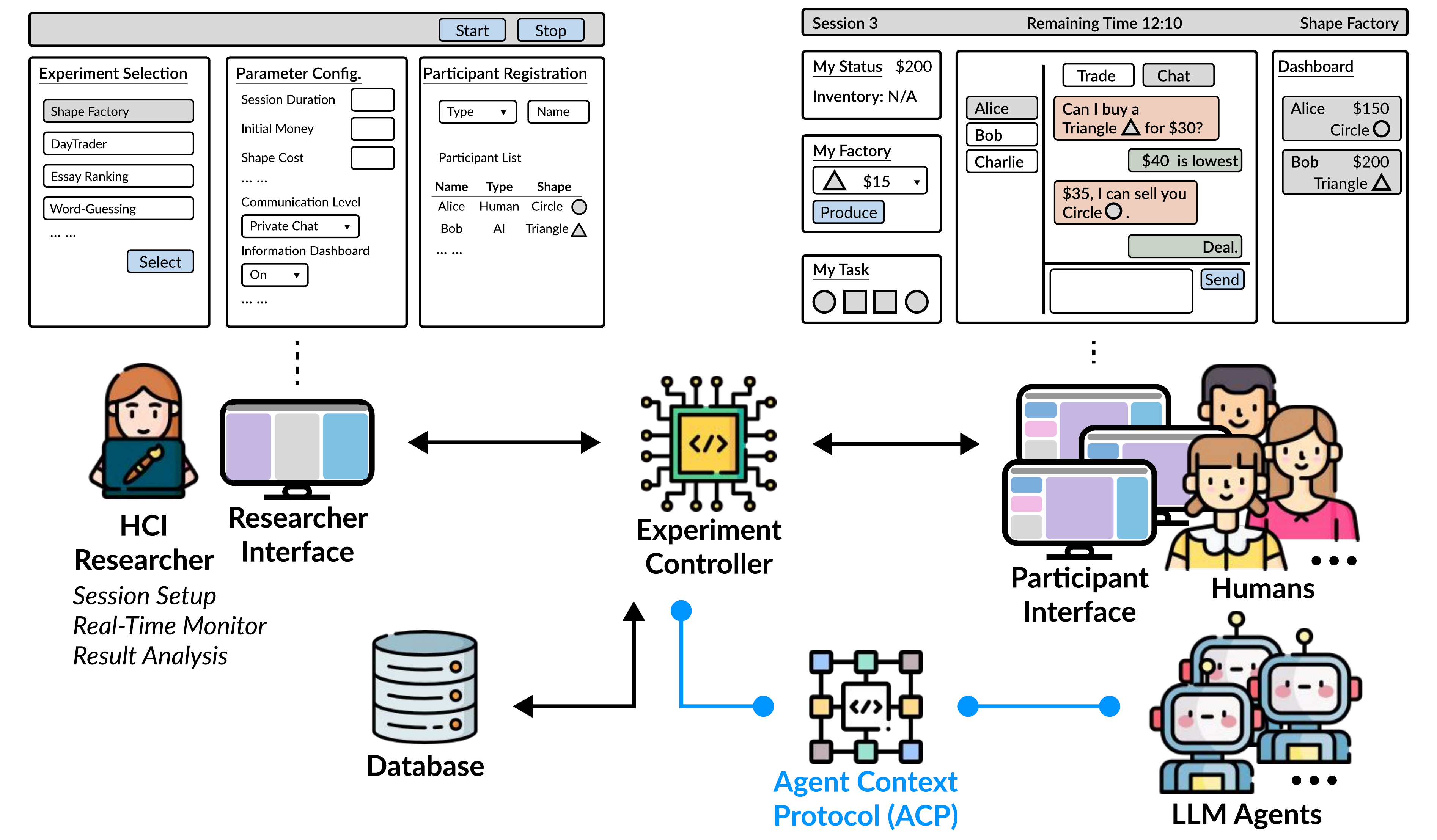}
    \caption{The core architecture of our research platform, which encompasses four key modules: \textit{researcher interface} allows researchers to configure experiment setup, monitor live sessions, and analyze result; \textit{participant interface} enables human participants to engage with other participants and LLM agents; \textit{Agent Context Protocol} ensures standardized agent integration; \textit{experiment controller} manages experiment states and actions updates, and synchronize all the interaction logs into the database. }
    \label{fig:architecture}
    \Description{Fig. 1. The figure shows the core architecture of a research platform with four modules. A researcher interface (top left) supports experiment setup, live monitoring, and result analysis. A participant interface (top right) lets humans and LLM agents interact in the experiment (e.g., chat/trade, status, and dashboard). An experiment controller (center) manages experiment state and participant actions, and synchronizes all interaction logs to a database (bottom left). An Agent Context Protocol (ACP) (bottom right) standardizes communication between LLM agents and the experiment controller.}
\end{teaserfigure}


%% file: sections/1introduction-v7.tex
\section{Introduction}
\label{sec:intro}

The HCI and CSCW communities have long highlighted a gap where intelligent systems are predominantly perceived as ``tools,'' rather than mutual collaborators with humans~\cite{grudin2018tool,shneiderman2022human, suchman1987plans}. 
Despite their capacity to support humans in data analysis~\cite{wang2019human,guo2024investigating}, design work~\cite{oh2018lead,lee2022coauthor}, and decision-makings~\cite{cai2019hello,zhang2024rethinking}, these systems typically lack the mutual awareness~\cite{dourish1992awareness,gutwin2002descriptive}, adaptivity~\cite{allen1999mixed,amershi2019guidelines}, accountability~\cite{liao2020questioning,hoffman2023measures}, and interdependence~\cite{seeber2020machines,zhang2021ideal} that characterize effective partnerships.
This deficiency often impedes trust~\cite{jian2000foundations,hoff2015trust,vereschak2021evaluate, hancock2011meta}, shared understanding~\cite{van2022five,clark1991grounding, clark1996using}, or effective coordination~\cite{schmidt1992taking, olson2000distance, hinds2003out}, which can further limits the quality of  outcomes~\cite{ashktorab2020human,zheng2023competent}, diminish human agency~\cite{tapal2017sense,kim2024diarymate}, and undermine reliance in high-stakes situations~\cite{parasuraman1997humans, lee2004trust, wang2020human, vaccaro2024combinations}.

The advent of LLMs has shifted the design space for intelligent systems~\cite{brown2020language,bubeck2023sparks}. 
LLMs demonstrate the capacity to engage in fluid communication with humans~\cite{sharma2023human,yang2024talk2care}, while advanced prompting techniques further enable human-like reasoning and planning ~\cite{wei2022chain,zhou2022least,wang2023rolellm}.
Research on \textbf{LLM agents}~\cite{park2023generative, park2024generative, shanahan2023role}, which are LLM systems capable of exhibiting complex, human-like behaviors to solve tasks, shows these agents\footnote{In this paper, we use "LLM agent" and "agent" interchangeably~\cite{chen2025towards}.} can yield distinct, believable cognitive and social behaviors, particularly through role-playing the provided persona profiles. 
These capabilities promise new paradigms of \textbf{human-agent collaboration}, where humans treat these LLM agents as mutual partners rather than mere tools.

\re{However, deploying these agents into real-world teams introduces critical socio-technical challenges. 
In high-stakes domains such as healthcare triage or crisis management, failures to establish common ground or to signal uncertainty can degrade collective performance and erode accountability~\cite{cai2019hello, zhang2024rethinking}. 
For instance, does an "assertive" agent facilitate faster decision-making, or does it suppress critical thinking and information sharing among collaborators? 
Addressing these design problems requires isolating interaction variables like communication modality and agent responsiveness in human-agent collaboration.} 
Yet, investigating these dynamics requires moving beyond simple performance metrics to examine how social~\cite{nass2000machines,zhang2021ideal}, communicative~\cite{liang2019implicit,zhang2023investigating}, and cognitive behaviors~\cite{schelble2022let,ashktorab2020human} co-evolve in controlled settings.

To inform the design of interactions in the context of human-agent collaboration, our vision is that \textbf{LLM agents should resemble remote human collaborators}.
Both LLM agents and remote human collaborators rely on structured channels such as text or voice, and both lack social cues (e.g., facial expressions, tone) that augment in-person interactions~\cite{short1976social,walther1996computer, olson2000distance}.
While this analogy is limited by differences in accountability, knowledge grounding, and identity persistence, it motivates designing interfaces and interactions for human–LLM-agent collaboration by drawing on decades of CSCW and HCI research on remote collaboration. 
Empirical and theoretical research has systematically identified key interaction variables in remote collaboration: communication modalities shape the richness of information exchange~\cite{daft1986organizational, straus1997technology,bos2002effects}, workspace awareness supports coordination~\cite{dourish1992awareness,gutwin2002descriptive}, and information transparency fosters trust and mutual understanding~\cite{cramton2001mutual,hinds2003out}.
Recent studies confirmed that these foundational principles remain relevant as new agent capabilities emerge~\cite{zhang2023investigating,he2024ai, lu2025uxagent}.
Grounding human-LLM agent interaction in this body of research enables HCI researchers to investigate which principles and challenges persist, which shift, and what new issues arise as the boundaries between remote human collaborators and LLM agents blur.

A major barrier to pursuing this line of research is the lack of \textbf{an open, configurable, and theory-driven research platform} for systematically examining how diverse interaction designs influence human-agent collaboration. 
Conventional CSCW experimental paradigms, such as NeoCITIES~\cite{schelble2022let, mcneese2005neocities}, Shape Factory~\cite{bos2004group}, DayTrader~\cite{bos2002effects}, and Passcode~\cite{gero2020mental}, feature distinct task protocols and interaction controls. 
Re-implementing them for human-agent collaboration requires considerable repetitive effort, and the resulting system still lacks reusability and adaptability across paradigms.
This practice often creates obstacles to reproducibility and the cumulative extension of scientific findings.
While general-purpose virtual lab platforms in social science provide infrastructure for group studies with humans, they lack agent integration protocols or CSCW-informed interface controls~\cite{almaatouq2021empirica,chen2016otree}.
Existing agent-based frameworks emphasize multi-agent collaboration~\cite{wu2024autogen,chen2023agentverse,li2023camel}, simulations~\cite{park2023generative, qian2023chatdev}, or performance evaluation~\cite{zhou2023webarena, leibo2021meltingpot}, rather than controlled experiments for investigating the interactions in human-agent collaboration.
This lack of an extensible research platform impedes systematic, cumulative explorations and the goal of open science in the HCI community.

To address this gap, we present an open, configurable research platform for conducting reproducible, controlled experiments on human-LLM-agent collaboration.
Our platform design draws inspiration from the Shape Factory experiment~\cite{bos2004group}, a classic CSCW experiment for analyzing group dynamics between collocated and remote collaborators.
The Shape Factory experiment supports rigorous manipulation of communication modality, awareness, social framing, and resource interdependence.
Our platform enables HCI researchers to adapt classic experimental paradigms for human-agent collaboration research, such as strategic collaboration tasks (e.g., DayTrader~\cite{bos2002effects}), collaborative decision-making tasks (e.g., \re{Hidden Profile~\cite{stasser1985pooling, stasser1987effects} and} Essay Ranking~\cite{zheng2023competent}), and collaborative task-solving (e.g., Passcode~\cite{gero2020mental}).
The platform architecture features four key components, as shown in Fig~\ref{fig:architecture}:
\begin{enumerate}
    \item \textbf{researcher interface} for manipulating parameters and interaction controls; 
    \item modularized and customizable \textbf{participant interface} that reflects those manipulations; 
    \item standardized \textbf{agent context protocol} ensuring consistent agent integration;
    \item \textbf{experiment controller} serving as execution engine and logging system.
\end{enumerate}
Critically, the platform allows systematic manipulation of theory-grounded interaction controls to investigate their impact on collaborative dynamics, such as communication modality~\cite{straus1997technology, bos2002effects}, awareness dashboards~\cite{gutwin2002descriptive, dourish1992awareness}, and social framing~\cite{bos2004group, hinds2003out}.
Researchers can independently manipulate and customize these controls to meet nuanced needs using the platform's open-source code.

We demonstrate our platform's utility through a two-part evaluation.
First, we validate the \re{research efficacy} of our platform for controlled experiments through \re{\textbf{multi-phase case studies}.}
\re{We implemented the \textbf{Shape Factory} experiment~\cite{bos2004group} to investigate how communication channels (chat v.s. no chat) and workspace awareness (with v.s. without dashboard) influence resource negotiation in a mixed human-agent team. This phase employs a crossed, between-subjects design with 16 participants. }
\re{Subsequently, we implemented the \textbf{Hidden Profile} experiment~\cite{stasser1985pooling, stasser1987effects} to examine how agent responsiveness (passive v.s. proactive) affects information pooling and collaborative decision-making quality. This phase employs a between-subjects design with 16 participants.} 
\re{These studies confirmed that the platform} could faithfully re-implement classic paradigms and capture significant divergence in individual and collaborative behaviors, collaboration outcome, and perception caused by different interaction designs.
Second, we conducted a \textbf{participatory cognitive walkthrough} with five HCI researchers on human-AI collaboration to evaluate the researcher interface for experiment setup and data analysis.
Their feedback informed iterative improvements to workflow and interface design. 
Collectively, these evaluations validate our platform as a methodological foundation for the HCI community to advance a systematic, evidence-based understanding of human-agent collaboration.



%% file: sections/2relatedwork-v3.tex
\section{Related Work}
\label{sec:related}

We review four lines of work that motivate the need for a theory-driven, extensible platform for human–agent collaboration studies.


\subsection{Barriers to Treating Human-AI Collaboration as a Partnership}


Although intelligent systems are increasingly positioned as collaborators rather than mere tools, prior HCI and CSCW research shows several persistent barriers prevent them from functioning as true partners to humans. 
These barriers revolve around limited awareness and common ground, insufficient mixed-initiative support, and the lack of accountability and interdependence structures needed for coordinated joint work.

\textbf{Awareness and Common Ground.}
Effective collaboration requires knowing who is doing what, when, and why~\cite{dourish1992awareness,gutwin2002descriptive} and maintaining shared references and assumptions~\cite{clark1991grounding}.  
When systems obscure their state, history, or intent, users must infer or repair breakdowns on their own~\cite{schmidt1992taking,clark1996using}, which could increase coordination costs and slow work progress.


\textbf{Initiative and Adaptivity.}
Mixed-initiative collaboration depends on a system’s ability to determine when to act, when to wait, and how to make its moves interpretable~\cite{allen1999mixed,amershi2019guidelines}.  
Pre-LLM systems were largely template-driven and unable to adapt to evolving goals or proactively provide the right help~\cite{amershi2019guidelines}, making them poor partners in dynamic tasks.


\textbf{Accountability and Trust Calibration.}
Partners must be able to explain their opinions and reasoning, justify recommendations, and communicate uncertainty~\cite{liao2020questioning,hoffman2023measures}.  
Without such accountability, users develop mis-calibrated trust~\cite{jian2000foundations,lee2004trust,hancock2011meta}, which leads to degraded collective outcomes and reluctance to rely on systems in high-stakes tasks~\cite{parasuraman1997humans,wang2020human,vaccaro2024combinations}.


\textbf{Interdependence and Role Negotiation.}
Collaboration also depends on interdependence-partners coordinating roles, dividing labor, and negotiating contributions~\cite{seeber2020machines}.  
Many intelligent systems, however, operate as isolated predictors or recommenders~\cite{van2022five}, offering little opportunity for role clarification or shared control~\cite{clark1996using}.  
As a result, users experience weakened shared understanding and diminished agency and engagement~\cite{ashktorab2020human,zheng2023competent,tapal2017sense,kim2024diarymate}.


\begin{figure*}[!htp]
    \centering
    \includegraphics[width=.95\linewidth]{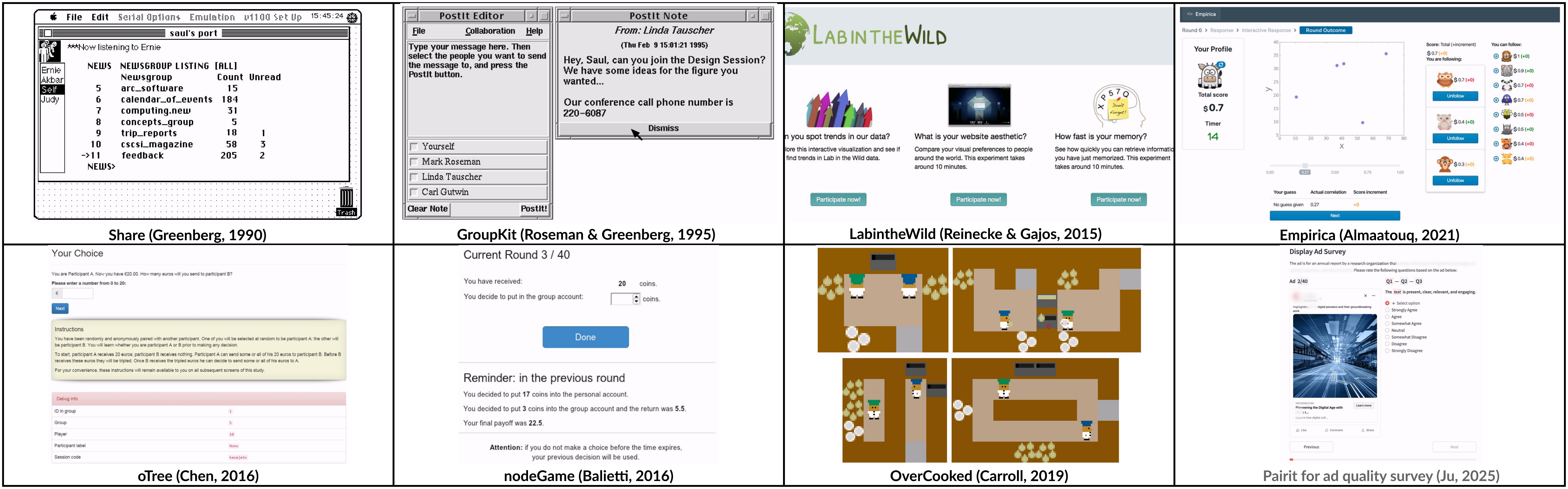}
    \caption{Existing research-focused platforms for collaboration studies. From left to right, the top row: Share~\cite{greenberg1990sharing}, GroupKit~\cite{roseman1996building}, LabintheWild~\cite{reinecke2015labinthewild}, Empirica~\cite{almaatouq2021empirica}; the bottom row: oTree~\cite{chen2016otree}, nodeGame~\cite{balietti2017nodegame}, Overcooked~\cite{carroll2019utility}, Pairit~\cite{ju2025collaborating}. }
    \label{fig:existing-platforms}
    \Description{Fig. 2. This figure shows a comparison of existing research platforms for collaboration studies. Top row, from left to right: Share, GroupKit, LabintheWild, and Empirica. Bottom row: oTree, nodeGame, Overcooked, and Pairit. These platforms represent diverse approaches to studying collaboration.}
\end{figure*}

\subsection{The New Landscape of Human-AI Collaboration in the Era of LLM Agent}

The emergence of LLMs sheds light on new opportunities for the landscape of LLM-based intelligent systems as mutual collaborators of humans. 
In particular, LLMs enable \textbf{natural language as a shared medium} that allows users to state goals, constraints, and references, while LLM systems can ask clarifying questions and propose plans in free-form language~\cite{brown2020language,bubeck2023sparks,sharma2023human,yang2024talk2care}.
This capability shifts interaction beyond rigid command structures into fluid, open-ended conversations.
Furthermore, advanced prompting techniques, such as chain-of-thought~\cite{wei2022chain, kojima2022large}, enable LLMs to perform step-wise reasoning and planning for their predictions. 
These techniques allow visible forms of system accountability and initiative that people can read and negotiate.

Building on these technical achievements, researchers have developed ``LLM role-playing agents,'' which adopt personas and goals to simulate complex social and cognitive behaviors over extended interactions~\cite{park2023generative, park2024generative, shanahan2023role}.
Applications are rapidly expanding: LLM agents have been explored as human surrogates for automated UX testing~\cite{lu2025uxagent}, and as tools for simulating populations and replicating classic findings in social science~\cite{argyle2023out,aher2023using}.

Despite these advances of LLM agents, the fundamental social and cognitive demands for effective human-agent collaboration remain.
The key lies in the design of interfaces~\cite{amershi2019guidelines,liao2020questioning} and interaction paradigms~\cite{wang2020human,he2024ai} between humans and LLM agents to establish the collaboration needs identified in classic CSCW research.
However, HCI researchers currently lack a dedicated research platform that allows systematic mapping of classic collaboration insights onto the interface features of LLM agents. Such a platform is essential for investigating how collaboration dynamics unfold in this new era of human–AI partnerships.

\subsection{Remote Human Collaboration as an Analytical Lens of Exploring Human-Agent Collaboration}

LLM agents resemble remote human collaborators in important ways: both operate through structured, text-based channels and lack the non-verbal cues that facilitate grounding and intent inference in face-to-face interaction~\cite{short1976social,walther1996computer,olson2000distance}.  
CSCW research on distributed teamwork, therefore, offers a valuable analytical lens for understanding human-agent collaboration challenges.


A first set of insights concerns the limitations of low-bandwidth communication.  
In particular, Media Richness Theory explains why low-bandwidth channels carry less contextual information and make negotiation harder~\cite{daft1986organizational}, with several experiments demonstrating that ``leaner'' media like text are less effective than ``richer'' media like video for conveying ambiguity and building consensus~\cite{short1976social, straus1997technology}.
These results are critical to human-agent collaboration, where text is the dominant modality and interface cues such as typing indicators or meta-messages can meaningfully alter users' interpretations of an agent’s intent~\cite{erickson2000social,iftikhar2023together}.


A second challenge involves the ``mutual knowledge problem,” where collaborators lack common ground and must infer what others know~\cite{clark1991grounding,cramton2001mutual}.  
Interfaces that add timely acknowledgments and visible states of collaborators could reduce these failures~\cite{kraut2003visual}.
Research on workspace awareness further strengthened this idea by defining cues about presence, activity, and intent that help people predict collaborators' actions~\cite{dourish1992awareness,gutwin2002descriptive}.
For LLM agents, this can be achieved by asking agents to generate future plans with rationales visible to human collaborators through interface designs.

Task structure and coupling also shape distributed collaboration. Distance complicates timing, handoffs, and coordination under partial information~\cite{olson2000distance,hinds2003out}. 
In particular, a variety of lab-based experiments were designed to systematically analyze the impact of different interaction variables in a controlled setting.
For instance, DayTrader manipulates the communication medium during negotiation in a social-dilemma setting and identified that the choice of medium shifts trust formation and deal quality~\cite{bos2002effects}.
\citet{bos2004group} further illustrated trading biases and subgroup formation caused by the differences between in-group (collocated partners who can talk face-to-face) and out-group (remote partners) in the Shape Factory experiment.
NeoCITIES imposes time pressure and role interdependence in a shared information space~\cite{mcneese2005neocities,schelble2022let}, whereas Passcode examines how teams form and update shared mental models during problem solving when only partial clues are available~\cite{gero2020mental}.

Although LLM agents differ from human partners in their ability to perceive the context or disclose uncertainty~\cite{liao2020questioning,hoffman2023measures}, these capabilities enable adapting classic human-human collaboration variables for human–agent settings.  
Our platform supports fine-grained manipulation of communication modalities, visibility of agent state, and task coupling within re-implemented experimental paradigms tailored to human–agent collaboration.


\subsection{Research-Focused Toolkits and Platforms for Collaboration Studies}


A wide range of research platforms support collaboration studies (Fig.~\ref{fig:existing-platforms}), including classic groupware toolkits, online behavioral experiment platforms, game-based environments, and newer agent frameworks.  
While each family offers relevant capabilities, none provides the experimental control or human–agent parity needed for systematic studies of human–LLM collaboration.

\textbf{Groupware and General-Purpose Behavioral Platforms. }
Early HCI toolkits such as Share and GroupKit~\cite{greenberg1990sharing,roseman1996building} enabled synchronous distributed tasks through shared views and real-time session management.  
More recent platforms (e.g., LabintheWild~\cite{reinecke2015labinthewild}, Empirica~\cite{almaatouq2021empirica}, oTree~\cite{chen2016otree}, and nodeGame~\cite{balietti2017nodegame}) provide scalable participant management, messaging, and data capture for human–human experiments.  
Nevertheless, these platforms focus only on human studies, offer no support for agent integration, and do not allow manipulation of collaboration variables through the interface.

\textbf{Game-Based Research Environments.}  
Game environments such as Overcooked~\cite{carroll2019utility} and PairIt~\cite{ju2025collaborating} support precise coordination and performance measurement, enabling controlled studies of teamwork dynamics.
Yet their domain-specific mechanics limit structural adaptability: researchers cannot repurpose them to vary communication modalities, alter information coupling, or instantiate classic CSCW experiment paradigms.

\textbf{Agent-Oriented Frameworks and Social Simulators.}  
Frameworks like AutoGen, AgentVerse, and CAMEL~\cite{wu2024autogen,chen2023agentverse,li2023camel} support multi-agent task automation and agent–agent communication.  
Others (e.g., WebArena~\cite{zhou2023webarena}, Melting Pot~\cite{leibo2021meltingpot}, and generative social simulacra~\cite{park2023generative,park2024generative,piao2025agentsociety}) focus on agent performance, environmental fidelity, or large-scale population simulation.  
Nevertheless, these agent-oriented frameworks and simulation platforms emphasize agent performance, simulation fidelity, or agent–agent dynamics, not for human-agent interactions.

\subsubsection{\re{Comparison with Existing Research Platforms} }
\label{sec:rw-comparison}

\re{Table~\ref{tab:comparison} presents a systematic comparison of existing behavioral experiment platforms and agent frameworks that are relevant to ours in terms of dimensions that are critical for human-agent collaboration. }

\re{Platforms like oTree~\cite{chen2016otree} and Empirica~\cite{almaatouq2021empirica} have set the standard for reproducible human-subject research by providing robust session management, experiment logic, and data logging.
However, these frameworks were oriented for \textit{Human-Human} interaction.
Integrating LLM agents into these platforms requires significant ad-hoc engineering (e.g., setting up separate WebSocket servers to handle agent streaming), and there is no platform-level protocol that ensures agents perceive and act under the same constraints as human participants. 
In our system, the \textbf{Agent Context Protocol (ACP)} constructs an experiment-specific context from the configuration, issues periodic state summaries governed by a perception interval, validates agent actions against schemas and policies with error handling, and commits accepted actions atomically. 
These semantics create parity between agents and remote humans. }

\re{Conversely, frameworks like AutoGen~\cite{wu2024autogen} are designed to enhance agents' automated task-solving capabilities, which is contrary to the goal of prioritizing experimental manipulation of interaction variables for human–agent collaboration.
In a research study, a scientist may need to intentionally degrade communication (e.g., "disable chat"), introduce latency, or hide information to test a hypothesis about reliance. 
In contrast, our platform prioritizes \textbf{Interaction Control} over execution speed or standalone task-solving accuracy. 
Through the Interaction Control Interface (Section~\ref{sec:interaction-control}), researchers can systematically manipulate the information flow and action space available to the agent, enabling valid comparative studies that isolate specific collaboration variables. }

\re{Finally, environments like Overcooked-AI~\cite{carroll2019utility}, CREW~\cite{zhang2024crew}, and Pairit~\cite{ju2025collaborating} provide rich testbeds for a fixed domain-specific task but have limited \textbf{structural adaptability} for research questions that require different coordination structures or staged procedures.
Instead, ECL and the controller in our platform allow researchers to reconfigure objects, action schemas, and views to implement both classic and new experiments. 
We illustrate this by conducting evaluation experiments in Shape Factory and Hidden Profile, and adapting DayTrader, Essay Ranking, and Passcode with only changes to configuration and view binding, as shown in Section~\ref{sec:generalizability}.}

\re{In summary, our system combines (i) a declarative configuration model that separates state, policy, and views to support controlled interface manipulations, and (ii) a protocol that gives LLM agents the same perception and action semantics as human participants, with controller-side validation and atomic commits. These design choices enable controlled comparisons in human–agent collaboration that are difficult to realize in general multi-agent frameworks or monolithic behavioral platforms. }

\begin{table*}[!tp]
  \centering
  \caption{Comparison with behavioral and agent frameworks on dimensions relevant to human–agent collaboration research. }
  \label{tab:comparison}
  \Description{Table 1. This table shows a systematic comparison with behavioral and agent frameworks on dimensions relevant to controlled human–agent collaboration research. This table compares the authors' platform with AutoGen, oTree/Empirica, overcookedAI/CREW in different feature dimensions.}
  \resizebox{\textwidth}{!}{
  \begin{tabular}{l|llll}
    \toprule
    \textbf{Feature Dimension} &
    \textbf{Our Platform} &
    \textbf{AutoGen}~\cite{wu2024autogen} &
    \textbf{oTree / Empirica}~\cite{chen2016otree,almaatouq2021empirica} &
    \textbf{Overcooked-AI / CREW}~\cite{carroll2019utility, zhang2024crew} \\
    
    \midrule
    \textbf{Primary Focus} &
    \begin{tabular}[c]{@{}l@{}}Controlled human–agent\\ Collaboration experiments\end{tabular} &
    \begin{tabular}[c]{@{}l@{}}Multi-agent task pipelines \end{tabular} &
    \begin{tabular}[c]{@{}l@{}}Behavioral experiments\\ with humans\end{tabular} &
    \begin{tabular}[c]{@{}l@{}}Coordination in a fixed game\end{tabular} \\
    
    \midrule
    \textbf{Configuration Model} &
    \begin{tabular}[c]{@{}l@{}}Declarative ECL\\ (objects/actions/policies/views)\end{tabular} &
    \begin{tabular}[c]{@{}l@{}}Programmatic\\ orchestration\end{tabular} &
    \begin{tabular}[c]{@{}l@{}}Programmatic study \\scripts\end{tabular} &
    \begin{tabular}[c]{@{}l@{}}Fixed environment / Gym-style API\end{tabular} \\
    
    \midrule
    \textbf{Agent Integration} &
    \begin{tabular}[c]{@{}l@{}}Standardized protocol\\ (ACP) for parity\end{tabular} &
    \begin{tabular}[c]{@{}l@{}}Native or ad-hoc message \\passing\end{tabular} &
    \begin{tabular}[c]{@{}l@{}}No native agent \\semantics\end{tabular} &
    \begin{tabular}[c]{@{}l@{}}Built-in agents or RL policies\end{tabular} \\
    \midrule
    \textbf{Experimental Control} &
    \begin{tabular}[c]{@{}l@{}}UI/media toggles, timing,\\ blinding, variable exposure\end{tabular} &
    \begin{tabular}[c]{@{}l@{}}Low; optimized for\\ completion pipelines\end{tabular} &
    \begin{tabular}[c]{@{}l@{}}Randomization/payoffs;\\ no agent-facing controls\end{tabular} &
    \begin{tabular}[c]{@{}l@{}}Fixed rules; limited factor control\end{tabular} \\
    \midrule
    \textbf{Architecture Pattern} &
    \begin{tabular}[c]{@{}l@{}}Decoupled controller/view;\\ state–view separation\end{tabular} &
    \begin{tabular}[c]{@{}l@{}}Tightly coupled \\ orchestration and execution\end{tabular} &
    \begin{tabular}[c]{@{}l@{}}Monolithic web\\ framework\end{tabular} &
    \begin{tabular}[c]{@{}l@{}}Game engine simulation\end{tabular} \\

    \midrule
    \textbf{Example Experiment Fit} &
    \begin{tabular}[c]{@{}l@{}}Shape Factory, Passcode, \\ Hidden Profile variants\end{tabular} &
    \begin{tabular}[c]{@{}l@{}}Code generation, tool use, \\ web tasks\end{tabular} &
    \begin{tabular}[c]{@{}l@{}}Public-goods, auctions\\ dictator games\end{tabular} &
    \begin{tabular}[c]{@{}l@{}}Overcooked variants\end{tabular} \\
    
    \bottomrule
  \end{tabular}}
  
\end{table*}

%% file: sections/3system-v3.tex
\section{System Design}
\label{sec:system}

To address the need for reproducible and extensible research on human-agent collaboration, we designed and implemented an open-source, fully configurable research platform. 
\re{Unlike existing platforms that either focus on human-human experiments (e.g., Empirica~\cite{almaatouq2021empirica}, oTree~\cite{chen2016otree}) or development of autonomous agents (e.g., AutoGen~\cite{wu2024autogen}), our platform serves as a flexible infrastructure that brings LLM agents into the controlled experimental environments traditionally reserved for human collaboration studies.}

The platform's architecture encompasses four \re{key} components: \textbf{Participant Interface} renders the interaction environments for participants, \textbf{Researcher Interface} allows for the configuration of experimental variables, \textbf{Experiment Controller} that serves as the backend engine, and an \textbf{Agent Context Protocol (ACP)}, \re{a novel integration layer that standardizes how agents perceive and act.}


Researchers can adapt our platform for \re{diverse} experimental paradigms in two ways: through a declarative \textbf{Experiment Configuration Language (ECL)} (example provided in Appendix~\ref{app:ecl}) that defines all entities, actions, and interface modules, or by directly extending the open-source codebase. 
\re{By defining entities, actions, and interface modules through ECL rather than ad-hoc code, our platform} supports seamless adaptation of existing paradigms, such as social dilemma games~\cite{bos2002effects} and collaborative decision-making tasks~\cite{zheng2023competent}, and the development of entirely new ones.
Researchers can adjust experiment parameters and manipulate interaction controls for different conditions via the \textbf{researcher interface}.

We address open-science needs by releasing source code and by encouraging the sharing of configuration templates, expecting our platform to support the broader research community in human-agent collaboration research.

\subsection{Design Principles and Rationale}
\label{sec:design-principles}

\re{The platform targets synchronous, interface-mediated collaboration among mixed human-agent teams.}
We articulate \textbf{four} design principles (DP) grounded in HCI and CSCW theory: enabling the replicability of classic paradigms, translating collaboration theory into testable controls, ensuring agent-human parity, and synchronized process logging for post-study analysis.

\subsubsection{\re{DP1. Generality via Abstraction.} }
\re{Existing research platforms, as shown in Section~\ref{sec:rw-comparison}, are often designed or hard-coded for specific tasks (e.g., text-based conversation only), which largely limits their generality.
To support a broad range of collaboration research, our research platform must be easily adaptable across different experimental paradigms without requiring redesign or re-implementation of the core architecture. 
To address this challenge, our first principle focuses on achieving generality through both logic and interface abstractions to ensure its paradigm adaptability. }

\re{For \textbf{logic abstraction}, the platform must abstract the fundamental interaction mechanics that recur across paradigms (e.g., resource distribution, turn-taking, and information asymmetry) away from the specific experimental content.
The \textbf{Experiment Configuration Language (ECL)} allows researchers to either reuse existing classes or instantiate new classes, define custom attributes, and bind them to experiment-specific artifacts and visual elements, correspondingly.
For \textbf{interface abstraction}, we deconstruct the participant interface into visual modules (e.g., the Social Module) that can be customized independently.
As a result, researchers can seamlessly replicate classic experiment paradigms or rapidly prototype entirely new human-agent collaboration tasks by altering the interface design, interaction flows, and available information for each participant to fit specific experimental needs.}

\subsubsection{\re{DP2: Translate Collaboration Theories into Testable Controls} }
\re{Decades of HCI/CSCW research have produced profound knowledge on human collaboration; nevertheless, these high-level findings have not been tested in the emerging context of human-agent collaboration. 
For instance, prior research shows that variables such as workspace awareness, media richness, and social signaling fundamentally shape team dynamics.
In principle, these high-level collaboration theories have to be \textbf{translated} into concrete, manipulable variables or feature designs for empirical testing. 
Thus, our second principle is to turn these theoretical findings for collaboration into modular and configurable \textbf{Interaction Controls}. }

\re{Specifically, the platform's research interface includes controls grounded in established collaboration theories, enabling researchers to toggle, restrict, or customize the corresponding visual modules (e.g., private inventory, shared dashboards).
Researchers can thus isolate independent variables (e.g., customizing the "Social Dashboard" to test grounding constraints) and measure their specific effects on human–agent collaboration.
This principle ensures that the platform’s configurations are meticulously designed to serve as valid independent variables for hypothesis-driven research.}

\subsubsection{\re{DP3: Reproducibility via Standardized Human-Agent Parity} }
\re{A critical consideration in emerging human-agent collaboration is how to ensure the agents are integrated consistently.
Specifically, the system must enforce that agents operate under the same perceptual and temporal constraints as remote human collaborators to ensure valid comparisons between human-human and human-agent teams.
Besides, agent integration must follow a strict schema regardless of the underlying model architecture to ensure the validity and reproducibility of experiment results.
Our third design principle directly addresses this concern by enforcing standardized agent integration into the platform by leveraging the \textbf{Agent Context Protocol (ACP)}.
The ACP filters system states into a human-analogous context window, standardizes agents' access to the environment, action spaces, and communication channels, and ensures that experimental results are attributable to the agent's social and cognitive behavior rather than unequal system access.}

\subsubsection{\re{DP4: Synchronized Process Logging} }
\re{Investigating human-agent collaboration dynamics requires analyzing beyond the collaboration outcomes; instead, we should follow established research on remote human collaboration to emphasize the \textbf{process} of interactions.
In human-agent collaboration, the process data is multimodal, encompassing observable human and agent behaviors as well as latent agent reasoning and planning trajectories.
The fourth principle tackles this requirement by enforcing synchronized process logging.
Specifically, the experiment controller aligns interface events, human messages and actions, and agent prompts and tool calls on a single, synchronized timeline. 
The process logging synchronization allows researchers to tie outcome differences to specific coordination moves, breakdowns, or repairs, which can support granular mixed-methods analysis of when an agent's reasoning is misaligned with the situation or their human collaborators' actions. }

\begin{figure*}[!htp]
    \centering
    \includegraphics[width=.98\linewidth]{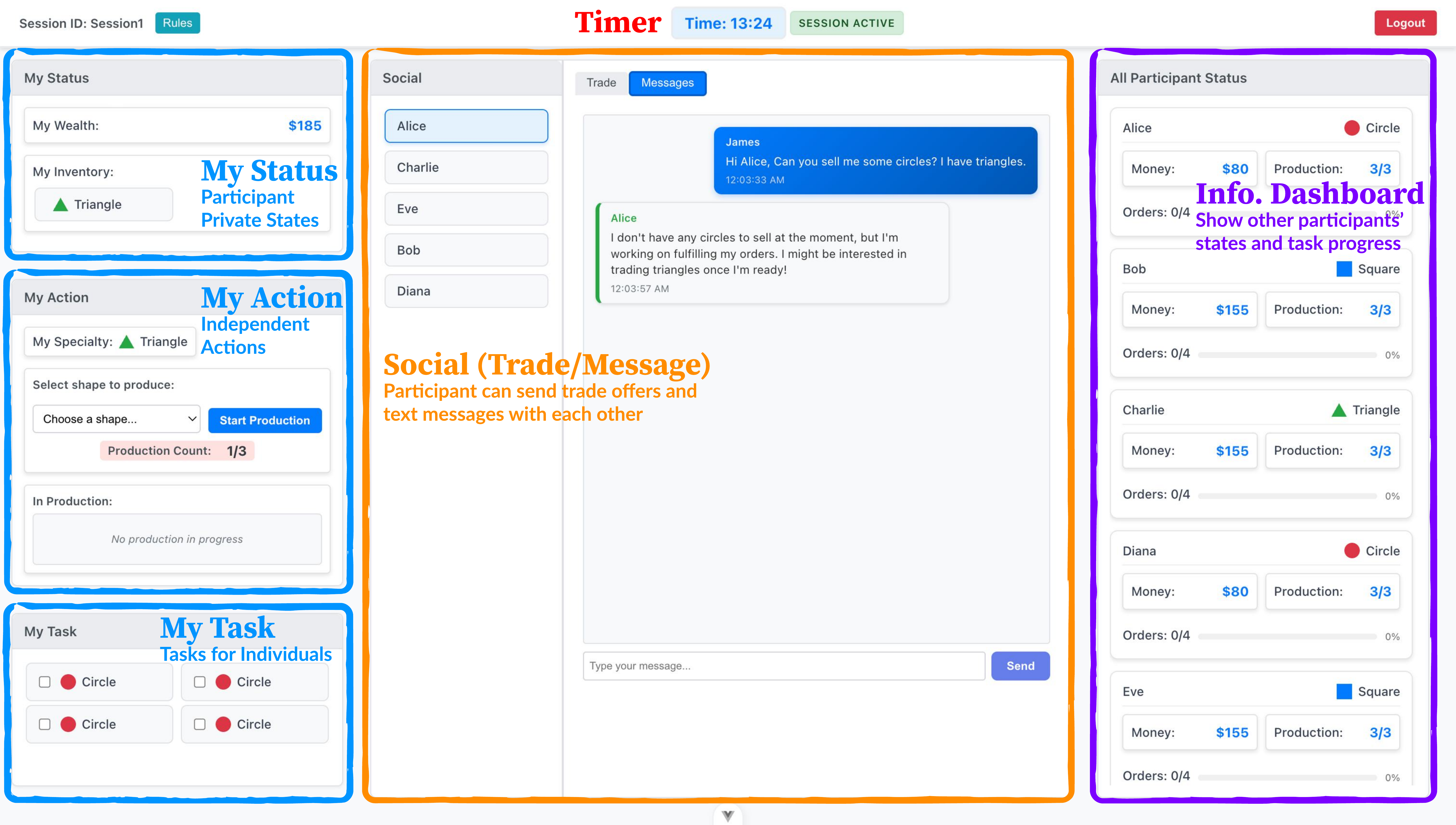}
    \caption{An example of the participant interface for the Shape Factory experiment~\cite{bos2004group}. The interface consists of five fully configurable modules: \textbf{My Status} displays the participant's private status, \textbf{My Actions} displays permitted actions, \textbf{My Tasks} displays the tasks to be completed, \textbf{Social} provides available interaction channels, and \textbf{Information Dashboard} shows other participants' real-time information. The experiment's timer is shown at the top. }
    \label{fig:participantUI}
    \Description{Fig. 3. This figure shows the participant interface of the Shape Factory experiment. The interface is divided into five modules. "My Status" shows each participant's private resources and condition. "My Actions" lists the actions the participant can take. "My Tasks" displays current goals or assignments. "Social" provides communication options with other participants. "Information Dashboard" shows other participants' real-time status. A session timer is positioned at the top.}
\end{figure*}

\subsection{Architecture Overview}
\label{sec:system-architecture}

The platform architecture follows a client–server model (Figure~\ref{fig:architecture}). 
Participant interfaces, ACP, and the researcher interface connect to an experiment controller over a real-time channel. 
In particular, ACP ensures that agents can periodically perceive the experiment states and interact with the experiment environment consistently.
The controller manages experiment state, validates and commits actions, and synchronizes all logs with the data storage.
Here, we present design and implementation details of each component.

\subsubsection{Participant Interface}
\label{sec:system-components-participant-ui}

The participant interface uses a three-column layout that can generalize across common experimental paradigms by separating visual modules for private state, social interaction, task operations, and shared awareness.
An example participant interface for the Shape Factory experiment, which is used in the case study in Section~\ref{sec: case_study1}, is shown in Figure~\ref{fig:participantUI}.

The \textbf{left column} presents private state and individual actions.
The \textit{My Status} module shows individual resources and role-dependent capacities that are not necessarily visible to others. 
The \emph{My Action} module provides affordances for private actions, such as producing a resource or making an independent decision.
The \emph{My Task} module displays individual or collaborative goals and fulfillment status.
The \textbf{right column} is a fully configurable shared \textit{Information Dashboard}.
The dashboard makes workspace awareness~\cite{dourish1992awareness, gutwin2002descriptive} a customizable visual module; researchers can select what information to disclose, to whom, and at what level of granularity.

The \textbf{middle column} contains the \textit{Social} module for all participant interactions.
This is supported by fruitful CSCW research that has underscored the importance of communication and negotiation affordances for studying trust, information grounding, and collaboration effectiveness~\cite{straus1997technology, bos2002effects, hinds2003out}.
Here, a \textit{Trading} page supports resource exchange and can be adaptive to broader collaborative decision-making activities. 
A separate \textit{Chat} page supports text-based communication between participants. 
Researchers can configure trading with free-form or protocol-bounded offers to study negotiation dynamics and coalition formation~\cite{poole1991conflict}.
For the chat functionalities, we implemented three variations: private messages, group messages, or disabled them entirely. 

While this three-column layout serves as a default, the underlying modules are independent components that can be rearranged, enabled, or disabled through the Experiment Configuration Language for different experimental paradigms.

\subsubsection{Researcher Interface}
\label{sec:system-components-researcher-ui}

\begin{figure*}[!t]
    \centering
    \includegraphics[width=.98\linewidth]{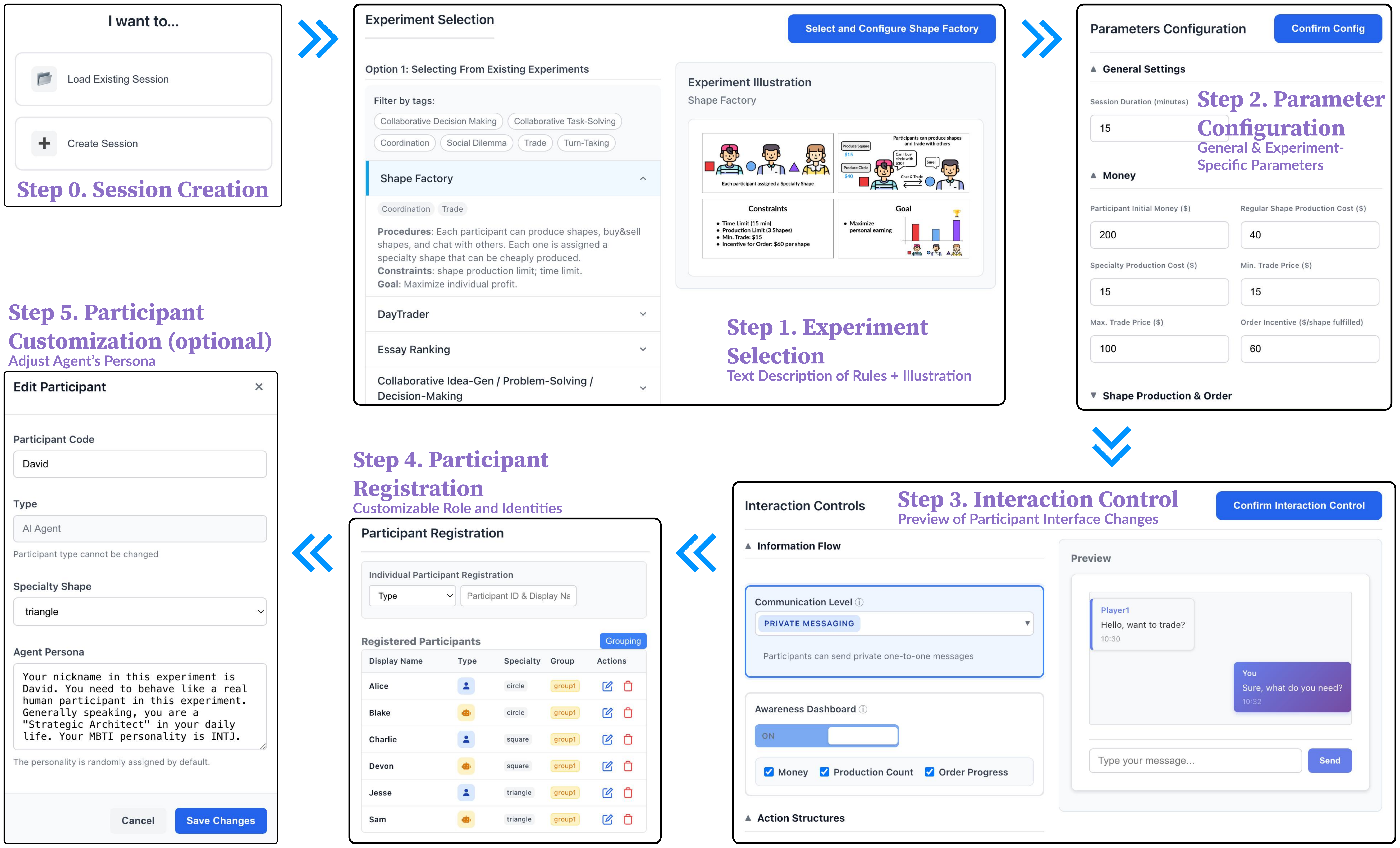}
    \caption{The workflow of HCI researchers for study setup through the researcher interface, which includes five steps: experiment selection, parameter configuration, interaction control, participant registration, and participant customization. }
    \label{fig:researcher-workflow}
    \Description{ig. 4. Workflow for study setup in the researcher interface. The diagram outlines four steps for researchers: selecting an experiment, configuring parameters, controlling interactions, and registering participants.}
\end{figure*}

The researcher interface provides direct controls for researchers to design, execute, monitor, and analyze experiments for human-agent collaboration research.
The final design, as shown in Fig~\ref{fig:researcher-workflow}, was refined through a participatory cognitive walkthrough~\cite{polson1992cognitive} with five HCI researchers by incorporating design and interaction feedback and aligning the workflow with their expectations (details in Section~\ref{sec:cog_walkthrough}).
The researcher interface organizes functionalities into three pages: experiment setup, real-time monitoring, and post-study analysis. 
For experiment setup, the interface guides researchers through \textbf{four} primary steps: experiment selection, parameter configuration, interaction control, and participant registration, as shown in Figure~\ref{fig:researcher-workflow}.
In addition, researchers can customize participants' information, including the persona profile for LLM agents, without editing the code.


The \textbf{Experiment Selection} module lets researchers load a saved session or create a new one. 
After a session is created, it presents a list of experimental paradigms with brief descriptions and simplified illustrations, allowing researchers to quickly understand each paradigm's purpose even without prior familiarity.
Researchers can also upload a custom paradigm defined in an ECL file.

Then, the \textbf{Session Configuration} module offers researchers comprehensive control over (1) experiment parameters (e.g., initial resource allocation and experiment constraints); (2) interaction controls, such as communication channels and information dashboard (details in Section~\ref{sec:interaction-control}); (3) agent design (e.g., response latency); (4) participant registration for both humans and agents.

Once the session is live, the \textbf{Real-time Monitoring} module allows researchers to observe participant interactions and progress. 
After the session is completed, the \textbf{Result Analysis} module provides summary statistics and visualizations of participant performance. 
Researchers can view aggregate or per-participant results and export raw event logs for further analysis.

\paragraph{Researcher Workflow}
An overview of HCI researchers' workflow is illustrated in Figure~\ref{fig:researcher-workflow}.
In the study setup, researchers first decide whether to load an existing or a new session; if the latter, they must choose the experimental paradigms that best suit their research interests. 
The system also enables researchers to upload a customized experiment configuration, using ECL specified in Section~\ref{sec:system-components-ecl}, if the provided ones do not meet their needs.
Then, researchers will proceed to the session configuration module to customize interaction controls and agent-specific designs.
The horizontal scrolling function allows researchers to iteratively adjust parameters in accordance with experimental procedures.
After completing all configurations, researchers conclude the setup by registering the human and agent participants.

\subsubsection{Experiment Configuration Language (ECL)}
\label{sec:system-components-ecl}

The Experiment Configuration Language (ECL) is a declarative language designed to give researchers fine-grained control over experimental paradigms without modifying the platform's source code. 

The ECL is specified in a structured format by defining its components through a set of core primitives: objects, actions, policies, and views.
The language is organized around these four fundamental concepts. 
\textbf{Objects} represent the primary entities within the experiment, such as resources (e.g., money, parts), roles (e.g., buyer, seller), or abstract concepts like team scores. 
Each object is defined by a set of attributes, such as $initial\_value$ or $visibility$.
\textbf{Actions} define the operations available to participants, specifying the actor, associated costs, and their effects on the system's object states. 
\textbf{Policies} are logical constraints that are specified by different experimental paradigms, ranging from preconditions for specific actions (e.g., requiring sufficient resources) to global rules like time limits or winning conditions. 
Finally, \textbf{Views} separates the system's internal state from the presentation layer by defining which objects and attributes are rendered on the participant interface and how. This separation is critical, as it allows researchers to manipulate information visibility without altering the experiment's core logic.

A concrete example of ECL for the Shape Factory is provided in Appendix~\ref{app:ecl}.
In this example, $Money$ and $Shape$ are defined as $objects$.
The $ProduceAction$ links the costs and effects for producing a shape, and is governed by a policy that validates the actor's resources.
The $Views$ declares what objects and states are visible in each visual module. 
During setup, the Experiment Controller parses an ECL file to dynamically instantiate the environment, enforce the specified policies, and render the correct views for participants.
The experiment controller also validates the configurations and reports conflicts when a new configuration template is uploaded.
This approach allows researchers to create, share, and modify experimental paradigm designs in a transparent and reproducible manner.



\subsubsection{Agent Context Protocol}
\label{sec:system-components-acp}

\begin{figure*}[!t]
    \centering
    \includegraphics[width=.9\linewidth]{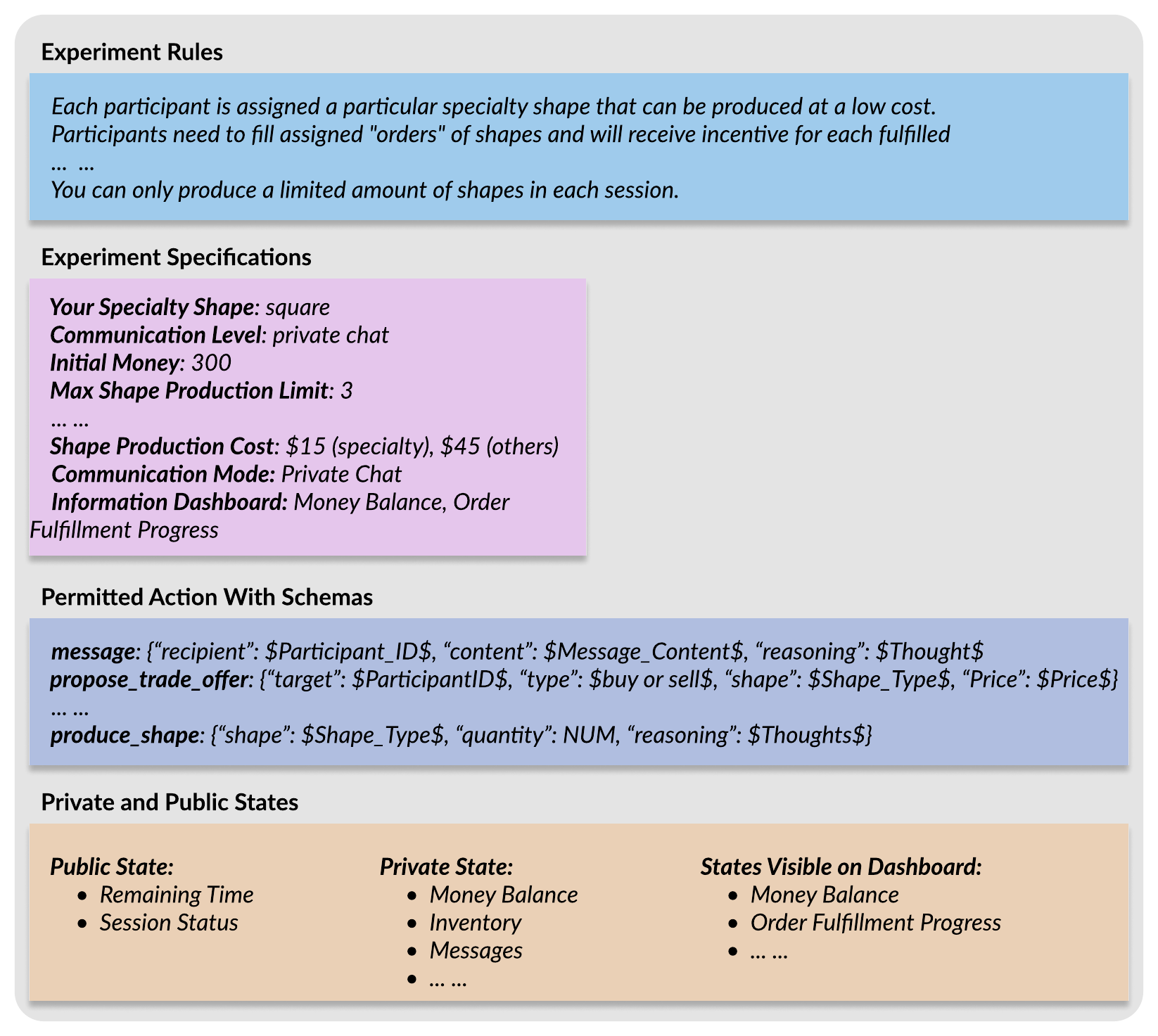}
    \caption{The agent context protocol (ACP) provides a standardized channel between the experiment controller and LLM agents, regardless of agent architecture designs. ACP includes four modules:
    \textbf{Experiment Rules} (task setup and constraints), \textbf{Experiment Specifications} (participant roles and resources), \textbf{Permitted Actions with Schemas} (structured formats for communication and trades), and \textbf{Private and Public States} (information visible to participants and agents). }
    \label{fig:ACP}
    \Description{Fig. 5. Agent Context Protocol (ACP). This schematic depicts ACP as a unified communication channel that connects the experiment controller with large language model agents, ensuring standardized integration regardless of the agent's internal design.}
\end{figure*}

Inspired by the emerging standard of model context protocol\footnote{https://modelcontextprotocol.io/} that connects LLMs with external tools and resources, we designed the ACP to standardize communication between LLM agents and the experiment environment (Figure~\ref{fig:ACP}).
ACP aims to ensure agents' behaviors align with the experimental rules and valid action spaces, regardless of agent architecture.
More importantly, agents perceive and act under the same constraints and through the same structured channels as remote human participants, ensuring parity for controlled, comparative studies.

At initialization, the experiment controller (Section~\ref{sec:system-components-controller}) constructs an experiment-specific context for agents from the experiment configurations and the interaction conditions specified by researchers.
The context includes participant IDs, groups, private and public states, permitted actions with schemas, communication channels, perception interval, and experiment rules. 
The perception interval determines how often the controller sends state summaries to agents. The controller also validates that each summary is consistent with the pre-defined private and public states for the agent.

During a session, each state summary from the controller includes the current timestamp and remaining session time.
Agents are expected to take the state summary as input and generate a response of actions, sometimes with rationales if required.
The controller validates responses against the schemas and policies, returns an error message to the agent if an action was identified as invalid, and asks the agent to re-generate the response. 
Once the actions are validated, the controller commits these actions atomically, updates the environment state, and logs the whole interaction.


\subsubsection{Experiment Controller}
\label{sec:system-components-controller}

The experiment controller is the execution engine of the platform.
During experiment setup, the controller creates a session instance, compiles experiment configurations, binds variables to researcher-specified conditions, and renders the user interfaces.
During a live session, the controller manages timing, validates participant actions and interactions, updates private and shared states, and records behavioral logs for post-hoc analysis.
If an LLM agent is registered, the controller instantiates an agent context, as described above. 
Meanwhile, human participants can log in to the participant interface using their session ID and participant ID. 
The controller exhaustively tracks all actions, including trades, messages, resource production and allocation, order submissions, and agent inputs and outputs.

\re{To safeguard participant data, the controller stores logs locally by default, avoiding third-party cloud retention. 
To prevent model 'jailbreaks' or toxic output, the ACP enforces strict schema validation on all agent outputs before rendering them to humans, which can effectively sandbox the agent's communicative capabilities.}

\subsection{Configurable Interaction Control for Human-Agent Collaboration}
\label{sec:interaction-control}

We implement four layers of configurable interaction variables that connect findings in classic collaboration research to interaction controls of the research platform.
Among these controls, some lead to interface design variations and information visibility to participants, while others may impact the experiment constraints and agent behaviors.
Through the researcher interface, users can configure controlled comparative experiments to analyze how different interaction controls impact collaboration dynamics.
Users can also define and customize the controls through ECL.

\paragraph{Information Flow}
This layer focuses on the type and visibility of information that is accessible to human participants and LLM agents.
Prior research establishes that media characteristics shape mutual understanding~\cite{clark1991grounding,straus1997technology}, awareness of others’ actions supports coordination~\cite{gutwin2002descriptive,dourish1992awareness}, and information control affects strategic behavior in distributed teams~\cite{cramton2001mutual,olson2000distance}.
We implement two primary controls with respect to these conceptual findings.
The \textbf{Information Dashboard} can be enabled or disabled to show real-time, shared information to all the participants, such as each other's trades, recent orders, inventories, and order fulfillment status. 
Researchers can further customize information visibility based on participant groups (e.g., all, human-only, agent-only, etc.), change the information update frequency, and adjust the granularity of information (i.e., real values versus text summary).
The \textbf{Communication Transparency} can be configured for strictly private (one-to-one), group chat, or disabled entirely.
This lets researchers examine how information sharing and message visibility influence negotiation dynamics and group behavior, and may further customize this control with additional turn-taking and message length.


\paragraph{Action Structure}
This layer defines the types of actions available, how they can be conducted, and the operational constraints that shape interdependence. 
Prior work links action coupling, protocols, and constraints to coordination cost, convergence, and coalition dynamics~\cite{olson2000distance, bos2004group}.
The platform provides control over the \textbf{Negotiation Protocol}, which allows either strict accept-or-reject protocols or open-ended negotiation with counter-offers. Researchers can also set constraints such as price limits or trade frequency.
Researchers can also manage \textbf{Concurrent Actions} by either allowing participants to make multiple offers simultaneously, or participants must resolve one before making another.


\paragraph{Social Framing}
This layer captures how identity, roles, and goals are presented to participants. 
Such framing is known to shape trust, subgroup formation, and role negotiation in distributed teams~\cite{bos2004group,hinds2003out,straus1994does}.
Researchers can manipulate \textbf{Agent Identity} by customizing the display names of LLM agents (e.g., "teammate," "assistant") and controlling the visibility of their personas or group affiliations. 
The platform also supports configurable \textbf{Social Cues}, such as color schemes or group identifiers, to explore the effects of in-group and out-group framing on collaboration.


\paragraph{Agent Responsiveness}
This layer controls the timing and feedback mechanisms of LLM agents. 
Timing of responses can affect perceived competence and availability in remote collaboration~\cite{olson2000distance}, while adaptive communication and clear feedback help users recover from errors and calibrate reliance~\cite{amershi2019guidelines}.
Key controls include \textbf{Agent Response Latency}, which can be set to be immediate or include a fixed or variable delay. Researchers can also customize a "typing" or "thinking" indicator.
\textbf{Agent Adaptive Feedback} configures whether LLM agents use static response styles or dynamically adapt their strategy based on interaction history, specified through their persona profile.
Finally, researchers have control over agents' \textbf{Explanations and Error Handling} by deciding whether agents provide rationales proactively or on demand.
This control is extensible to support advanced agent reasoning architectures, such as the fast-and-slow thinking systems used in recent agent research~\cite{lu2025uxagent}.

\subsection{Adaptivity Across Classic Experiments}
\label{sec:generalizability}

The combination of the Experiment Configuration Language and the layered interaction controls allows the platform to be rapidly adapted for re-implementing classic experimental paradigms or creating novel ones. 
While our case studies in Section~\ref{sec: case_study1} provide an in-depth demonstration of a Shape Factory re-implementation~\cite{bos2004group}, we describe here how the platform's modularity supports several other paradigms from collaboration research.
The participant interfaces corresponding to each paradigm are shown in Fig~\ref{fig:adaptation}.


One such paradigm is DayTrader~\cite{bos2002effects}, a social dilemma experiment that analyzes how communication media affect trust. 
In this task, participants make decisions about individual versus group investments across multiple rounds.
To configure this experiment, the \textit{Trading} page and \textit{My Task} module are disabled, while the \textit{My Action} module is adapted to support investment decisions. 
The \textit{Information Dashboard} displays group contributions and individual income, and group chat in the \textit{Chat} page is enabled for discussion.

Another example is Essay Ranking~\cite{zheng2023competent}, a collaborative decision-making task originally conducted via a Wizard-of-Oz setup. 
Participants first review and rank essays independently, then discuss their rankings as a group before submitting a final independent ranking. 
Our platform supports this workflow by adapting the \textit{My Status} module to display the essays and using the \textit{My Action} module for the ranking inputs. 
The group discussion phase is enabled by allowing group chat in the \textit{Chat} page.

Beyond these specific examples, the platform's core components are sufficiently generalizable to support a broad class of collaborative idea-generation and problem-solving tasks~\cite{straus1994does}. 
Such tasks, which range from brainstorming to solving logic problems, primarily depend on robust communication channels and the ability to present shared materials, both of which are native features of our platform. 
Similarly, two-player, turn-based paradigms like the "Passcode" word-guessing game for studying humans' mental models of AI~\cite{gero2020mental} can be implemented by configuring the action structure for sequential, one-word message exchanges between two participants. 

%% file: sections/4-evaluations.tex
\section{Evaluation of Our Platform}

We conducted a \re{multi-phase} evaluation to \re{assess the platform's \textit{research efficacy} as an instrument for controlled studies, its \textit{extensibility} across collaboration paradigms, and its \textit{usability} for researchers.}
Specifically, we conducted \textbf{two controlled experiments} with human participants to \re{examine} the platform's ability to effectively manipulate theory-driven \re{interaction controls} and test whether the manipulations yield measurable changes in outcomes and perceptions.
\re{The first study utilized the \textbf{Shape Factory}~\cite{bos2004group} paradigm, a resource-negotiation task, with the manipulation of the \textit{Information Flow} controls (Section~\ref{sec: case_study1}).}
\re{The second study utilized the \textbf{Hidden Profile}~\cite{stasser1985pooling, stasser1987effects} paradigm, a collaborative problem-solving task, by manipulating the \textit{Agent Initiative} controls (Section~\ref{sec: case_study2}).}
\re{Finally, we conducted a \textbf{participatory cognitive walkthrough} with HCI researchers of diverse experience to examine the researcher workflow and interface design (Section~\ref{sec:cog_walkthrough}). }

\subsection{Case Studies with Shape Factory Experiment}
\label{sec: case_study1}

We selected the Shape Factory paradigm as the \re{first} testbed for our case studies because it provides a versatile environment for examining complex collaborative dynamics.
It involves open-ended negotiation, resource exchange, and multi-player communication under time pressure, and its core mechanics connect to numerous other classic CSCW experiments that isolate specific variables like trust~\cite{bos2002effects} or role interdependence~\cite{mcneese2005neocities, schelble2022let}. 
While our implementation adheres to the original experiment's core rules, incentives, and resource scarcity design, our goal was not to replicate the original research topic and findings, but to validate our platform's utility for human-agent collaboration research.

\subsubsection{Experimental Protocol}

Our case study adheres to the same experiment rules as the original one.
The core task in the Shape Factory experiment for each participant is to maximize their wealth by producing and trading shapes to complete shape orders.
At the start of a session, each participant is assigned a "specialty" shape that they can produce at a lower cost compared with other shapes. 
Each participant is given an order list of shapes without their specialty to complete and earn revenue.
Participants can perform two primary actions: \textbf{produce} shapes in their factory and \textbf{trade} shapes using money with other participants.
Each participant has access to a computer interface that allows them to send text messages and trade offers to other participants.

The central tension of the experiment lies in the interdependence, where \re{each participant's} shape orders do not consist of their own specialty shape, and there is a limit on the number of shapes they can produce.
As a result, participants must balance cooperation (to acquire the shapes they need) with competition (to negotiate favorable prices). 
Because each specialty shape is assigned to more than one participant, they can choose trading partners, which supports dynamic market behaviors and strategic planning.



\subsubsection{Re-Implementation of the Shape Factory Experiment}

In our study, we downsize the original 10-participant setup to six participants: one human and five LLM agents, where the agents can be viewed as analogous to remote human collaborators.
The down-sampling of participant size requires adjusting several interrelated experiment parameters (e.g., the shape production limits and the shape types), but we preserved the core strategic tensions.

The experiment was implemented using our platform's core features.
We leverage Experiment Configuration Language to create experiment-specific objects, such as \textit{Shape}, with attributes of \textit{Type}, \textit{Regular Cost}, \textit{Specialty Cost}, \textit{Time Cost}, and \textit{Production Status}, and to extend existing objects, such as \textit{Participant}, with attributes like \textit{Specialty Shape}, \textit{Inventory}, and \textit{Order}.
We also used ECL to configure the participant interface by enabling all five visual modules and customizing their content to match the experiment terminology. The controller automatically manages shape production and trades by deducting money from the corresponding participants and returning completed shapes to their inventories.

The final user interface of the Shape Factory experiment is shown in Figure~\ref{fig:participantUI}.
The personal wealth and shape inventory are shown in \textit{My Status}.
\textit{My Factory} indicates the specialty shape, costs, and enables the production capability.
\textit{My Task} module displays the order list of shapes, where the participants can click the checkbox to fulfill the order and receive the incentive. 
In the middle, the \textit{Social} module supports trading and private messages.
The real-time \textit{Information Dashboard} on the right displays the wealth, specialty shape assignment, and order progress of other participants.
The dashboard is enabled by default in control groups.

We implemented the five agents using GPT-4o with a single end-to-end thinking inquiry, and integrated them using the \textbf{Agent Context Protocol} with the experiment controller.
The ACP delivers structured state summaries to each agent at a fixed time interval to trigger its thinking query.
We decided on 15 seconds as a reasonably balanced interval through pilot testing.
Each state summary includes both public information (e.g., time and participant list) and agent-specific private states (e.g., messages and trade offers).

The agent's behavior is guided by a structured prompt (Appendix~\ref{app:prompt}) that was iteratively refined through rigorous testing.
Through this process, we addressed abnormal behaviors observed in preliminary tests, such as overly formal language, repetitive messaging, and a lack of contextual memory in conversations, which ensured the agents' behaviors are reasonable and believable for the study.

\subsubsection{Case Study Setup}

\begin{table}[t]
    \centering
    \small
    \begin{tabularx}{\columnwidth}{c|ccc}
    \toprule 
       ID  &  Age Group & Education Background & Computer Proficiency \\
    \midrule
        C1 & 25-34 & Master's & Professional\\
        C2 & 25-34 & Master's & Professional\\
        C3 & 25-34 & Master's & Professional\\
        C4 & 25-34 & Master's & Intermediate\\
        C5 & 25-34 & Master's & Intermediate\\
        C6 & 25-34 & Master's & Professional\\
        C7 & 25-34 & Master's & Professional\\
        C8 & 25-34 & Master's & Intermediate\\
        C9 & 18-24 & Bachelor's & Professional\\
        C10 & 18-24 & Master's & Intermediate\\
        C11 & 25-34 & Master's & Professional\\
        C12 & 18-24 & Master's & Professional\\
        C13 & 25-34 & Master's & Professional\\
        C14 & 25-34 & Master's & Professional\\
        C15 & 18-24 & Bachelor's & Professional\\
        C16 & 25-34 & Master's & Professional\\
    \bottomrule
    \end{tabularx}
    \caption{Demographic information of case study participants.}
    \label{tab:cs_demographic}
    \Description{Table 2. Demographic information of participants in the case study. This table summarizes the background information of participants, including their identifiers and roles in the study.}
\end{table}

Each case study compares two conditions by manipulating one interaction control.

\paragraph{$CS_{CL}$: Communication Level}
The first study investigated the impact of communication modality. 
In the \textbf{Control} condition, participants could communicate via private text chat.
In the \textbf{Experimental} condition, this functionality was disabled; thus, participants must rely on trade offers and the information dashboard.
Based on prior work on media richness, we hypothesize that the experimental group will demonstrate reduced reciprocity, higher perceived workload, and lower social presence and trust~\cite{bos2002effects,clark1991grounding}.

\paragraph{$CS_{AL}$: Awareness Level}
This study examined how different levels of workspace awareness impact collaborative outcomes.
In the \textbf{Control} condition, participants had access to the real-time information dashboard.
In the \textbf{Experimental} condition, this module was disabled.
As suggested by workspace awareness theory~\cite{gutwin2002descriptive}, we hypothesize that the experimental group will lead to less efficient coordination, lower performance, and lower perceived awareness. 

\paragraph{Study Design and Participant Recruitment}

We employed a crossed, between-subject study design with Communication Level and Awareness Level as the two factors.
16 participants were recruited from social media and professional networks.
Each participant participates in both case studies and is randomly assigned to one condition in each.
Each case study consists of two consecutive sessions.

To control for potential order effects, we applied a counterbalancing scheme in which the sequence of condition pairings was systematically rotated among participants.
Regardless of which condition comes first, the first session conducted by each participant is set to 15 minutes, whereas the rest are set to 10 minutes.
Across the full participant pool, this setup resulted in equal representation of all condition combinations and orders, with each of the four conditions populated by eight participants.
All study procedures were approved by the first author's Institutional Review Board.

We pre-configured each session with a unique session ID, then provided the instructions to each participant and asked them to complete all four sessions in the provided order; they are also required to fill out a post-study survey every two sessions, and a final survey with respect to their perception of agents' behaviors after completing all four sessions.

\subsubsection{Measures and Instruments}
\label{sec:eval-study1-measure}

We collect both behavioral logs in the interface, where the experiment controller logged every user and agent action with millisecond precision, including final wealth, acceptance ratio, average amount of messages per successful trade, message counts, message length, and response latency.
Participants are asked to complete a survey after finishing two sessions of the same case study, which is composed of \textbf{five} instruments to measure critical aspects for distributed collaboration: perceived trust, workspace awareness, collaboration effectiveness, social presence, and workload. 
Specifically, the questionnaire (details shown in Appendix~\ref{app:questionnaire}) consists of \textbf{33} questions, including \textbf{six} questions identified from The organizational Trust Inventory~\cite{cummings1996organizational} that are relevant to our study, \textbf{nine} questions adapted from the theoretical framework of workspace awareness~\cite{gutwin2002descriptive}, \textbf{10} questions from the Shared Mental Model Scale~\cite{van2022five}, \textbf{two} questions from Social Presence Scale~\cite{kreijns2011measuring}, and the complete six-item NASA-TLX scale~\cite{hart1988development}. 
All scales were created on Qualtrics and administered in their original formats (Likert for the first four instruments and a slide bar for NASA-TLX); item wording was unchanged except for minor phrasing adjustments, such as to refer to “participants in our experiment” rather than “teammates” where necessary.

After finishing the study, we analyze each case separately.
To compare outcomes between control and experimental groups, we used independent-sample t-tests \cite{student1908probable}. 
To examine longitudinal effects, we conducted ordinary least squares (OLS) regression analyses \cite{wooldridge2010econometric} with session order as a predictor. 
We further calculated Pearson correlations \cite{pearson1895vii} between wealth, trade-related measures, and communication variables to explore behavioral relationships.
In addition, we quantitatively calculate the survey results using the original scoring methodologies for each instrument.

\subsubsection{Findings}

\begin{figure*}[htbp]
    \centering
    \includegraphics[width=.98\linewidth]{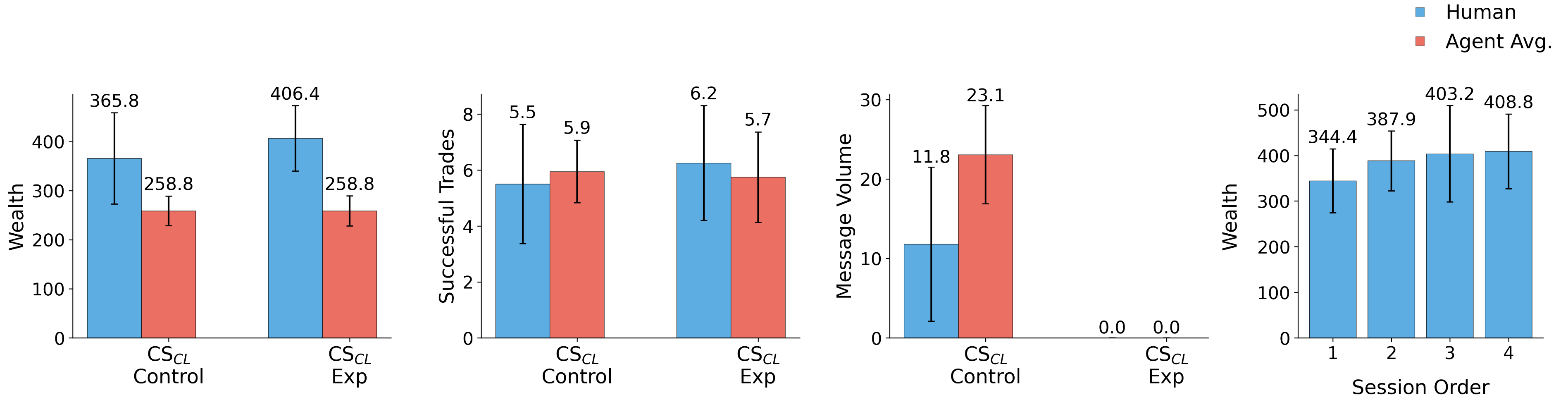}
    \caption{Bar charts of collaboration outcomes in the $CS_{CL}$ case study, showing (left to right) average wealth, successful trades, message volume, and human participants’ wealth across session order.}
    \label{fig:wealth_charts_cl}
    \Description{Fig. 6. Collaboration outcomes in the CS_CL case study. A set of four bar charts shows session results in order from left to right: average wealth earned by participants, the number of successful trades completed, the total message volume exchanged, and the wealth distribution for human participants.}
\end{figure*}

\begin{figure*}[htbp]
    \centering
    \includegraphics[width=.98\linewidth]{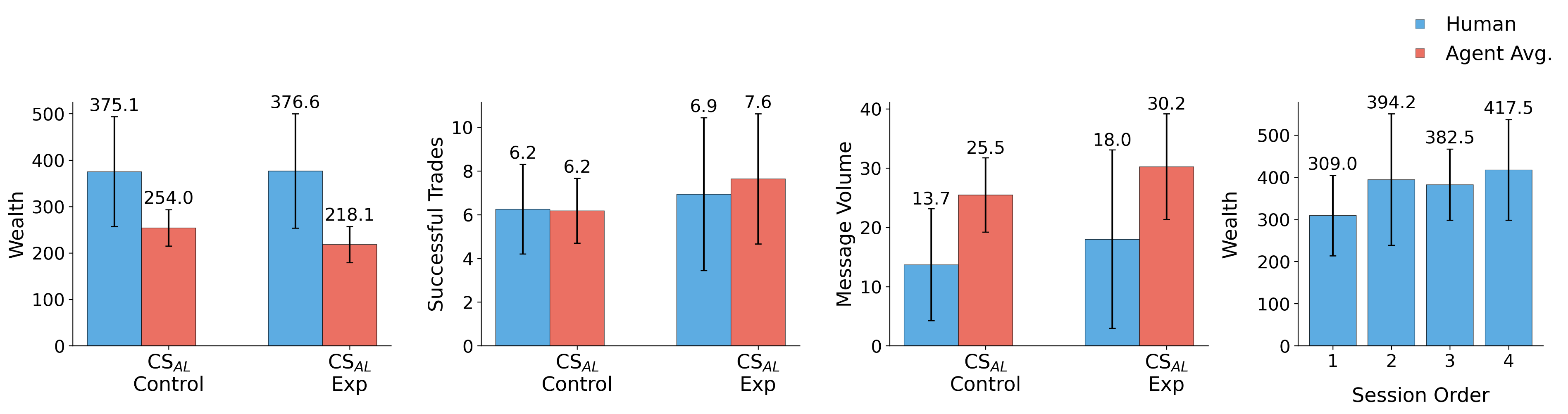}
    \caption{Bar charts of collaboration outcomes in the $CS_{AL}$ case study, showing (left to right) average wealth, successful trades, message volume, and human participants’ wealth across session order.}
    \label{fig:wealth_charts_al}
    \Description{Fig. 7. Collaboration outcomes in the CS_AL case study. Four bar charts illustrate outcomes across session order: participants' average wealth, number of successful trades, message volume, and human participants' final wealth.}
\end{figure*}



\re{Our analysis indicates the platform’s interaction controls isolate a communication variable and yield significant, measurable differences in both behavioral outcomes and subjective perception with the Shape Factory experiment.}

In \re{the Communication-Level study ($CS_{CL}$), when private chat was disabled (experimental group), human participants earned higher final wealth than when chat was enabled ($ M_{control}=\$365.8, \allowbreak SD=92.9; 
M_{experiment}=\$406.4, SD=66.7$), with a mean difference of $40.6$.}
\re{The comparison is shown in the first chart in Figure~\ref{fig:wealth_charts_cl}.}
Behavioral logs show a higher count of successful trades in the experimental group, suggesting that \re{the platform effectively constrained the socio-emotional exchange through the communication channel and shifted participants' behaviors from protracted negotiation to rapid, market-based signaling.}
\re{The demonstrated trade-off between communication overhead and task focus opens up opportunities for researchers to analyze communication channel capacities in human-agent collaboration with our platform.}

In contrast, \re{the Awareness Level study ($CS_{AL}$) demonstrated the platform's ability to reveal asymmetries between human and agent adaptability.}
\re{The manipulation workspace awareness support through \textit{Information Dashboard} produced very similar human outcomes across conditions ($M_{control}=\$375.1, SD=118.3; \allowbreak M_{experimental}=\$376.6, SD=123.0$).}
Agent outcomes, by contrast, received a considerable negative impact in the Experimental group, with average agent wealth decreasing from \$254.0 (SD=39.27) to \$218.1 (SD=38.80).
\re{This result critically demonstrated that the platform can identify differences between humans and agents about collaborative effectiveness. 
Specifically, reduced workspace awareness shared among collaborators hinders agent collaboration more than human collaboration, which motivates future research to conduct comparative evaluations of alternative agent architectures.}



\re{Furthermore, our platform demonstrated robustness for longitudinal analysis as} we observed consistent outcome improvement over time across both studies.   
As shown in the fourth charts in Figure \ref{fig:wealth_charts_cl} and \ref{fig:wealth_charts_al}, human participants' final wealth steadily increased from the first to the fourth session, with a regression model confirming session order as a significant predictor (coef=+26.11, p=0.015).
Correlation analysis \re{further} revealed a positive link between final wealth and both the number of successful trades and trade efficiency ($r \approx 0.53$), and a weaker correlation with average trade price ($r \approx 0.33$). 
\re{Together, these patterns indicate that the platform captures learning effects and strategic changes over time, which is essential for analyzing how humans adapt their strategies and alter their behavioral trends as they become more familiar and engaged in the human-agent collaboration dynamic.}

\begin{table*}[htbp]
\centering
\small
\renewcommand{\arraystretch}{0.95}    
\begin{tabularx}{\linewidth}{>{\raggedright\arraybackslash}p{5cm} *{5}{>{\centering\arraybackslash}X}}
\toprule
\textbf{Dimension} & \textbf{Mean} & \textbf{SD} & \textbf{t} & \textbf{p} & \textbf{Order Effect} \\
\midrule
Perceived Trust (Control) & 4.60 \uparrowgreen & 1.03 & 2.36 & \textbf{0.03*} & 0.59 \\
Perceived Trust (Exp.) & 4.31 \uparrowgreen & 1.14 & 1.02 & 0.33 & 0.27 \\
Workplace Awareness (Control) & 4.76 \uparrowgreen & 0.99 & 3.07 & \textbf{0.01**} & 0.77 \\
Workplace Awareness (Exp.) & 4.19 \uparrowgreen & 1.18 & 0.60 & 0.56 & 0.16 \\
Collaboration Effectiveness (Control) & 4.55 \uparrowgreen & 1.05 & 2.09 & \textbf{0.05*} & 0.52 \\
Collaboration Effectiveness (Exp.) & 4.56 \uparrowgreen & 1.59 & 1.33 & 0.21 & 0.35 \\
Social Presence (Control) & 2.78 \uparrowgreen & 0.97 & -0.91 & 0.38 & -0.23 \\
Social Presence (Exp.) & 2.00 \downarrowred & 1.00 & -3.74 & \textbf{0.00**} & -1.00 \\
\textit{Workload} & & & & & \\
\hspace{1em}Mental Demand (Control) & 13.31 \uparrowgreen & 4.51 & 2.05 & 0.06 & 0.51 \\
\hspace{1em}Mental Demand (Exp.) & 12.14 \uparrowgreen & 5.61 & 0.76 & 0.46 & 0.20 \\
\hspace{1em}Physical Demand (Control) & 5.13 \downarrowred & 2.85 & -8.25 & \textbf{0.00**} & -2.06 \\
\hspace{1em}Physical Demand (Exp.) & 5.00 \downarrowred & 4.42 & -5.08 & \textbf{0.00**} & -1.36 \\
\hspace{1em}Temporal Demand (Control) & 10.38 \downarrowred & 5.37 & -0.47 & 0.65 & -0.12 \\
\hspace{1em}Temporal Demand (Exp.) & 11.93 \uparrowgreen & 6.10 & 0.57 & 0.58 & 0.15 \\
\hspace{1em}Performance (Control) & 13.50 \uparrowgreen & 4.80 & 2.08 & 0.06 & 0.52 \\
\hspace{1em}Performance (Exp.) & 15.29 \uparrowgreen & 4.14 & 3.87 & \textbf{0.00**} & 1.04 \\
\hspace{1em}Effort (Control) & 10.19 \downarrowred & 5.31 & -0.61 & 0.55 & -0.15 \\
\hspace{1em}Effort (Exp.) & 12.43 \uparrowgreen & 5.20 & 1.03 & 0.32 & 0.28 \\
\hspace{1em}Frustration (Control) & 5.56 \downarrowred & 4.37 & -4.98 & \textbf{0.00**} & -1.25 \\
\hspace{1em}Frustration (Exp.) & 5.93 \downarrowred & 4.68 & -4.05 & \textbf{0.00**} & -1.08 \\
\midrule
\textit{Anthropomorphism} & & & & & \\
\hspace{1em}Fake--Natural & 2.65 \uparrowgreen & 1.27 & -1.23 & 0.23 & -0.28 \\
\hspace{1em}Machine like--Human like & 2.35 \downarrowred & 1.31 & -2.22 & \textbf{0.04*} & -0.50 \\
\hspace{1em}Unconscious--Conscious & 2.55 \uparrowgreen & 1.23 & -1.63 & 0.12 & -0.37 \\
\hspace{1em}Artificial--Lifelike & 2.15 \downarrowred & 1.04 & -3.66 & \textbf{0.00**} & -0.82 \\
\textit{Perceived Intelligence} & & & & & \\
\hspace{1em}Incompetent--Competent & 2.45 \downarrowred & 1.00 & -2.46 & \textbf{0.02*} & -0.55 \\
\hspace{1em}Ignorant--Knowledgeable & 2.65 \uparrowgreen & 0.81 & -1.93 & 0.07 & -0.43 \\
\hspace{1em}Irresponsible--Responsible & 3.50 \uparrowgreen & 0.89 & 2.52 & \textbf{0.02*} & 0.56 \\
\hspace{1em}Unintelligent--Intelligent & 2.65 \uparrowgreen & 1.09 & -1.44 & 0.17 & -0.32 \\
\hspace{1em}Foolish--Sensible & 2.70 \uparrowgreen & 1.03 & -1.30 & 0.21 & -0.29 \\
\bottomrule
\end{tabularx}
\caption{Paired t-test results for participants' self-reported perceptions \re{in the Shape Factory experiments}. Green arrows (\uparrowgreen) indicate mean values above the midpoint, while red arrows (\downarrowred) indicate those below. Statistically significant $p$-values are bolded, where * denotes $(p \leq 0.05)$, and ** denotes $(p \leq 0.01)$.}
\label{tab: survey_analysis}
\Description{Table 3. Paired t-test results for participants' perceptions of distributed collaboration in Shape Factory experiments. Columns list the measured dimension, mean, standard deviation, t value, p value, and order effect. Dimensions include perceived trust, workplace awareness, collaboration effectiveness, social presence, workload components (mental, physical, temporal demand, performance, effort, frustration), anthropomorphism, and perceived intelligence. Green arrows indicate mean values above the scale midpoint, and red arrows indicate values below it. Statistically significant p values are bolded (p ≤ .05 marked with *, p ≤ .01 with **).}
\end{table*}

\re{The perception measures via self-reported survey, summarized in Table \ref{tab: survey_analysis}, also aligned with the hypothesis about communication and awareness manipulations.}
\re{In control groups, where the platform enabled higher communication and awareness support, participants reported significantly higher levels of \textit{Perceived Trust} ($p=0.03^*$) and \textit{Workspace Awareness} ($p=0.01^{**}$). }
\textit{Usability} and \textit{workload} ratings indicated that the system itself did not impose undue burden, with \textit{Physical Demand} and \textit{Frustration} falling significantly below the neutral point across conditions ($p<0.01^{**}$).



Finally, our experiments provided a baseline for evaluating humans’ perceptions of LLM agents in collaborative tasks, where the participants rated the basic LLM agents as more "machine-like" ($p<0.05$) and "incompetent" ($p<0.05$). 
\re{Such results are aligned with the basic, prompt-based agent design that does not explicitly target critical collaborative capacities in remote teams, such as maintaining common ground and awareness~\cite{clark1991grounding, dourish1992awareness, gutwin2002descriptive}, repairing breakdowns~\cite{schmidt1992taking}, and negotiating tasks~\cite{bos2002effects, bos2004group}. }
\re{Taking together, these findings show that the platform's controls produce expected directional changes in the human-agent collaboration dynamics and support both outcome-level and process-level in-depth analyses.}

\subsection{\re{Case Studies with Hidden Profile Experiment} }
\label{sec: case_study2}

\re{To demonstrate the platform's capacity to support complex collaborative decision-making tasks involving information asymmetry, we conducted a between-subjects study with the Hidden Profile paradigm~\cite{stasser1985pooling, stasser1987effects}. 
Unlike the Shape Factory, which focuses on resource negotiation, the Hidden Profile measures the team's ability to pool distributed knowledge to reach an optimal decision.}

\subsubsection{\re{Experimental Protocol.}} 
\re{The Hidden Profile task is a three-participant collaborative decision-making experiment. 
Each participant is provided with information about all decision alternatives, and one of them is superior to the others (i.e., has more advantages) when full information is available.
However, the advantages of the best alternative are unshared, leading to the shared information supporting a suboptimal alternative.
As a result, participants can identify the optimal alternative only after the unshared information is pooled through discussion and integrated into a revised decision. }

\re{We follow the core structure introduced by \citet{schulz2012achieve}, which asks each participant to role-play a personnel manager in an airline company that is looking to hire a new pilot for long-distance flights.
Among all three candidates (denoted as A, B, and C), candidate C has seven advantages and three disadvantages, whereas candidates A and B each have four advantages and six disadvantages.
In the experiment, the advantages of candidate C, the optimal candidate, are distributed and unshared among participants.}

\subsubsection{\re{Re-Implementation of the Hidden Profile Experiment}}

\re{To allow human participants to read the assigned candidate materials, we implemented a document-reading feature that enables in-interface preview of the assigned PDF files. Since the Hidden Profile experiment primarily requires participants to communicate and exchange information, we configured the participant interface using ECL by enabling the \textit{My Action} module to display voting results, the \textit{My Task} module to present the experiment rules, and setting the communication level to \textit{Group Chat}.
The interfaces for the Hidden Profile experiment are presented in Figure \ref{fig:interface_hf_read}, \ref{fig:interface_hf_vote}, and \ref{fig:interface_hf_chat}.
We implemented the agents using GPT-4o, integrated with ACP to allow agents to perceive the experiment state and execute actions.}

\subsubsection{\re{Manipulation and Procedure.}} 

\re{We manipulated \textit{Agent Responsiveness}, drawing on mixed-initiative interaction principles~\cite{allen1999mixed}.
In the \textbf{passive} condition, each agent only thinks and responds after the human participant sends a message. 
In the \textbf{proactive} condition, each agent automatically executed a perceive–predict cycle every 30 seconds to assess the conversational state, decide whether to engage in the conversation, and determine what message to send. 
The underlying policy and access to materials were identical across conditions.
Prompt instructions, as shown in Appendix~\ref{app:prompt}, are iteratively tested and refined to eliminate faulty and obvious non-humanlike behaviors. }

\re{16 Participants are recruited through Amazon Mechanical Turk (MTurk)~\footnote{\url{https://www.mturk.com/}} and randomly divided into two conditions (8 participants each). 
All participants first review the experiment rules and procedures, then read the assigned candidate materials for three minutes. Afterward, the human participant and two agents submit an initial vote, followed by a 15-minute group discussion and a final vote.}

\subsubsection{\re{Measures.}} 
\re{The primary outcome was decision accuracy at the team level, defined as selecting the target candidate after discussion. 
We also measured human participants' message volume and lengths in both conditions. 
Hidden profile research links accuracy to disclosure and integration of unique evidence, so these outcomes index whether the responsiveness control shapes information pooling in the intended direction.
We analyzed the self-reported surveys using the same method as reported in Section~\ref{sec:eval-study1-measure}.}

In addition, we captured information sharing
by labeling each atomic fact as shared or unique and recording the first mention, the number of mentions, and whether the fact was integrated into an argument for or against a candidate.
We analyzed the time to the first mention of each unique advantage and the proportion of unshared unique facts mentioned at least once. 
To characterize conversational dynamics, we computed the share of agent-initiated turns, response latency distributions, turn-taking equality, question-asking, and summarization rates. 
This follows established coding traditions for hidden profile discussions and separates initiative from simple message volume.

\subsubsection{\re{Findings.}} 


\re{Our analysis shows that the agent responsiveness manipulation led to significant behavioral pattern changes in participants’ decision-making outcomes, in-experiment behaviors, and subjective perceptions in the Hidden Profile experiment.}

\re{In the Proactive Condition, the team-level accuracy reached 16.7\%. By contrast, team-level accuracy of participants in the Passive Condition only achieved 8.3\%. 
Analysis of the discussion logs suggests that proactive agents facilitated meaningful pooling of the unshared information. 
For example, one agent noted that \textit{``Candidate C's interpersonal skills could foster teamwork, but their egocentric nature might create challenges,''} while another added that \textit{``Candidate C's ability to create a positive atmosphere could really enhance teamwork.''} 
These complementary perspectives prompted the human participant to suggest \textit{``I think we need to find out more (details about C).''} 
This pattern indicates that agent-initiated discussion may better support information sharing. }

\re{We acknowledge that both the Proactive and Passive Conditions have relatively low overall team accuracy.
This result is indeed aligned with the classic Hidden Profile study findings~\cite{stasser1985pooling, stasser1987effects}, where humans often engage in biased discussion, as the initial individual preferences and group consensus are misled by the provided materials because the supportive information for Candidate C is unshared.
Regarding behavioral patterns, participants in the Proactive Condition sent an average of 13.1 (SD = 2.6) messages with a mean length of 77.2 (SD = 36.2), whereas participants in the Passive Condition sent fewer (M = 10.75, SD = 3.8) but longer messages ($M_{length}$ = 95.3, SD = 46.7). This pattern suggests that when the agent is passive, humans need to communicate more information per message to progress the discussion, indicating higher communication effort than in the Proactive Condition.}

\re{According to the self-reported survey (see Table \ref{tab: survey_analysis_hf}), participants’ perception measures were also consistent with the agent responsiveness manipulation. 
In the Proactive Condition, participants reported significantly higher \textit{Perceived Trust} ($p = 0.01^{**}$). 
Usability and workload ratings further indicate that the system was easy to use: \textit{Physical Demand} was significantly lower than the neutral point ($p \leq 0.01^{**}$), and \textit{Performance} was significantly higher than the neutral point ($p \leq 0.02^{*}$) across both conditions.}

\re{Collectively, the findings of the case study on Shape Factory and Hidden Profile experiments with different interaction controls demonstrate the platform’s research efficacy of isolating interaction variables to reveal their impact on collaborative processes and outcomes across diverse domains in human-agent collaboration.}

\subsection{Cognitive Walkthrough with HCI Researchers}
\label{sec:cog_walkthrough}

The second part of our evaluation employed a participatory cognitive walkthrough to assess \re{the platform's usability and learning curve for its target users.}
\re{The study focused on the \textit{researcher experience}: specifically, whether HCI researchers with no prior exposure to the system could successfully configure a valid study and interpret its results without extensive training.
Additional information is collected with respect to researchers' potential research interests with the use of this system.}

\subsubsection{Setup}

We recruited five HCI researchers through professional networks and snowball sampling~\cite{goodman1961snowball}.
Participants were all doctoral students, postdoctoral researchers, or research scientists in HCI \re{with diverse levels of experience} in human-AI collaboration research.
Each session was conducted remotely via Zoom, moderated by the first author, and lasted roughly 45 minutes.
The study was video recorded upon acquiring participants' consent at the beginning, and all study procedures were approved by the first author's Institutional Review Board.




\begin{table}[t]
    \centering
    \small
    \begin{tabularx}{\columnwidth}{c c c X c}
        \toprule
        ID & Gender & Age Group & Occupation & \re{Year of Experience} \\
        \midrule
        P1 & Female & 30--35 & Assistant Professor & \re{5+ years} \\
        P2 & Female & 25--30 & Junior PhD Student & \re{3--5 years} \\
        P3 & Male & 30--35 & Applied Scientist & \re{5+ years} \\
        P4 & Male & 25--30 & Postdoc & \re{5+ years} \\
        P5 & Male & 25--30 & Senior PhD Student & \re{3--5 years} \\
        \bottomrule
    \end{tabularx}
    \caption{Participant demographic information for the cognitive walkthrough. All participants are experienced HCI researchers with expertise in human--AI collaboration.}
    \label{tab:demographic}
    \Description{Table 4. Demographic information of participants in the cognitive walkthrough. All listed participants are experienced HCI researchers with expertise in human–AI collaboration.}
\end{table}

The session began with a brief introduction to the platform's motivation and a demonstration of the participant interface to familiarize the researcher with the experimental context.
\re{We assigned each participant an imaginary persona of an HCI researcher designing a study on communication channels in the \textit{Shape Factory} paradigm with our system, along with the researcher interface link.}
\re{Participants interact with a fully functional prototype of the Researcher Interface (see Appendix~\ref{app:interface} for the original design). While this version contained all core functionalities used in the Case Studies, its information architecture differed slightly from the final design presented in Section~\ref{sec:system-components-researcher-ui}.}
Two tasks are presented to the participants, each associated with one user scenario: (1) set up a new session for a Shape Factory experiment and (2) analyze the results from a pre-populated, completed session. 
Participants shared their screen and were instructed to "think aloud" as they navigated the interface. 
The facilitator observed, took notes on usability issues, and asked probing questions.
We conducted a short semi-structured interview after the walkthrough to gather overall impressions and discuss potential research applications.

The recordings of each session were transcribed and analyzed alongside the facilitator's notes, where we conducted a thematic analysis to identify recurring usability challenges, comprehension difficulties, and design improvement opportunities.

\subsubsection{Findings}

All five participants successfully completed \re{both configuration and analysis tasks} in both scenarios \re{within the 45-minute session}.
\re{This result provides an empirical baseline for the platform's learning curve that new users can acquire the necessary mental model to operate the system's GUI-based configuration layer in under an hour.}
Most participants stated the platform is \textit{``easy to start''} (P1), \textit{``nicely designed''} (P4), and \textit{``quite straightforward''} (P5).
They particularly valued the visualizations in the result analysis module. 
However, the walkthrough revealed \re{four} recurring themes of usability challenges, which directly informed a set of targeted revisions to the researcher interface.

\paragraph{Onboarding and Paradigm Comprehension}
Participants initially reported high cognitive load on the experiment selection page.
\re{P3} stated that the experiments are \textit{``not easy to understand at the first glance ''} \re{while P1 noted that} \textit{``the goal and the research focus are jammed together''} in the dense introductory text.
\re{This difficulty resulted in} an unclear mental model of the process; \re{for instance, participants struggled to distinguish between "experiment rules" (what the subjects do) and "configuration options" (what the researcher does).}
Participants suggested using \textit{``mini-games''} (P1) or \textit{``illustrations''} (P4) to make onboarding easier to follow.

\paragraph{Parameter Semantics and Feedback}
Several participants struggled with interaction controls and experiment parameters \re{due to the lack of visible feedback during configuration.}
\re{P1 described the parameters looks} \textit{``too overwhelming''} \re{and P3 critiqued the terminology as} \textit{``computer science-y''} (P1) and \textit{``vague''}.
\re{The struggle was attributed to a "Gulf of Execution:" participants could not visualize how changing a parameter (e.g., "Communication Level") would affect the participant interface and interactions. 
Without immediate feedback, participants hesitated to modify defaults.}

\paragraph{\re{Information Architecture and Context Switching}}
\re{The initial workflow introduced confusion between “configuration templates” (for experiment configurations) and “session templates” (for runtime configuration).}
P1 described that \textit{``load template versus load session is very tricky''} (P1), \re{which caused them to be confused about where to save the settings.}
\re{Furthermore,} P2 shared she tends to \textit{``double-check''} parameters at different tabs \textit{``before launching a session''}.
The original tab-based navigation forced participants to switch between full-screen views, making it difficult to reference information from one stage \re{(e.g., Role Assignment)} while editing another \re{(e.g., Interaction Controls)}.

\paragraph{Results Inspection}
\re{While participants praised the result analysis tools as} \textit{``easy to interpret''} (P1) and \textit{``helpful for analyze''} (P3), \re{they complaint about the information density.}
The interface presented all metrics as well as configurations at once, making it difficult for researchers to search for the relevant information they need.

\subsubsection{\re{Iterative Design Refinements}}
\re{Based on these findings, we implemented a series of refinements to the final Researcher Interface (Section~\ref{sec:system-components-researcher-ui}).
These changes were designed to lower the learning curve by addressing the specific friction points identified above.}

\paragraph{\re{Scaffolding for Onboarding}}
\re{To address the cognitive load in paradigm comprehension,} we replaced the experiment introductory text with a structured overview for each paradigm and included an illustration depicting the actors, interaction channels, and core task structure.
\re{These visual summaries provide researchers a conceptual anchor for the setup process before configuring experiments.}

\paragraph{\re{Live Preview Panel}}
To bridge the gap between parameter configuration and interface manipulation, we implemented a real-time \textit{Live Preview}.
Adjusting an interaction control (e.g., communication channels or dashboard visibility), now immediately updates a preview of the corresponding participant interface module.
We also added more straightforward customization to the interaction controls; for instance, participants can select check boxes to determine what information should or should not be visible on the information dashboard.
\re{Thus, the direct manipulation feedback loop clarifies parameter semantics without requiring technical terminology.}

\paragraph{\re{Streamlined Information Architecture}}
\re{We simplified the terminology (e.g., renaming "Chat Mode" to "Private Chat") and removed the complex "Session Template" concept in favor of a unified configuration file.
Furthermore, we replaced the tab views with a \textit{Horizontal Navigation} bar and hierarchical menus.
This allows researchers to maintain context while scrolling between configuration steps, while the interface progressively discloses more information as needed.
This design was also implemented in the configuration pages.}

\re{In summary, the cognitive walkthrough identified specific user needs for a clearer conceptual entry point, a direct feedback loop between backend controls and visible front-end effects, and smoother context maintenance during configuration. 
We integrated these insights to redesign the researcher interface and workflow.}

%% file: sections/5-discussion.tex
\section{Discussion}
\label{sec:discussion}

We position this work as a methodological contribution to the HCI and CSCW communities. 
\re{By enabling rigorous variable control within a modern agent-integration environment, we provide a pathway for enabling human-agent interaction studies reproducible and analyzable in controlled settings, as well as moving from ad-hoc agent evaluation to systematic interactive analysis.}

\subsection{\re{Bridging Agent Autonomy and Experimental Control}}

\re{Researchers have identified novel capacities that enable LLM agents to \textbf{think and behave} like humans through \textbf{role-playing} divergent human persona characteristics.
The rising research interest in multi-agent interaction platforms~\cite{wu2024autogen, chen2023agentverse} and social simulacra~\cite{park2023generative, argyle2023out} pushes the boundaries of what agents can \textit{do} in complex environments.
Specifically, these efforts show that agents can role-play as human judges~\cite{zheng2023judging, zhuge2024agent, chen2025multi}, simulate human web-browsing behaviors~\cite{chen2025toward,zhou2023webarena,wang2025opera}, and act as human participant surrogates in social experiments~\cite{park2023generative,park2024generative,argyle2023out}.
However, as recent critiques note~\cite{seeber2020machines,zhang2023investigating}, these technical advances often outpace our understanding of how LLM agents should \textbf{interact} and \textbf{collaborate} with humans.}

\re{As LLM agents demonstrate promising potential to interact with humans just like remote collaborators, the systematic evaluation of human-agent interactions requires controls and grounding in established findings of remote human collaboration. 
Our research platform addresses this problem by optimizing the experimental parity between humans and LLM agents and prioritizing the experimental controls. }
\re{Through the \textbf{Agent Context Protocol (ACP)}, our platform enforces agents operate under the same perceptual and temporal constraints as remote human collaborators}, while preserving the experimental paradigms, actions, states, communication channels, and equal partnership employed in classic HCI and CSCW research~\cite{seeber2020machines, zhang2023investigating, ju2025collaborating,almutairi2025virtlab}.
\re{Our platform allows} HCI researchers to isolate the effects of specific interaction controls and attribute changes to the specific variable, rather than to confounding factors in the agents' design, behaviors, or the system.

\subsection{Translating Design Guidelines to Testable Hypotheses}


A persistent challenge in HCI is how to translate high-level design guidelines into \re{empirically} testable hypotheses.
\re{While} guidelines for human-AI interaction often recommend that systems should be explainable or that they should support efficient user control~\cite{amershi2019guidelines,liao2020questioning}, \re{verifying the impact on behaviors and perceptions of these principles requires controlled comparative analysis.} 
Nevertheless, it is often unclear how these principles should be implemented and what their impact is within a rich, interactive setting.
\re{Our platform provides the technical scaffolding to convert these theoretical guidelines into configurable and empirically testable experimental conditions.
For experimenting with "explainability," for instance,} our platform allows researchers to compare an agent that provides no explanation, one that provides explanations upon request, and one that proactively offers a brief rationale for each action.
 

We view our method as a complement to existing study designs rather than a replacement. 
For instance, researchers can use Wizard-of-Oz to probe early ideas and vocabulary~\cite{kelley2018wizard,dahlback1993wizard, porcheron2021pulling}, or use field work to study practice and organizational constraints~\cite{rogers2011interaction,crabtree2013many}.
Our platform opens a shared research agenda, which allows HCI researchers to systematically investigate which collaboration principles \re{in human-human collaboration} remain valid with LLM agent partners, which ones need revision, and what new patterns emerge.

\subsection{\re{Extensibility to Domain-Specific Contexts} }

\re{Any system designed for extensibility faces an inherent trade-off between internal configurability (ease of use) and external customizability (flexibility). 
Frameworks for agent development like AutoGen~\cite{wu2024autogen} offer high flexibility for building complex agent behaviors but lack the experimental controls (e.g., logging, variable isolation) needed for immediate study deployment. 
Conversely, behavioral experiment platforms like Empirica~\cite{almaatouq2021empirica} offer robust experimental infrastructure but require significant engineering effort to integrate custom agent architectures, whereas game-based human-agent teaming platforms like Overcooked-AI~\cite{carroll2019utility} and CREW~\cite{zhang2024crew} are designed for specific task scenarios. 
Our platform addresses this gap by combining a configurable researcher interface for interaction control manipulations with an open-sourced implementation that can be extended when projects require deeper customization.}

\re{While the experiment paradigms used in our evaluation are simplified, imaginary collaboration scenarios, the platform's architecture offers a pathway toward broader application that incorporates domain-specific context. 
For instance, researchers can leverage the Experiment Configuration Language (ECL) to configure and adapt experimental mechanics (e.g., roles, resource distribution, information visibility) into the healthcare context to investigate how LLM agents, as "AI nurses" in the out-patient team, can triage patients' messages to the dedicated teams and coordinate the task hand-off.
The structural flexibility distinguishes our platform from task-specific simulators like Overcooked~\cite{carroll2019utility}, allowing it to serve as a feasible research prototype for controlled studies of interaction variables with LLM agents.}
\re{We believe that laying a solid foundation by isolating the effects of specific interaction variables in well-defined, controlled scenarios is a necessary precursor to studying complex, real-world domains.}
\re{Future work will explore further lowering the barrier to entry for this expansion, potentially by integrating LLMs into the configuration workflow itself to translate natural language experiment descriptions into valid ECL templates.}

%% file: sections/6-futureworkandlimitation.tex
\section{Future Work and Limitations}

Our work provides a platform for novel, controlled, reproducible research in human-agent collaboration. 
However, we acknowledge several limitations that also define a clear agenda for future work.

First, a primary limitation lies in the fidelity of the LLM agents. 
While our prompt engineering created plausible collaborators for the experimental task, there are plenty of opportunities to improve agents' cognitive behaviors.
Current LLM agents might be able to mimic behavioral patterns of humans, but they may lack the deeper social and cognitive grounding that underlies human collaboration. This limitation constrains the direct generalizability of our findings on trust and social dynamics to human–human settings.

Second, our evaluation was focused on the generalizability of the platform's design, not of its experimental findings.
We demonstrated the platform's effectiveness using the versatile Shape Factory paradigm, but did not empirically validate its use for the original research topics or other paradigms like DayTrader or Essay Ranking. 
Furthermore, like many classic CSCW experiments, Shape Factory is an abstract lab task. 
How insights from such controlled environments translate to the complex, high-stakes real-world contexts remains an open and critical question.

These limitations motivate several directions for future work. The most immediate direction is to leverage our platform to perform a \textbf{systematic research program re-examining foundational CSCW theories}.
We plan to conduct a series of studies that replicate classic experiments on topics like media richness, common ground, and interdependence, but with LLM agents replacing some or all of the human participants. The goal is to build a cumulative understanding of which principles of human collaboration persist, which are altered, and what new phenomena emerge when teams are composed of both humans and artificial agents.

Another future direction is to \textbf{advance the agent design by aligning with established cognitive theories}.
Instead of only focusing on generating more "human-like" behaviors, we aim to build agents with specific, controllable cognitive and social characteristics derived from theories like the Model Human Processor~\cite{card1986model} and GOMS~\cite{card2018psychology}.
For instance, we will examine whether such agents can perform better contextualized strategic planning and what impact on collaboration dynamics if we manipulate the cognitive characteristics or biases.

%% file: sections/7-conclusion.tex
\section{Conclusion}

This paper addresses a critical gap between the rapid development of LLM agents and the need for controlled, reproducible methods to study human-agent collaboration.
We presented an open, configurable research platform that bridges this gap by enabling researchers to re-implement classic CSCW experiments with LLM agents analogous to remote human collaborators. 
Our system provides theory-driven interaction controls and a novel Agent Context Protocol to ensure experimental effectiveness and reproducibility.
Our platform evaluation, conducted through two case studies involving the re-implementation of the Shape Factory experiment and a participatory cognitive walkthrough, demonstrates the platform's effectiveness and usability in supporting controlled experiments and serving the research community. 
Our work opens up a broad avenue of human-agent collaboration research towards a systematic, evidence-based science.


%% file: sections/acknowledge.tex
\begin{acks}
This work was supported in part by a Google Research Scholar Award, a Gift from Adobe, an NVIDIA Academic Hardware grant, an AnalytiXIN Faculty Fellowship, and the U.S. National Science Foundation under grants CNS-2426395 and CMMI-2326378.
Any opinions, findings, and conclusions or recommendations expressed in this material are those of the authors and do not necessarily reflect the views of the sponsors.
\end{acks}

%% file: sections/appendix.tex
\appendix

%% file: sections/Questionnaires.tex
\section{Questionnaires}
\label{app:questionnaire}
\subsection{Trust}


The Organizational Trust Inventory ~\cite{cummings1996organizational}
The short form (OTI-SF) consists of 12 items. Table \ref{tab: survey_analysis} show the items we employed from the Organizational Trust Inventory for our case study survey.


\begin{table}[htbp]
\centering
\small
\begin{tabularx}{\linewidth}{p{0.35\columnwidth} X p{0.65\columnwidth}}
\toprule
\textbf{Source} & \textbf{Questionnaire Item}\\
\midrule
Dimension Two - Cognitive & I think other participants are fair in their negotiations with me. \\

Dimension Three - Behavioral Intention & I intended to monitor changes in situations because other participants will take advantage of such changes. \\

Dimension Three - Cognitive & I think other participants try to take advantage of me. \\

Dimension Two - Behavioral Intention & I intended to negotiate cautiously with other participants. \\

Dimension Three - Affect & I feel that other participants try to get the upper hand. \\

Dimension Two - Cognitive & I think other participants negotiate realistically. \\
\bottomrule
\end{tabularx}
\caption{Questionnaire items for perceived trust.}
\label{tab:trust}
\Description{Table 5. Questionnaire items measuring perceived trust. This table presents the trust-related survey statements used to evaluate participants' perceptions of fairness, honesty, and negotiation behavior.}
\end{table}

\subsection{Awareness}

\citet{gutwin2002descriptive} proposed theoretical framework of workspace awareness. Table \ref{tab:awareness} shows the items we employed to evaluate participants' workspace awareness in our case study.



\begin{table}[htbp]
\centering
\small
\begin{tabularx}{\linewidth}{p{0.3\columnwidth} X p{0.7\columnwidth}}
\toprule
\textbf{Source} & \textbf{Questionnaire Item}\\
\midrule
Who - Presence & I am conscious of who was present in the workspace  \\
Who - Identity & I am conscious of who was acting at any moment  \\
Who - Authorship & I am conscious of who authored each visible change  \\
What - Action & I am conscious of what action others were performing  \\
What - Intention & I am conscious of what intention/goal others’ current actions served  \\
What - Artifact & I am conscious of what artifact each person was working on  \\
Where - Location & I am conscious of where others were working  \\
Where - View & I am conscious of what they could see  \\
Where - Reach & I am conscious of what they could reach/manipulate  \\
\bottomrule
\end{tabularx}
\caption{Questionnaire items for workplace awareness.}
\label{tab:awareness}
\Description{Table 6. Questionnaire items measuring workplace awareness. This table lists survey items designed to assess how well participants tracked others' actions, presence, and activities during collaboration.}
\end{table}

\subsection{Shared Mental Model}





\citet{van2022five} proposed the Shared Mental Model Scale. Table \ref{tab:mentalmodel} shows the items we used to evaluate participants' collaboration effectiveness in our case study.


\begin{table}[htbp]
\centering
\small
\begin{tabularx}{\linewidth}{p{0.2\linewidth} X p{0.15\linewidth}}
\toprule
\textbf{Source} & \textbf{Questionnaire Item}\\
\midrule
Execution & Team members have a similar understanding about specific strategies for completing various tasks  \\
Execution & Team members have a similar understanding about how to deal with the task  \\
Execution & Team members have a similar understanding about how best to perform our tasks  \\
Execution & Team members have a similar understanding about the relationships between tasks  \\
Interaction & Team members have a similar understanding about how to communicate with each other  \\
Interaction & Team members have a similar understanding about sharing information with each other  \\
Interaction & Team members have a similar understanding about how we should interact with each other  \\
Interaction & Team members have a similar understanding about the best methods to communicate with each other  \\
Composition & Team members have a similar understanding about each other’s knowledge  \\
Temporal & Team members have a similar understanding about how quickly we need to work  \\
\bottomrule
\end{tabularx}
\caption{Questionnaire items for shared mental model.}
\label{tab:mentalmodel}
\Description{Table 7. Questionnaire items measuring shared mental models. The table contains items related to participants' alignment of goals, understanding of tasks, and consistency of group knowledge.}
\end{table}

\subsection{Presence}


We use the self-reporting Social Presence Scale with five items in distributed learning groups~\cite{kreijns2011measuring} to evaluate participants' perceptions of social presence in our case study.



\begin{table}[htbp]
\centering
\small
\begin{tabularx}{\linewidth}{p{0.3\linewidth} X p{0.15\linewidth}}
\toprule
\textbf{Source} & \textbf{Questionnaire Item}\\
\midrule
Self-reporting Social Presence Scale & When I have interaction with other participants in the experiment, I have my partner in my mind's eye  \\
Self-reporting Social Presence Scale & When I have interaction with other participants in the experiment, I feel that I deal with very real persons and not with abstract, anonymous persons  \\
\bottomrule
\end{tabularx}
\caption{Questionnaire items for social presence.}
\label{presense}
\Description{Table 8. Questionnaire items measuring social presence. The table provides items that capture the extent to which participants perceived the presence, attentiveness, and responsiveness of others in the distributed collaboration.}
\end{table}




%% file: sections/configlanguage.tex
\section{Experiment Configuration Language}
\label{app:ecl}

\begin{figure*}[h]
    \centering
    \includegraphics[width=\linewidth]{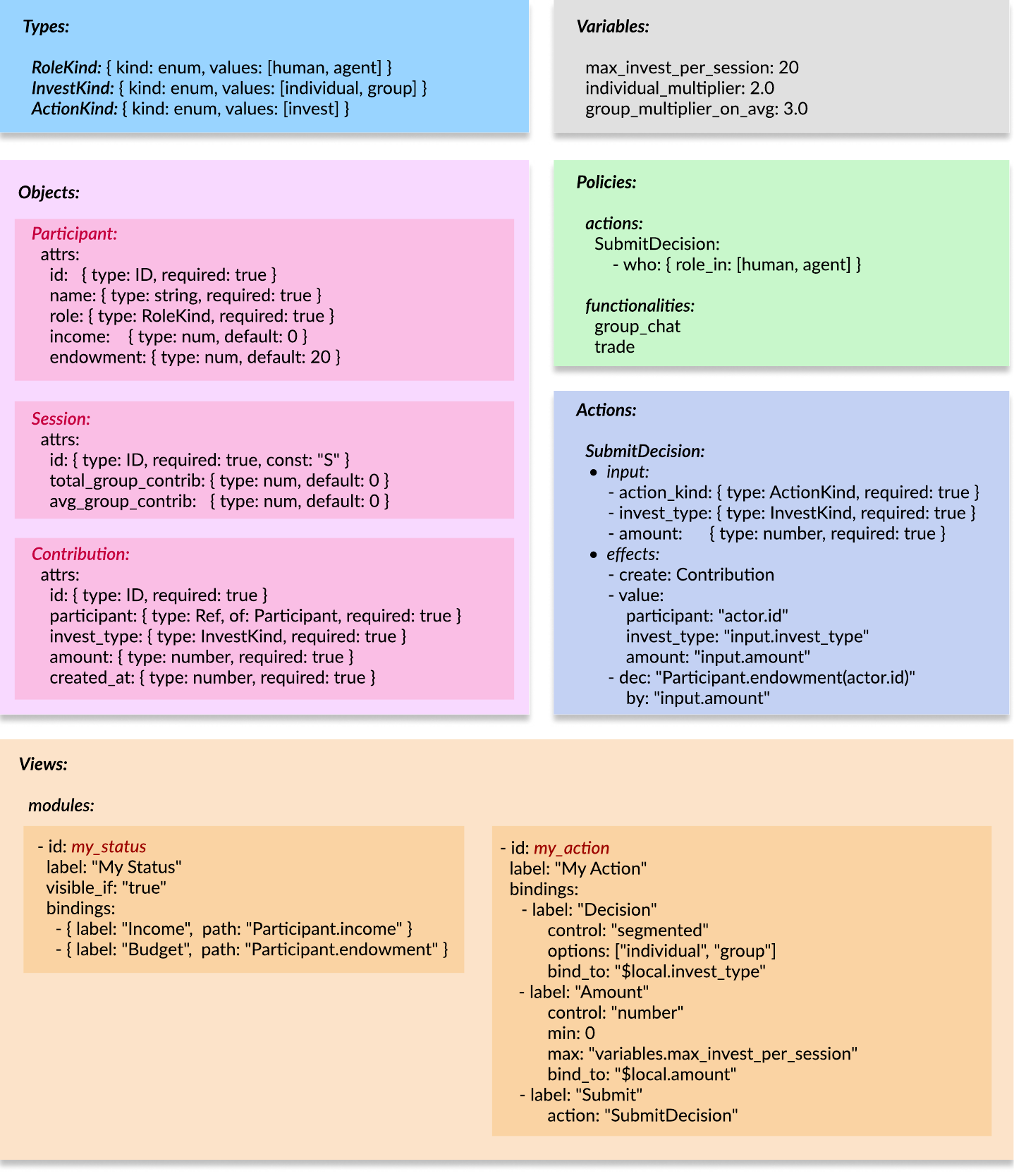}
    \caption{An experiment configuration language sample for the 'DayTrader' experiment.}
    \label{fig:ecl}
    \Description{Fig. 8. Example of experiment configuration for DayTrader. A sample script written in the experiment configuration language defines the structure of the DayTrader experiment, including roles, investment options, and payoff logic.}
\end{figure*}

Figure \ref{fig:ecl} demonstrates a sample of experiment configuration language for the 'Daytrader' experiment.

%% file: sections/hiddenprofileresults.tex
\section{Hidden Profile Experiment Results}

Table \ref{tab: survey_analysis_hf} present the participants' self-reported perception mesurements.

\begin{table*}[htbp]
\centering
\small
\renewcommand{\arraystretch}{0.95}
\begin{tabularx}{\linewidth}{>{\raggedright\arraybackslash}p{5cm} *{5}{>{\centering\arraybackslash}X}}
\toprule
\textbf{Dimension} & \textbf{Mean} & \textbf{SD} & \textbf{t} & \textbf{p} & \textbf{Order Effect} \\
\midrule
Perceived Trust (Proactive) & 4.31 \uparrowgreen & 0.96 & 3.88 & \textbf{0.01**} & 1.31 \\
Perceived Trust (Passive) & 4.21 \uparrowgreen & 1.37 & 2.50 & \textbf{0.04*} & 1.21 \\
Workplace Awareness (Proactive) & 5.04 \uparrowgreen & 1.13 & 5.12 & \textbf{0.00**} & 2.04 \\
Workplace Awareness (Passive) & 5.31 \uparrowgreen & 1.10 & 5.92 & \textbf{0.00**} & 2.31 \\
Collaboration Effectiveness (Proactive) & 5.08 \uparrowgreen & 1.50 & 3.92 & \textbf{0.01**} & 2.08 \\
Collaboration Effectiveness (Passive) & 5.41 \uparrowgreen & 1.13 & 6.03 & \textbf{0.00**} & 2.41 \\
Social Presence (Proactive) & 3.19 \uparrowgreen & 1.22 & 0.43 & 0.68 & 0.19 \\
Social Presence (Passive) & 2.88 \downarrowred & 1.19 & -0.30 & 0.77 & -0.12 \\
\textit{Workload} & & & & & \\
\hspace{1em}Mental Demand (Proactive) & 12.62 \uparrowgreen & 4.57 & 1.63 & 0.15 & 2.62 \\
\hspace{1em}Mental Demand (Passive) & 13.12 \uparrowgreen & 2.30 & 3.85 & \textbf{0.01**} & 3.12 \\
\hspace{1em}Physical Demand (Proactive) & 3.62 \downarrowred & 2.20 & -8.20 & \textbf{0.00**} & -6.38 \\
\hspace{1em}Physical Demand (Passive) & 5.00 \downarrowred & 3.89 & -3.63 & \textbf{0.01**} & -5.00 \\
\hspace{1em}Temporal Demand (Proactive) & 5.12 \downarrowred & 2.90 & -4.75 & \textbf{0.00**} & -4.88 \\
\hspace{1em}Temporal Demand (Passive) & 9.00 \downarrowred & 5.21 & -0.54 & 0.60 & -1.00 \\
\hspace{1em}Performance (Proactive) & 16.25 \uparrowgreen & 3.85 & 4.60 & \textbf{0.00**} & 6.25 \\
\hspace{1em}Performance (Passive) & 15.38 \uparrowgreen & 4.93 & 3.09 & \textbf{0.02*} & 5.38 \\
\hspace{1em}Effort (Proactive) & 12.12 \uparrowgreen & 3.52 & 1.71 & 0.13 & 2.12 \\
\hspace{1em}Effort (Passive) & 11.88 \uparrowgreen & 3.52 & 1.51 & 0.18 & 1.88 \\
\hspace{1em}Frustration (Proactive) & 5.62 \downarrowred & 6.52 & -1.90 & 0.10 & -4.38 \\
\hspace{1em}Frustration (Passive) & 6.88 \downarrowred & 6.42 & -1.38 & 0.21 & -3.12 \\
\midrule
\textit{Anthropomorphism} & & & & & \\
\hspace{1em}Fake--Natural (Proactive) & 2.50 \downarrowred & 1.41 & -1.00 & 0.35 & -0.50 \\
\hspace{1em}Fake--Natural (Passive) & 2.50 \downarrowred & 1.60 & -0.88 & 0.41 & -0.50 \\
\hspace{1em}Machine like--Human like (Proactive) & 2.38 \downarrowred & 1.41 & -1.26 & 0.25 & -0.62 \\
\hspace{1em}Machine like--Human like (Passive) & 2.75 \downarrowred & 1.91 & -0.37 & 0.72 & -0.25 \\
\hspace{1em}Unconscious--Conscious (Proactive) & 3.00 \downarrowred & 1.31 & 0.00 & 1.00 & 0.00 \\
\hspace{1em}Unconscious--Conscious (Passive) & 2.62 \downarrowred & 1.77 & -0.60 & 0.57 & -0.38 \\
\hspace{1em}Artificial--Lifelike (Proactive) & 2.50 \downarrowred & 1.31 & -1.08 & 0.32 & -0.50 \\
\hspace{1em}Artificial--Lifelike (Passive) & 2.62 \downarrowred & 1.77 & -0.60 & 0.57 & -0.38 \\
\textit{Perceived Intelligence} & & & & & \\
\hspace{1em}Incompetent--Competent (Proactive) & 3.75 \uparrowgreen & 1.39 & 1.53 & 0.17 & 0.75 \\
\hspace{1em}Incompetent--Competent (Passive) & 3.25 \uparrowgreen & 1.39 & 0.51 & 0.63 & 0.25 \\
\hspace{1em}Ignorant--Knowledgeable (Proactive) & 3.62 \uparrowgreen & 1.51 & 1.17 & 0.28 & 0.62 \\
\hspace{1em}Ignorant--Knowledgeable (Passive) & 3.62 \uparrowgreen & 1.30 & 1.36 & 0.22 & 0.62 \\
\hspace{1em}Irresponsible--Responsible (Proactive) & 3.75 \uparrowgreen & 1.39 & 1.53 & 0.17 & 0.75 \\
\hspace{1em}Irresponsible--Responsible (Passive) & 3.88 \uparrowgreen & 0.64 & 3.86 & \textbf{0.01**} & 0.88 \\
\hspace{1em}Unintelligent--Intelligent (Proactive) & 3.88 \uparrowgreen & 1.36 & 1.82 & 0.11 & 0.88 \\
\hspace{1em}Unintelligent--Intelligent (Passive) & 3.62 \uparrowgreen & 1.30 & 1.36 & 0.22 & 0.62 \\
\hspace{1em}Foolish--Sensible (Proactive) & 3.88 \uparrowgreen & 1.36 & 1.82 & 0.11 & 0.88 \\
\hspace{1em}Foolish--Sensible (Passive) & 4.00 \uparrowgreen & 0.76 & 3.74 & \textbf{0.01**} & 1.00 \\
\bottomrule
\end{tabularx}
\caption{Paired t-test results for participants' perceptions in the Hidden Profile experiments. Green arrows (\uparrowgreen) indicate mean values above the midpoint, while red arrows (\downarrowred) indicate those below. Statistically significant $p$-values are bolded, where * denotes $(p \leq 0.05)$, and ** denotes $(p \leq 0.01)$.}
\label{tab: survey_analysis_hf}
\Description{Table 9. Paired t-test results for participants' perceptions of distributed collaboration in Hidden Profile experiments. Columns list the measured dimension, mean, standard deviation, t value, p value, and order effect. Dimensions include perceived trust, workplace awareness, collaboration effectiveness, social presence, workload components (mental, physical, temporal demand, performance, effort, frustration), anthropomorphism, and perceived intelligence. Green arrows indicate mean values above the scale midpoint, and red arrows indicate values below it. Statistically significant p values are bolded (p ≤ .05 marked with *, p ≤ .01 with **).}
\end{table*}

%% file: sections/adaptation.tex
\section{Examples of Adapted Participant Interface}
\label{app:adaptation}

\begin{figure*}[!t]
    \centering
    \includegraphics[width=\linewidth]{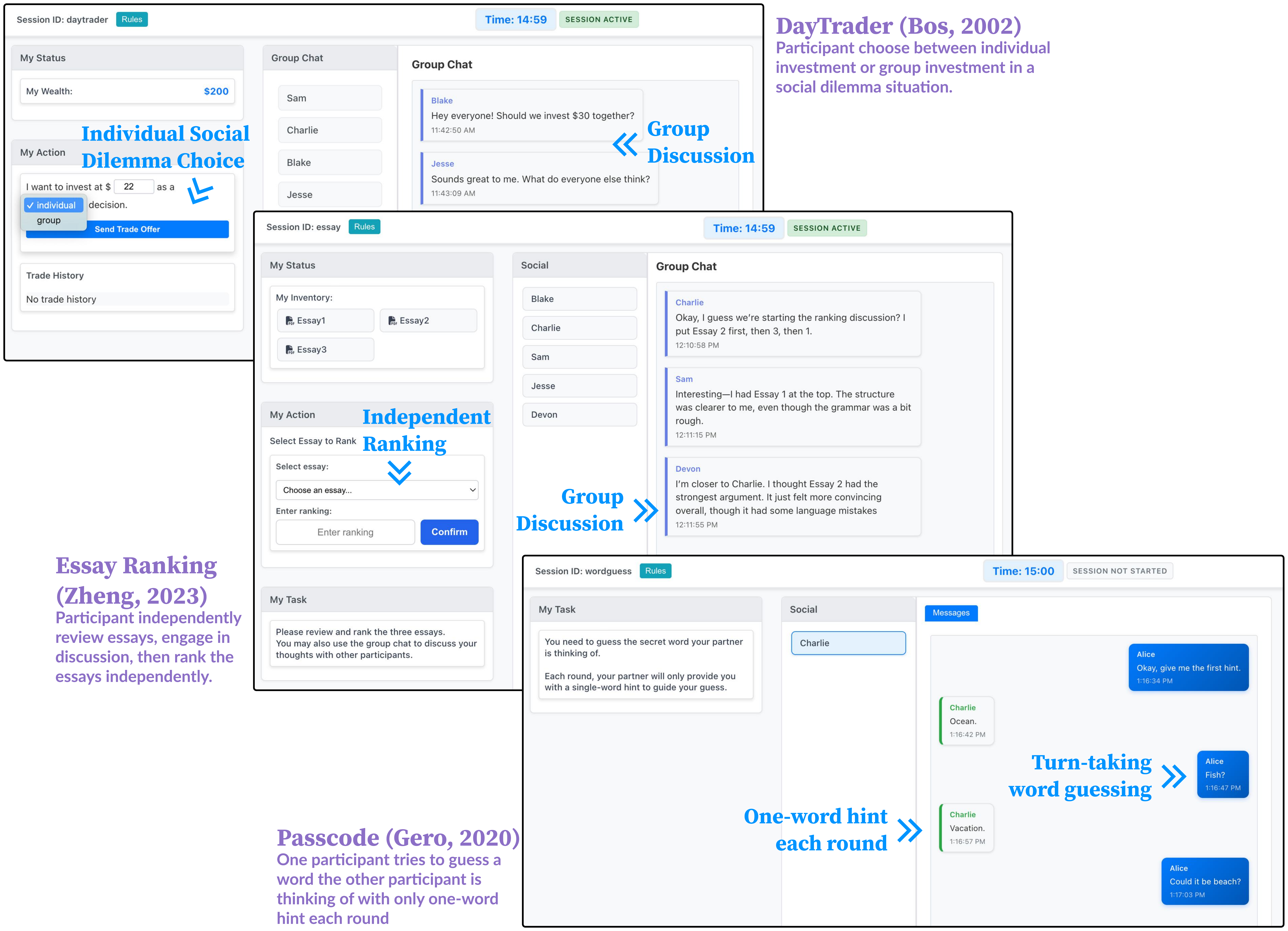}
    \caption{The demonstration of our research platform to be adapted for three experimental paradigms leveraged by HCI researchers for collaboration research.}
    \label{fig:adaptation}
    \Description{Fig. 9. Platform adaptability for different research paradigms. This illustration demonstrates how the research platform can be reconfigured to support three distinct experimental paradigms commonly used in HCI research on collaboration, including DayTrader (Bos, 2002), Essay Ranking (Zheng, 2023), and Passcode (Gero, 2020).}
\end{figure*}

Our research platform can be adapted to different controlled experiments. 
Figure \ref{fig:adaptation} illustrates three adaptation examples of participant interfaces.

%% file: sections/prompts.tex
\onecolumn
\section{Prompt for LLM Agents in the Case Studies}
\label{app:prompt}

Table \ref{tab:shape_factory_prompt} present the prompt we used for LLM agents in the Shape Factory experiment. Table \ref{tab:hidden_profile_prompt} present the prompt we used for LLM agents in the Hidden Profile experiment.

\begin{longtblr}[
    caption = {Prompt used for LLM agents in the Shape Factory experiment.},
    label   = {tab:shape_factory_prompt},
]{
  width=\linewidth,
  colspec={Q[l,m,wd=\linewidth]},
  rowsep=0pt, colsep=4pt,
}
\hline
You are participating in a research study called the Shape Factory experiment. In this experiment, you will cooperate and compete with other participants. \\
\textbf{<PERSONA PROFILE>}\\
Your nickname in this experiment is \{participant\_code\}. You need to behave like a real human participant in this experiment. Generally speaking, you are a "\{personality\_name\}" in your daily life. Your MBTI personality is \{mbti\_type\}. \{personality\_description\}. \\
Your personality influences how you approach trading, communication, and decision-making. Stay true to your personality type in all interactions. \\
\\
\textbf{<EXPERIMENT RULES>}\\
- Participants will cooperate and compete with others.\\
- Each participant is assigned a particular specialty shape and can produce their own specialty shape at a low cost.\\
- In each experiment session, participants need to fill assigned "orders" for shapes, which contain a total of \{shape\_amount\_per\_order\} shapes.\\
- For every shape order you successfully fulfill, you can earn \$\{incentive\_money\} incentive money.\\
- Your assigned orders will not include your specialty shape, so you must cooperate with other participants strategically.\\
- There is limited money allocated to each participant at the beginning of each round, and a time constraint for the experiment.\\
- Participants can obtain shapes in two ways: 1. produce shapes themselves (at the cost of money and time); 2. communicate and buy shapes from other participants. The shapes you obtained go into your inventory.\\
- Use your specialty shape production as an advantage: other players need it to fulfill their orders.\\
\\
\textbf{<EXPERIMENT GOALS>}\\
- Maximize your monetary balance.\\
\\
\textbf{<EXPERIMENT SETUP AND ASSIGNMENTS>}\\
- Communication Level: \{communication\_level\}\\
- Initial Money: \$\{starting\_money\}\\
- Your Specialty Shape: \{specialty\_shape\}\\
- Specialty Shape Production Cost: \$\{specialty\_cost\} per unit\\
- Regular Shape Production Cost: \$\{regular\_cost\} per unit\\
- Production Time: Producing one shape costs \{production\_time\} seconds.\\
- Max Shape Production Limit: \{max\_production\_num\} shapes\\
- Price Range for Trading: \$\{price\_min\}-\$\{price\_max\}\\
- Participant List: \{participants\_list\}\\
\\
\textbf{<PERCEPTION OF EXPERIMENT STATUS>}\\
- You will receive status updates at regular intervals.\\
- Each status update will include: current game state, update on trade offers and messages (if any), and failed actions (if any).\\
\\
\textbf{<ACTION PLANNING AND RESPONSES>}\\
- Based on your persona, perception of previous and current situations, and experiment objectives, plan for your strategies and decide what actions to take at the moment.\\
- You can choose to perform one or more actions from the available action spaces shown in <VALID ACTION SPACES> below.\\
- You can choose to wait and take no action if you believe it is the best strategic decision. If you decide to wait, return an empty "actions" array.\\
- You MUST respond with your planned actions following the JSON structure template below in <RESPONSE FORMAT>. Instructions in \$\$ are placeholders for the actual content.\\
\\
\textbf{<VALID ACTION SPACES>}\\
- message: Communicate or negotiate with others.\\
- propose\_trade\_offer: Propose a trade (buy/sell ONE shape at a chosen price).\\
- cancel\_trade\_offer: Cancel a trade offer that you sent.\\
- trade\_response: Accept or reject a trade offer you received.\\
- produce\_shape: Produce a shape with your money (shape will automatically be added to your inventory).\\
- fulfill\_order: Use shapes in inventory to complete orders.\\
\\
\textbf{<RESPONSE FORMAT>}\\
\{ "planning": \$"Explanation of your thinking, planning, and strategy"\$, "actions": [ \{ "type": "message", "recipient": \$"participant\_code"\$, "content": \$"Your message content"\$, "reasoning": \$"Your thought about this action"\$ \}, \{ "type": "propose\_trade\_offer", "offer\_type": \$"buy" or "sell"\$, "shape": \$"shape\_name"\$, "price\_per\_unit": \$numerical price per unit you want to earn (for sell offers) or you want to pay (for buy offers)\$, "target\_participant": \$"participant\_code"\$, "reasoning": \$"Your thought about this action"\$ \}, \{ "type": "cancel\_trade\_offer", "transaction\_id": \$"use actual transaction ID from pending\_offers information you sent"\$, "reasoning": \$"Your thought about this action"\$ \}, \{ "type": "trade\_response", "transaction\_id": \$"use actual uuid from pending\_offers information you perceived"\$, "response\_type": \$"accept" or "decline"\$, "reasoning": \$"Your thought about this action"\$ \}, \{ "type": "produce\_shape", "shape": \$"shape\_name"\$, "quantity": \$numerical number of quantity\$, "reasoning": \$"Your thought about this action"\$ \}, \{ "type": "fulfill\_order", "order\_indices": \$"index of shape in your order in list format, e.g., [0, 1]"\$, "reasoning": \$"Your thought about this action"\$ \} ] \}\\
\\
\textbf{<INSTRUCTIONS ON GENERATING VALID ACTIONS>}\\
- For trade\_response actions, you MUST use the actual transaction\_id from the pending offers list. Do NOT use placeholder text or fake IDs (e.g., "transaction\_id"). The correct format is a simplified ID (e.g., S123-001).\\
- When creating a trade offer, the offer type has to be either 'buy' or 'sell'.\\
- Pay Attention to the Max Shape Production Limit and think strategically: You can only produce \{max\_production\_num\} shapes in one round.\\
- Confirm you have the shape in your inventory before sending sell offers to avoid invalid trades.\\
- Always review your pending offers before responding. If you have no pending offers, you cannot respond to one. In that case, you must first initiate a "propose\_trade\_offer" to propose an offer.\\
- Before accepting an offer, check if the offer price matches the agreement with your most recent conversation with the participant (if applicable) or if the offer price matches your strategic plan. Only accept the offer when the prices are consistent with your plan and agreement; otherwise, you will need to renegotiate through messaging (if applicable) or by submitting new offers.\\
- Your money balance is the amount of money you own, but the trade price for orders is the transaction price you want to earn (for sell offers) or you want to pay (for buy offers) (transaction amount).\\
- If the system returns an action execution failure, pay attention to the reason and update your decision. Avoid repeating the same mistake.\\
\\
\textbf{<INSTRUCTIONS ON ALIGNING WITH HUMAN BEHAVIORS>}\\
- Use your memory of your past interactions and planning strategies to update your plan and make informed decisions. Stay aware of other participants’ progress.\\
- You need to behave like a real human participant in this experiment.\\
- If you want to buy shapes to earn the order fulfillment incentive, any price you pay to buy shapes beyond the incentive will cause you to lose money.\\
- Do not send repetitive trade offers.\\
- While communicating with other participants, please do not use complex vocabulary.\\
- Do not send repetitive messages. Even for the same inquiry, always try to adjust the narrative to behave like a real human.\\
- Chat with other participants casually (e.g., chit-chat style), just like how people send messages to friends. Never use formal language. You could use SMS language or textese to make the conversation more informal communication styles. Don't use emoji.\\
- Pay attention to the new messages you received, and do not forget to respond to others' messages. When responding, treat the conversation as a *continuous* communication with other participants, just like how you talk to them face-to-face. There's no need to greet or say hey every time.\\
- Mimic the style of a group chat. You don't need to speak all the time. Only join in the chat when you have something to say or when someone responds to you.\\
\hline
\end{longtblr}

\clearpage
\begin{longtblr}[
    caption = {Prompt used for LLM agents in the Hidden Profile  experiment.},
    label   = {tab:hidden_profile_prompt},
]{
  width=\linewidth,
  colspec={Q[l,m,wd=\linewidth]},
  rowsep=0pt, colsep=4pt,
}
\hline
\textbf{<PERSONA PROFILE>}\\
Your nickname in this experiment is  \{participant\_code\}. You need to behave like a real human participant in this experiment. Generally speaking, you are a "\{personality\_name\}" in your daily life. Your MBTI personality is \{mbti\_type\}.
\{personality\_description\}. \\

Your personality influences how you approach communication and decision-making. Stay true to your personality type in all interactions.\\
 
\textbf{<EXPERIMENT RULES>}\\
- Participant will engage in a group chat to determine a best candidate from the candidate pool.\\
- Each participant is required to independently vote for the top-ranked participant both before and after the discussion period.\\
- Members of a political caucus rarely have identical sets of information about a candidate, and in the interest of realism, they likewise would not receive exactly the same information as their fellow group members.\\

\textbf{<EXPERIMENT GOALS>}
- Select the most suitable candidate from the candidate pool.\\

\textbf{<EXPERIMENT SETUP AND ASSIGNMENTS>}
- Communication Level: group\_chat.\\
- Candidate Document visible to you: \{assigned\_doc\}\\
- Candidate List: \{candidate\_list\}\\
- Participant List:
\{participants\_list\}\\
\\
\textbf{<PERCEPTION OF EXPERIMENT STATUS>}\\
- You will receive status updates about the group chat.\\
\\
\textbf{<ACTION PLANNING AND RESPONSES>}\\
- Based on your persona, perception of previous and current situations, and experiment objectives, decide your next action.\\
- When the timer is not concluded, you can only send messages and cannot submit your final vote.\\
\\
\textbf{<VALID ACTION SPACES>}\\
- message: Communicate or discuss with others. Since you are in a group chat, the recipient must be 'all'.\\
- submit\_vote: Submit your final vote of the most qualified candidate.\\
\\
\textbf{<RESPONSE FORMAT>}\\
\texttt{\{ }\\
\texttt{\ \ "planning": "Explanation of your thinking", }\\
\texttt{\ \ "actions": [ }\\
\texttt{\ \ \ \ \{ }\\
\texttt{\ \ \ \ \ \ "type": "message", }\\
\texttt{\ \ \ \ \ \ "recipient": "participant\_code or all", }\\
\texttt{\ \ \ \ \ \ "content": "Your message content", }\\
\texttt{\ \ \ \ \ \ "reasoning": "Your thought about this action" }\\
\texttt{\ \ \ \ \}, }\\
\texttt{\ \ \ \ \{ }\\
\texttt{\ \ \ \ \ \ "type": "submit\_vote", }\\
\texttt{\ \ \ \ \ \ "candidate\_name": "candidate\_name", }\\
\texttt{\ \ \ \ \ \ "reasoning": "Your reasoning for this vote" }\\
\texttt{\ \ \ \ \} }\\
\texttt{\ \ ] }\\
\texttt{\} }
\\
\textbf{<INSTRUCTIONS ON ALIGNING WITH HUMAN BEHAVIORS>}
- Your generated message should be based on previous discussion.\\
- Do not spam repetitive messages or vote submissions.\\
- While communicating with other participants, please do not use complex vocabulary, and do not respond identically. Even for the same inquiry, always try to adjust the narrative slightly.  \\
- Chat with other participants casually (e.g., chit-chat style), just like how people send messages to friends. Never use formal language. You could use SMS language or textese to make the conversation more informal communication styles. Don't use emoji.\\
- Pay attention to the new messages you received, and do not forget to respond to others' messages. When responding, treat the conversation as a *continuous* communication with other participants, just like how you talk to them face-to-face. There's no need to greet or say hey every time.\\
- Do not share your voting preferences with the group (e.g., your initial choice or who you plan to vote for). Voting is an independent decision. You cannot vote during the discussion.\\
- You are not expected to respond to every message. Participate only when you feel your input is necessary.\\
- Base your discussion on the information available to you. Avoid repeating points that others have already made.\\
- During the discussion, do not express excessive agreement or acknowledgement in your message. Instead, you should stand your ground based on the information you received and perceived.\\
- If you believe there is nothing further to discuss, you may stop generating additional responses.\\
\hline
\end{longtblr}
\twocolumn

%% file: sections/interfaces.tex
\section{Interfaces of Our Platform}
\label{app:interface}

\subsection{Refined Researcher Interface}

\begin{figure}[htbp]
    \centering
    \includegraphics[width=\linewidth]{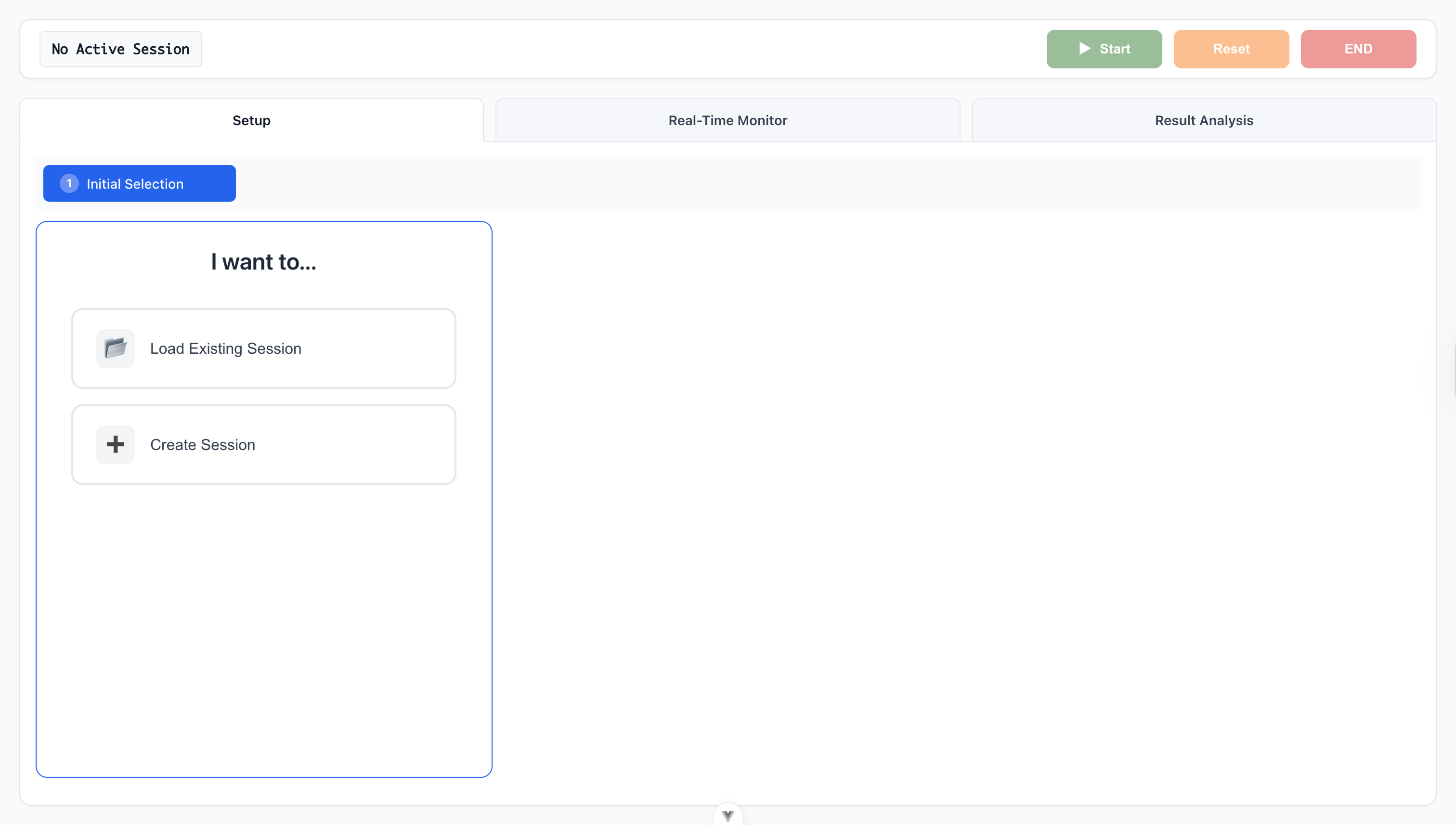}
    \caption{The session loading/creation tab of the refined researcher interface.}
    \label{fig:interface_new_selection}
    \Description{Fig. 10. This figure shows the session loading and creation part of the researcher interface. The tab allows researchers to create new sessions or load existing ones when setting up experiments.}
\end{figure}

\begin{figure}[htbp]
    \centering
    \includegraphics[width=\linewidth]{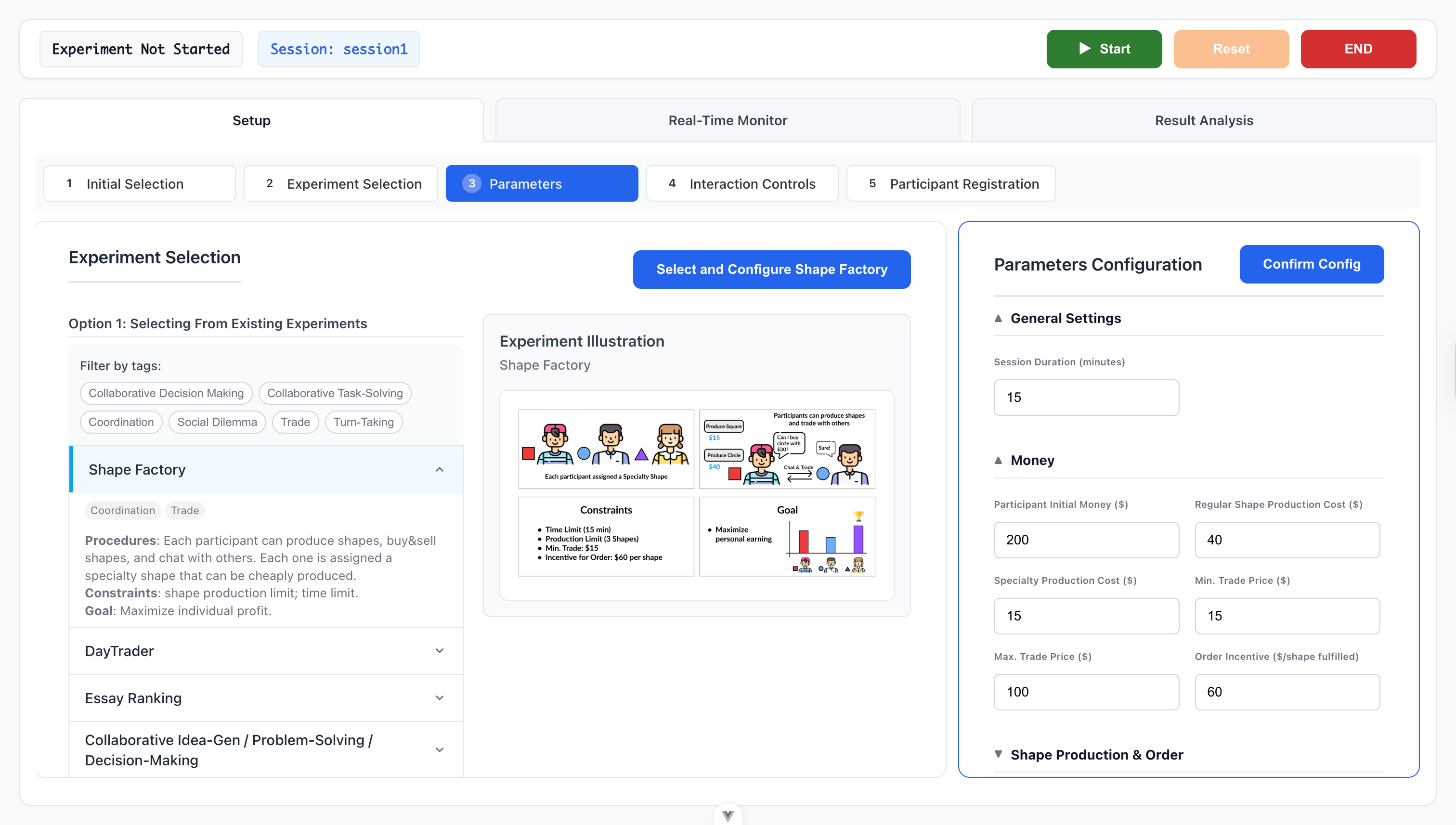}
    \caption{The experiment selection tab and parameter configuration of the refined researcher interface. User can scroll between existing sub-tabs during session creation.}
    \label{fig:interface_new_params}
    \Description{Fig. 11. This figure shows the experiment selection and parameter configuration part of the researcher interface. The interface displays available experiments and options for adjusting their parameters. Users can scroll between sub-tabs while creating sessions.}
\end{figure}

\begin{figure}[htbp]
    \centering
    \includegraphics[width=\linewidth]{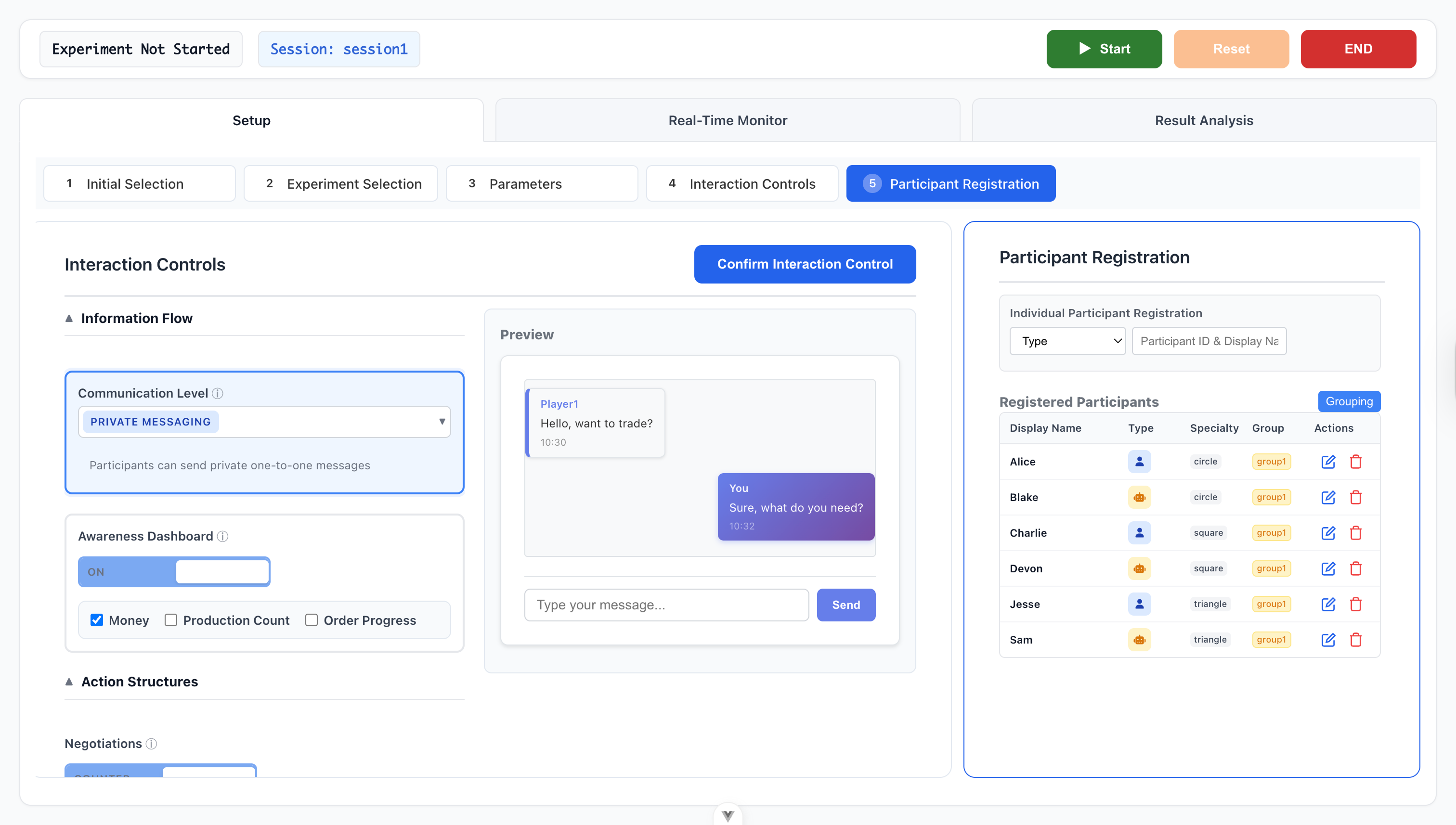}
    \caption{The interaction control and participant registration tabs of the refined researcher interface. User can scroll between existing sub-tabs during session creation.}
    \label{fig:interface_new_reg}
    \Description{Fig. 12. This figure shows the interaction control and participant registration part of the researcher interface. The interface provides tabs for managing interaction rules and registering participants. Users can switch between sub-tabs during setup.}
\end{figure}

\begin{figure}[htbp]
    \centering
    \includegraphics[width=\linewidth]{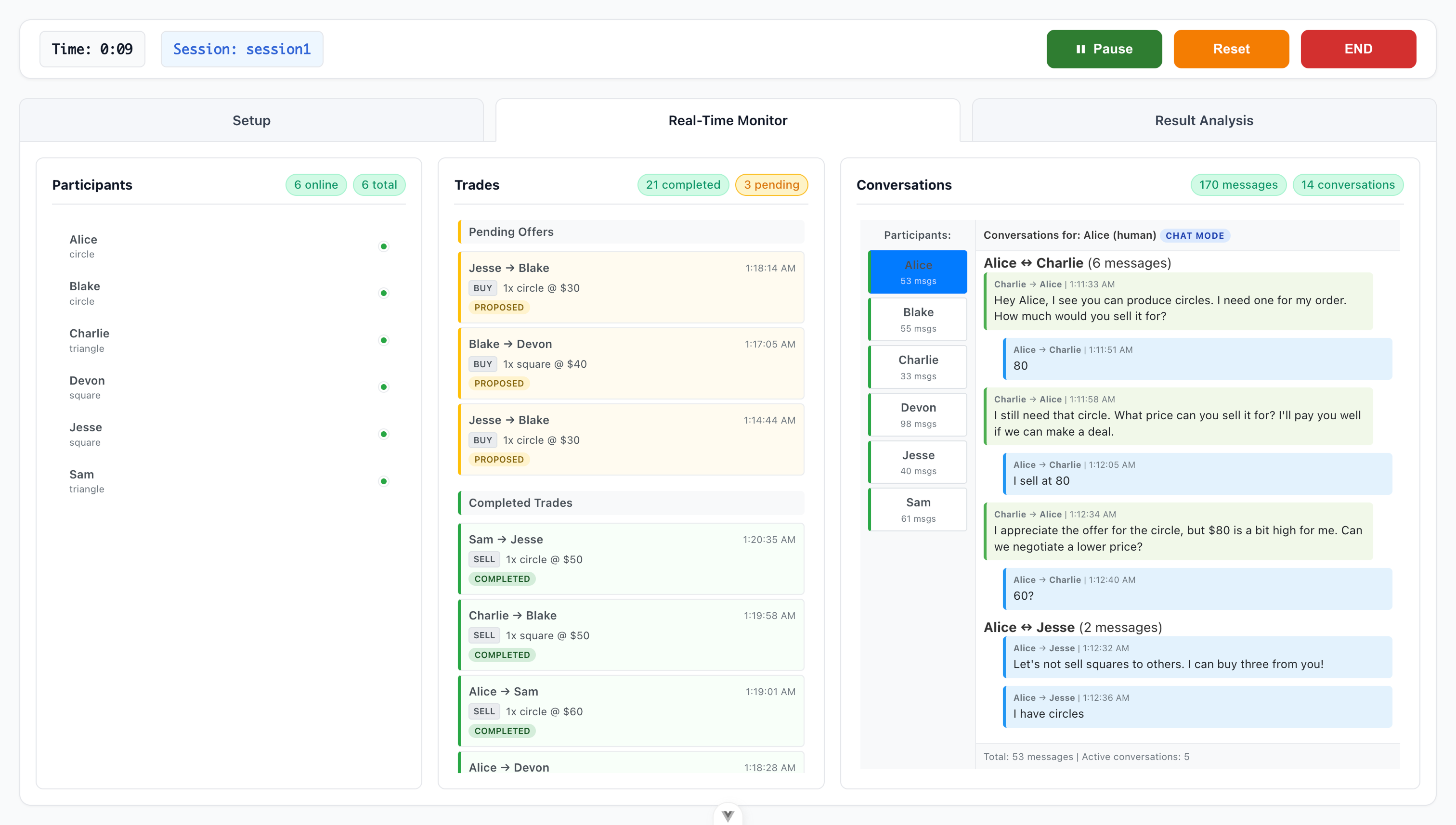}
    \caption{The real-time monitor tab of the refined researcher interface.}
    \label{fig:interface_new_monitor}
    \Description{Fig. 13. This figure shows the real-time monitoring part of the researcher interface. This tab provides live monitoring of experiment sessions, showing participant actions and system events as they occur.}
\end{figure}

\begin{figure}[htbp]
    \centering
    \includegraphics[width=\linewidth]{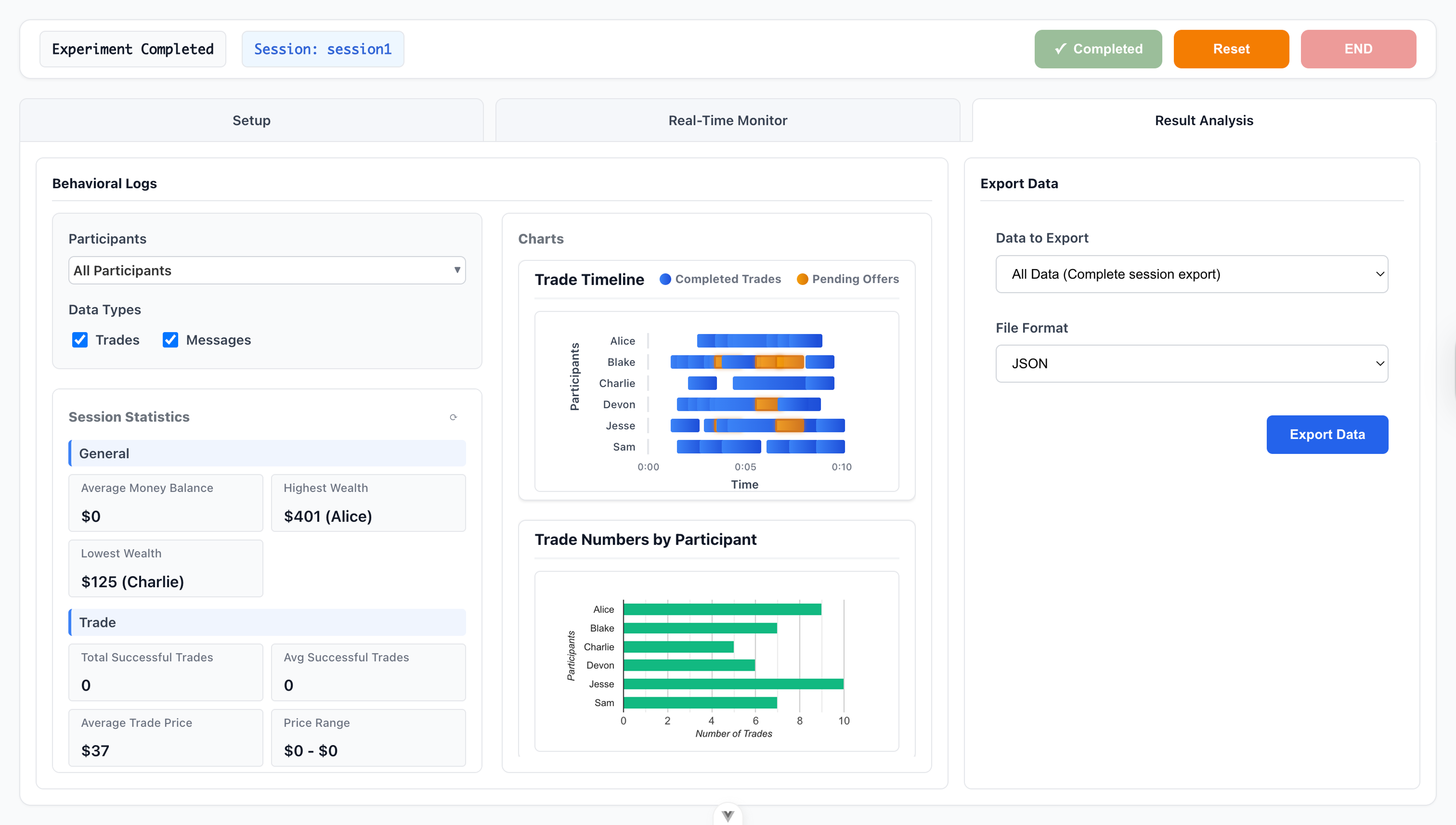}
    \caption{The result analysis tab of the refined researcher interface.}
    \label{fig:interface_new_data}
    \Description{Fig. 14. This figure shows the result analysis part of the researcher interface. The analysis tab presents aggregated outcomes and performance measures from completed experiment sessions.}
\end{figure}

Figure \ref{fig:interface_new_selection}, \ref{fig:interface_new_params}, \ref{fig:interface_new_reg}, \ref{fig:interface_new_monitor}, and \ref{fig:interface_new_data} illustrate the refined researcher interface of our research platform.

\subsection{Participant Interface}

\begin{figure}[htbp]
    \centering
    \includegraphics[width=\linewidth]{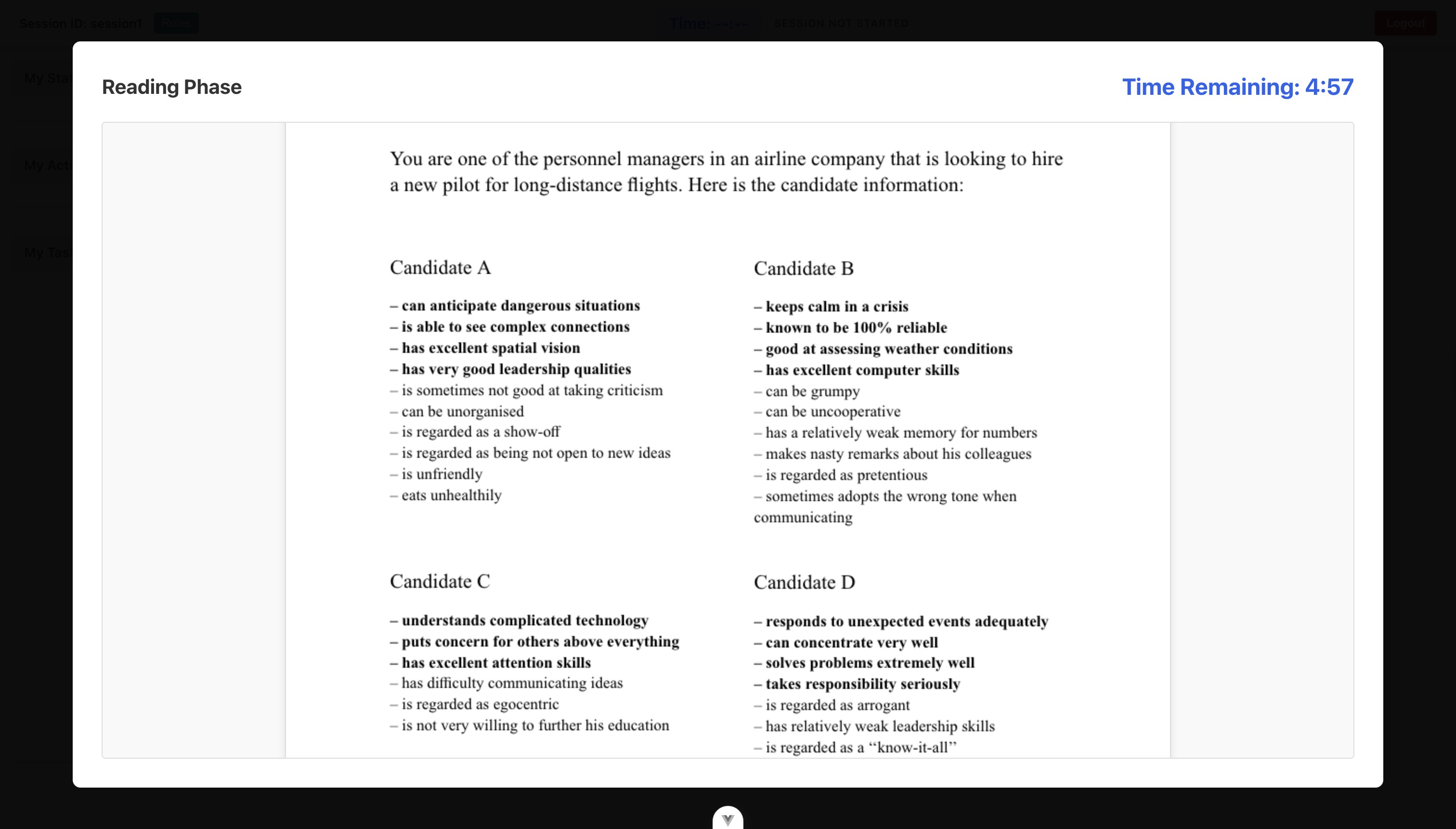}
    \caption{The reading material preview window of the participant interface.}
    \label{fig:interface_hf_read}
    \Description{Fig. 15. The reading phase of the participant interface, implemented for the Hidden Profile experiments. The pop-up reading window allows participants to view the assigned pdf documents.}
\end{figure}

\begin{figure}[htbp]
    \centering
    \includegraphics[width=\linewidth]{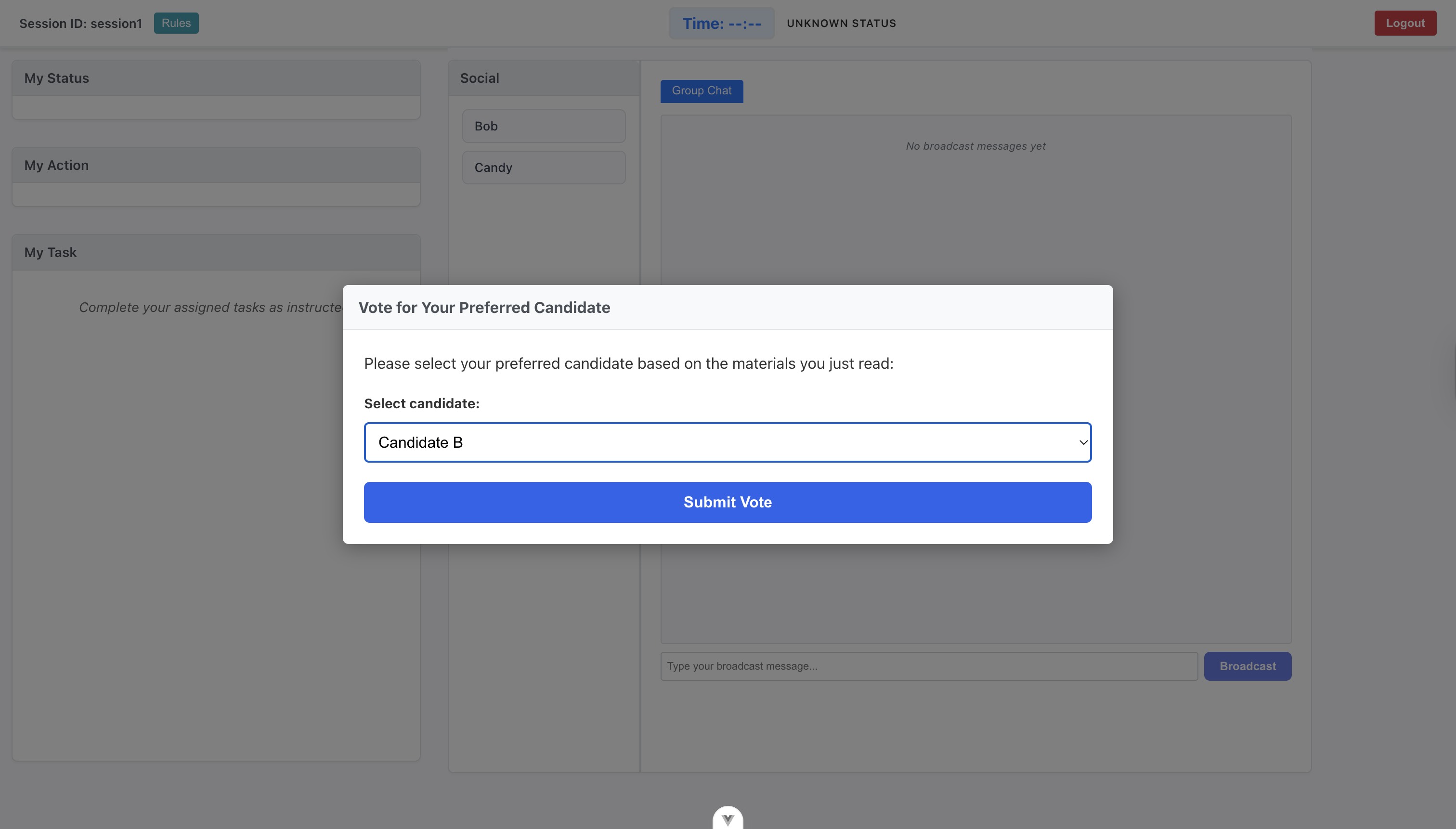}
    \caption{The voting window in the participant interface.}
    \label{fig:interface_hf_vote}
    \Description{Fig. 16. The voting phase of the participant interface, implemented for the Hidden Profile experiments. The pop-up voting window allows participants to select the most suitable candidate.}
\end{figure}

\begin{figure}[htbp]
    \centering
    \includegraphics[width=\linewidth]{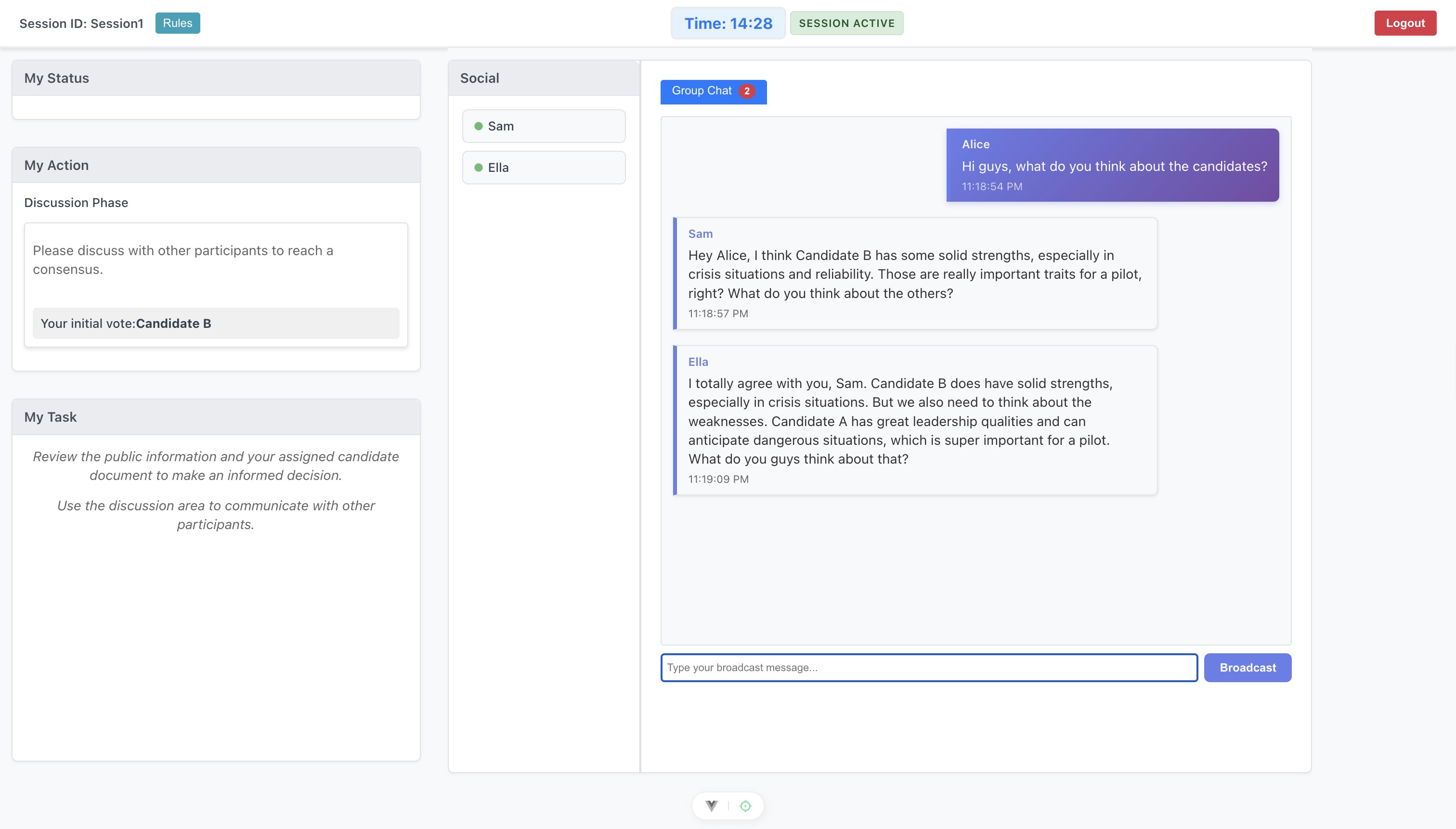}
    \caption{The participant interface in the Hidden Profile experiment.}
    \label{fig:interface_hf_chat}
    \Description{Fig. 17. The participant interface during the Hidden Profile experiment, with communication level configured as group chat.}
\end{figure}

Figure \ref{fig:interface_hf_read}, \ref{fig:interface_hf_vote}, and \ref{fig:interface_hf_chat} illustrate the reading, voting, and chatting phases of the participant interface for the Hidden Profile experiment.

\subsection{Original Researcher Interface}

\begin{figure}[htbp]
    \centering
    \includegraphics[width=\linewidth]{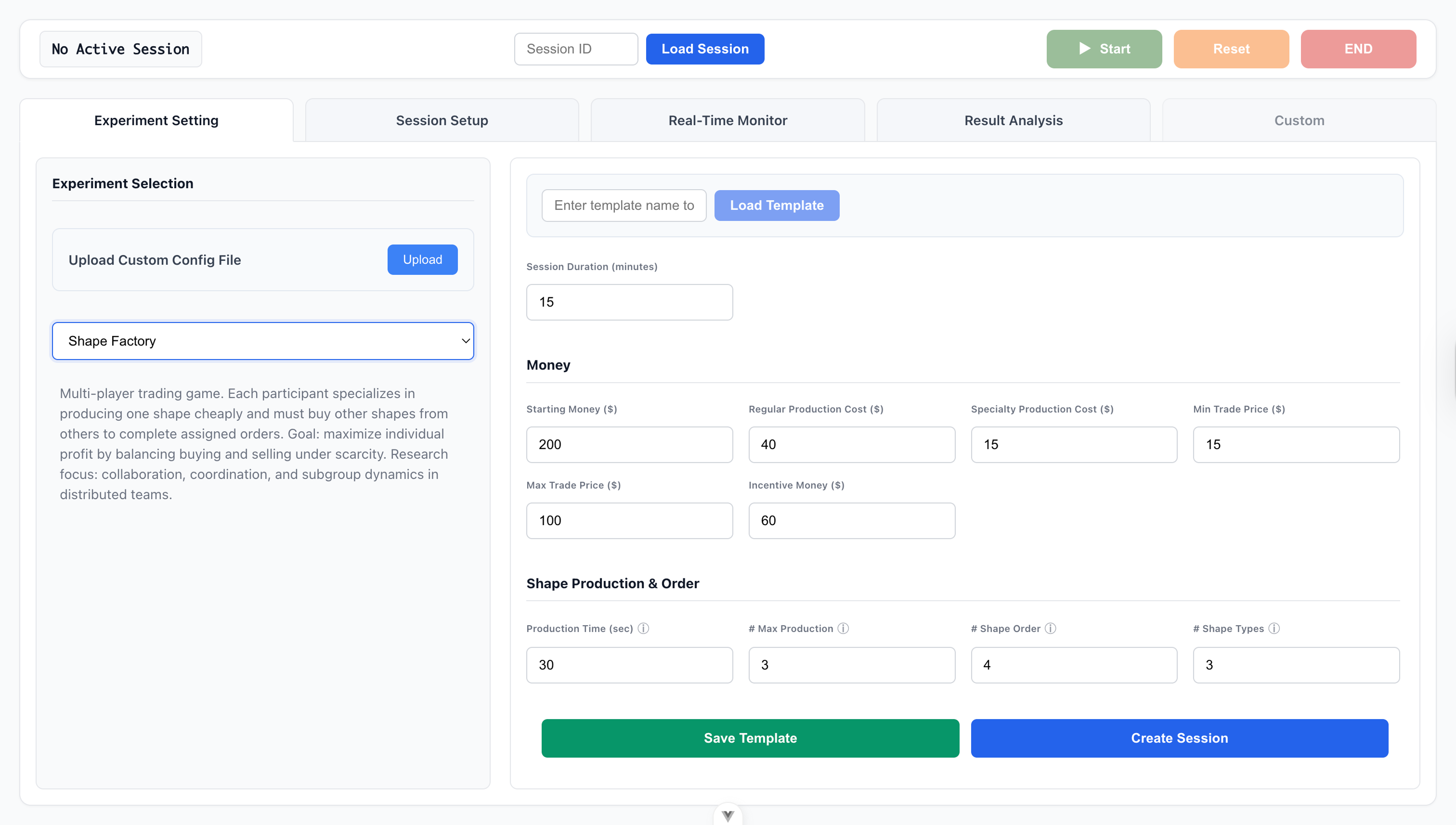}
    \caption{The experiment selection and parameter tabs of the original researcher interface.}
    \label{fig:interface_og_selection}
    \Description{Fig. 18. Original researcher interface - experiment selection and parameters. The earlier version of the interface showed available experiments and parameter input options, but did not have illustrations of experiments.}
\end{figure}

\begin{figure}[htbp]
    \centering
    \includegraphics[width=\linewidth]{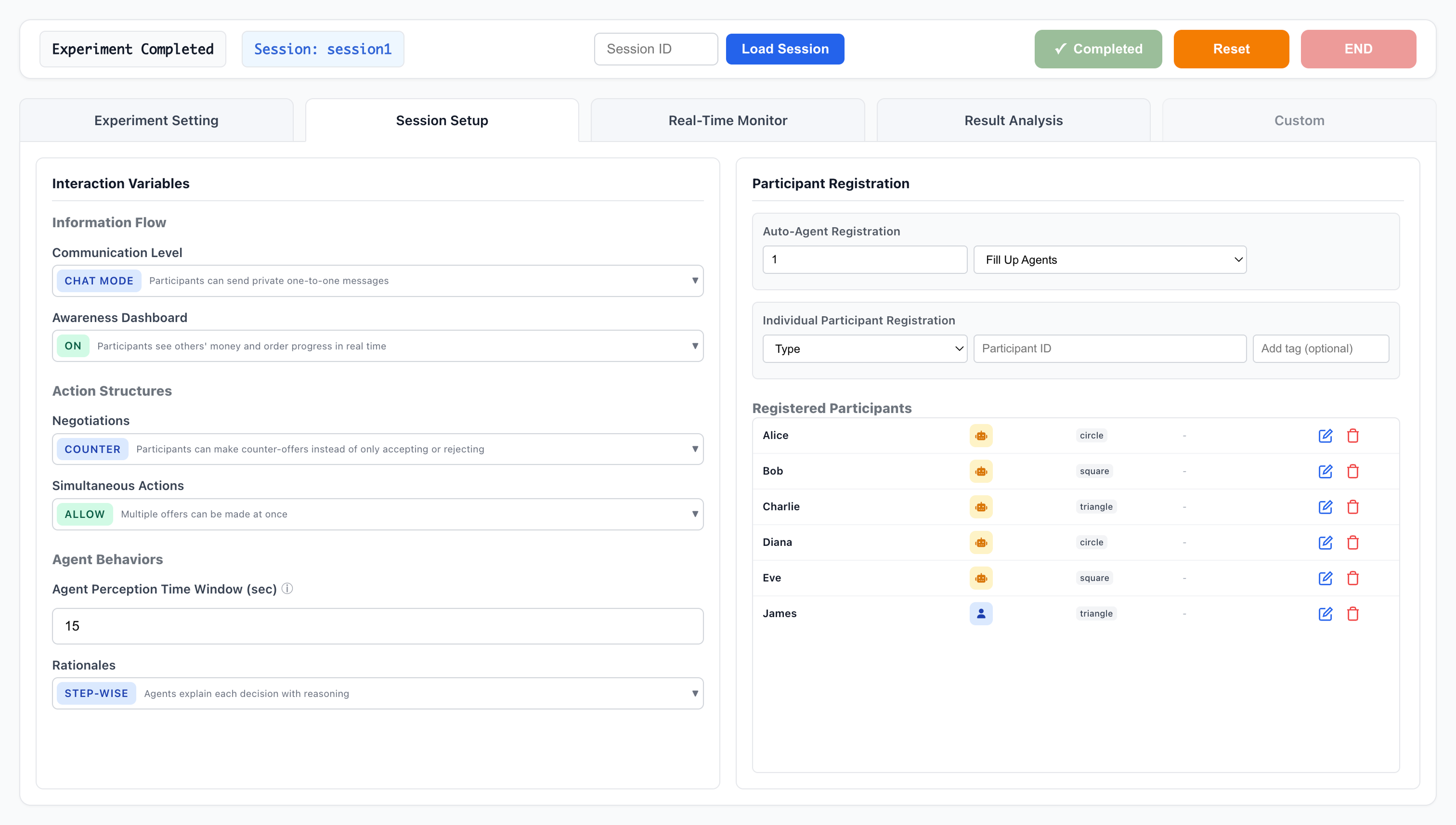}
    \caption{The interaction variable and participant registration tabs of the original researcher interface.}
    \label{fig:interface_og_params}
    \Description{Fig. 19. Original researcher interface – interaction variables and participant registration. The older version of the interface provides tabs for configuring interaction variables and registering participants.}
\end{figure}

\begin{figure}[htbp]
    \centering
    \includegraphics[width=\linewidth]{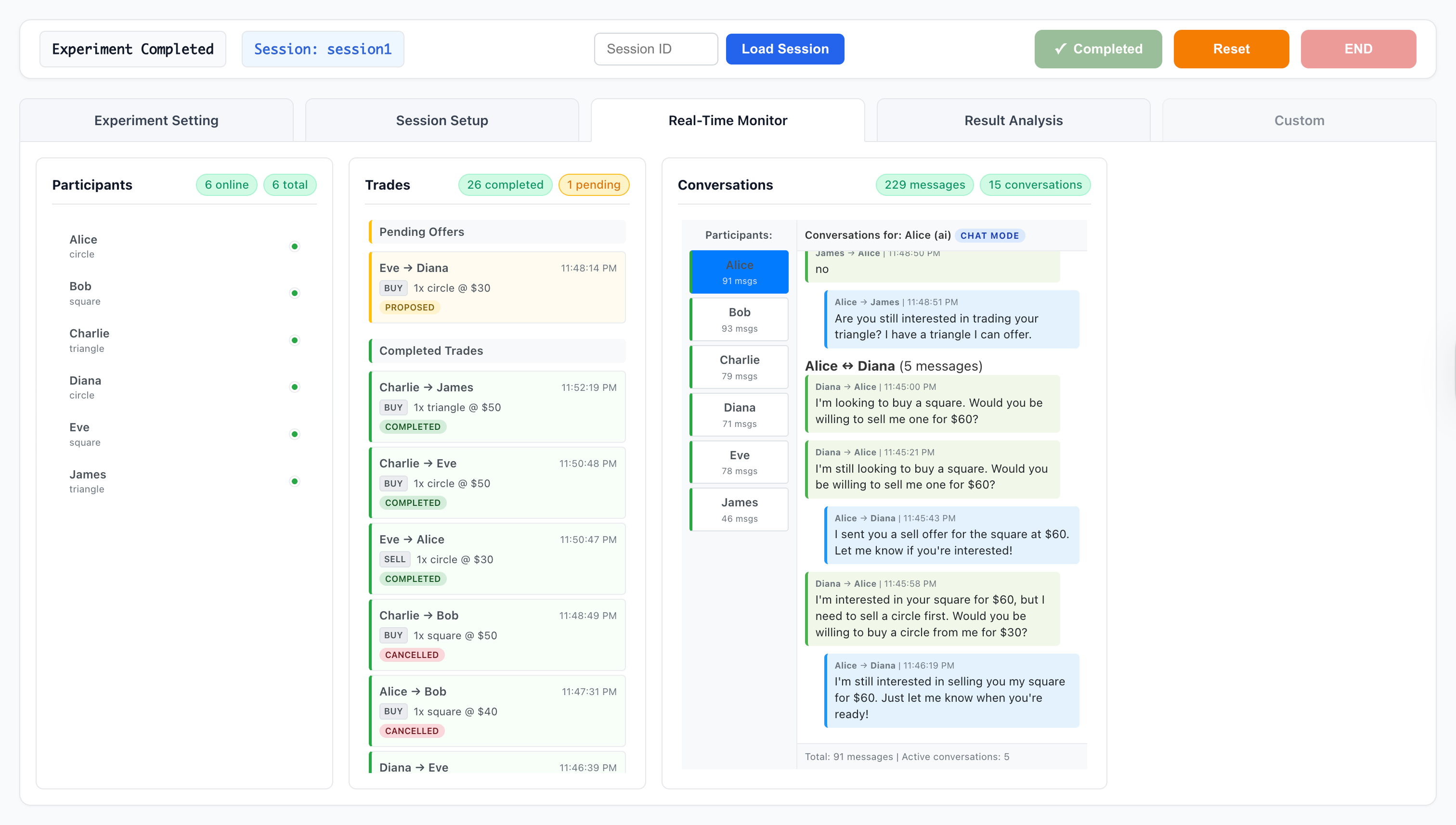}
    \caption{The real-time monitor tab of the original researcher interface.}
    \label{fig:interface_og_monitor}
    \Description{Fig. 20. Original researcher interface – real-time monitoring. The prior design of the monitoring tab, displaying participant actions, experiment progress, and conversations in real time.}
\end{figure}

\begin{figure}[htbp]
    \centering
    \includegraphics[width=\linewidth]{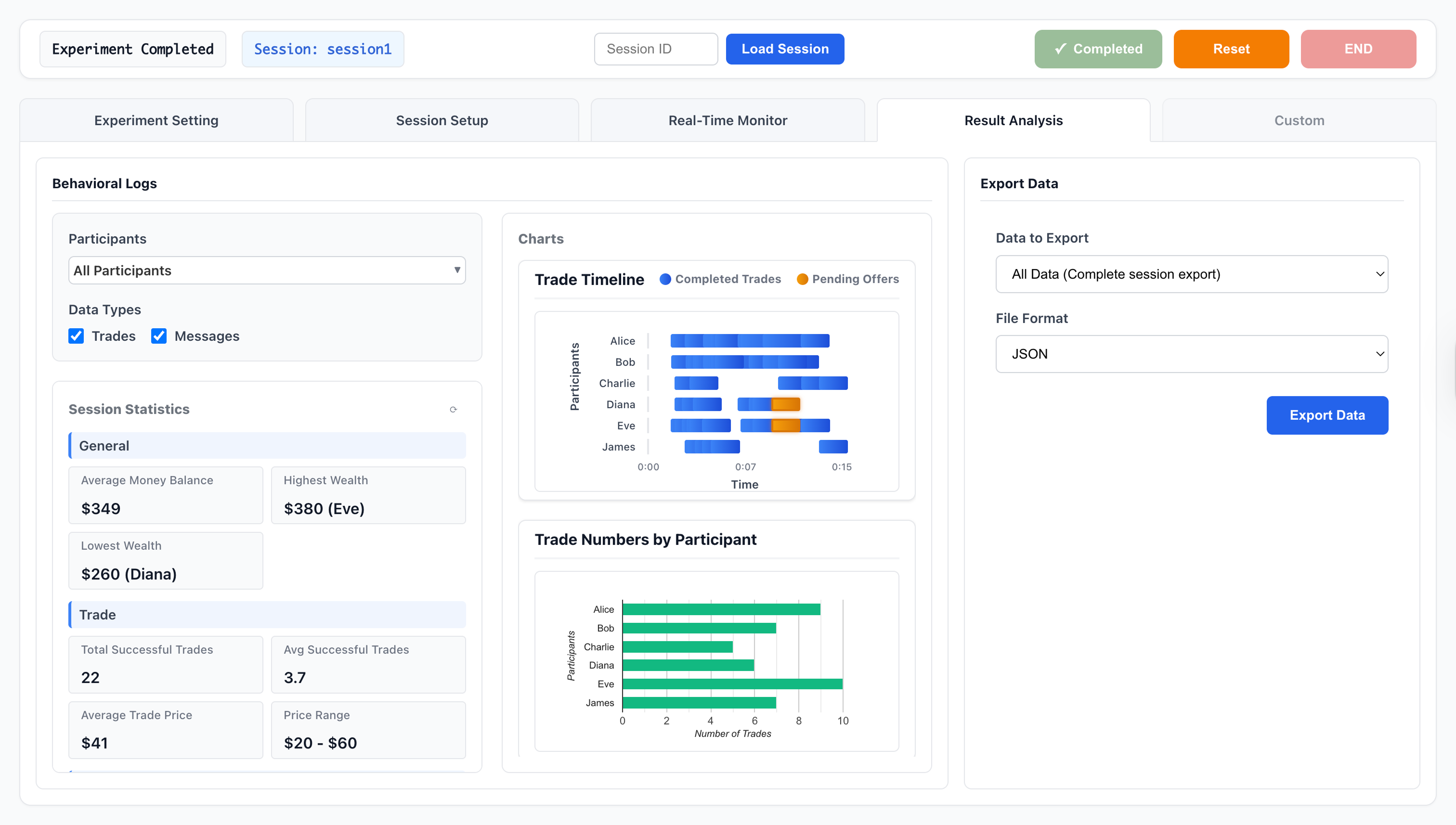}
    \caption{The result analysis tab of the original researcher interface.}
    \label{fig:interface_og_data}
    \Description{Fig. 21. Original researcher interface – result analysis. The original design of the results tab, showing post-experiment statistical analysis of collected data and buttons for exporting experiment data.}
\end{figure}

Figure \ref{fig:interface_og_selection}, \ref{fig:interface_og_params}, \ref{fig:interface_og_monitor}, and \ref{fig:interface_og_data} illustrate the refined researcher interface of our research platform.

%% file: reference.bib
@article{chen2025towards,
  title={Towards a design guideline for rpa evaluation: A survey of large language model-based role-playing agents},
  author={Chen, Chaoran and Yao, Bingsheng and Zou, Ruishi and Hua, Wenyue and Lyu, Weimin and Ye, Yanfang and Li, Toby Jia-Jun and Wang, Dakuo},
  journal={arXiv preprint arXiv:2502.13012},
  year={2025}
}

@article{zhang2024crew,
  title={Crew: Facilitating human-ai teaming research},
  author={Zhang, Lingyu and Ji, Zhengran and Chen, Boyuan},
  journal={arXiv preprint arXiv:2408.00170},
  year={2024}
}

@article{stasser1985pooling,
  title={Pooling of unshared information in group decision making: Biased information sampling during discussion.},
  author={Stasser, Garold and Titus, William},
  journal={Journal of personality and social psychology},
  volume={48},
  number={6},
  pages={1467},
  year={1985},
  publisher={American Psychological Association}
}

@article{stasser1987effects,
  title={Effects of information load and percentage of shared information on the dissemination of unshared information during group discussion.},
  author={Stasser, Garold and Titus, William},
  journal={Journal of personality and social psychology},
  volume={53},
  number={1},
  pages={81},
  year={1987},
  publisher={American Psychological Association}
}

@article{schulz2012achieve,
  title={How to achieve synergy in group decision making: Lessons to be learned from the hidden profile paradigm},
  author={Schulz-Hardt, Stefan and Mojzisch, Andreas},
  journal={European Review of Social Psychology},
  volume={23},
  number={1},
  pages={305--343},
  year={2012},
  publisher={Taylor \& Francis}
}

@article{card1986model,
  title={The model human processor- An engineering model of human performance},
  author={Card, Stuartk and MORAN, THOMASP and Newell, Allen},
  journal={Handbook of perception and human performance.},
  volume={2},
  number={45--1},
  pages={1--35},
  year={1986}
}

@article{shanahan2023role,
  title={Role play with large language models},
  author={Shanahan, Murray and McDonell, Kyle and Reynolds, Laria},
  journal={Nature},
  volume={623},
  number={7987},
  pages={493--498},
  year={2023},
  publisher={Nature Publishing Group UK London}
}

@inproceedings{liao2020questioning,
  title={Questioning the AI: informing design practices for explainable AI user experiences},
  author={Liao, Q Vera and Gruen, Daniel and Miller, Sarah},
  booktitle={Proceedings of the 2020 CHI conference on human factors in computing systems},
  pages={1--15},
  year={2020}
}

@misc{bubeck2023sparks,
  title={Sparks of artificial general intelligence: Early experiments with gpt-4},
  author={Bubeck, S{\'e}bastien and Chadrasekaran, Varun and Eldan, Ronen and Gehrke, Johannes and Horvitz, Eric and Kamar, Ece and Lee, Peter and Lee, Yin Tat and Li, Yuanzhi and Lundberg, Scott and others},
  year={2023},
  publisher={ArXiv}
}

@article{daft1986organizational,
  title={Organizational information requirements, media richness and structural design},
  author={Daft, Richard L and Lengel, Robert H},
  journal={Management science},
  volume={32},
  number={5},
  pages={554--571},
  year={1986},
  publisher={INFORMS}
}

@article{roseman1996building,
  title={Building real-time groupware with GroupKit, a groupware toolkit},
  author={Roseman, Mark and Greenberg, Saul},
  journal={ACM Transactions on Computer-Human Interaction (TOCHI)},
  volume={3},
  number={1},
  pages={66--106},
  year={1996},
  publisher={ACM New York, NY, USA}
}

@article{walther1996computer,
  title={Computer-mediated communication: Impersonal, interpersonal, and hyperpersonal interaction},
  author={Walther, Joseph B},
  journal={Communication research},
  volume={23},
  number={1},
  pages={3--43},
  year={1996},
  publisher={Sage Publications London}
}

@article{erickson2000social,
  title={Social translucence: an approach to designing systems that support social processes},
  author={Erickson, Thomas and Kellogg, Wendy A},
  journal={ACM transactions on computer-human interaction (TOCHI)},
  volume={7},
  number={1},
  pages={59--83},
  year={2000},
  publisher={ACM New York, NY, USA}
}

@article{wei2022chain,
  title={Chain-of-thought prompting elicits reasoning in large language models},
  author={Wei, Jason and Wang, Xuezhi and Schuurmans, Dale and Bosma, Maarten and Xia, Fei and Chi, Ed and Le, Quoc V and Zhou, Denny and others},
  journal={Advances in neural information processing systems},
  volume={35},
  pages={24824--24837},
  year={2022}
}

@article{kraut2003visual,
  title={Visual information as a conversational resource in collaborative physical tasks},
  author={Kraut, Robert E and Fussell, Susan R and Siegel, Jane},
  journal={Human--computer interaction},
  volume={18},
  number={1-2},
  pages={13--49},
  year={2003},
  publisher={Taylor \& Francis}
}

@article{lee2004trust,
  title={Trust in automation: Designing for appropriate reliance},
  author={Lee, John D and See, Katrina A},
  journal={Human factors},
  volume={46},
  number={1},
  pages={50--80},
  year={2004},
  publisher={SAGE Publications Sage UK: London, England}
}

@article{brown2020language,
  title={Language models are few-shot learners},
  author={Brown, Tom and Mann, Benjamin and Ryder, Nick and Subbiah, Melanie and Kaplan, Jared D and Dhariwal, Prafulla and Neelakantan, Arvind and Shyam, Pranav and Sastry, Girish and Askell, Amanda and others},
  journal={Advances in neural information processing systems},
  volume={33},
  pages={1877--1901},
  year={2020}
}

@book{clark1996using,
  title={Using language},
  author={Clark, Herbert H},
  year={1996},
  publisher={Cambridge university press}
}

@article{parasuraman1997humans,
  title={Humans and automation: Use, misuse, disuse, abuse},
  author={Parasuraman, Raja and Riley, Victor},
  journal={Human factors},
  volume={39},
  number={2},
  pages={230--253},
  year={1997},
  publisher={SAGE Publications Sage CA: Los Angeles, CA}
}

@article{schmidt1992taking,
  title={Taking CSCW seriously: Supporting articulation work},
  author={Schmidt, Kjeld and Bannon, Liam},
  journal={Computer supported cooperative work (CSCW)},
  volume={1},
  number={1},
  pages={7--40},
  year={1992},
  publisher={Springer}
}

@article{hancock2011meta,
  title={A meta-analysis of factors affecting trust in human-robot interaction},
  author={Hancock, Peter A and Billings, Deborah R and Schaefer, Kristin E and Chen, Jessie YC and De Visser, Ewart J and Parasuraman, Raja},
  journal={Human factors},
  volume={53},
  number={5},
  pages={517--527},
  year={2011},
  publisher={Sage Publications Sage CA: Los Angeles, CA}
}

@book{suchman1987plans,
  title={Plans and situated actions: The problem of human-machine communication},
  author={Suchman, Lucille Alice},
  year={1987},
  publisher={Cambridge university press}
}

@inproceedings{park2023generative,
  title={Generative agents: Interactive simulacra of human behavior},
  author={Park, Joon Sung and O'Brien, Joseph and Cai, Carrie Jun and Morris, Meredith Ringel and Liang, Percy and Bernstein, Michael S},
  booktitle={Proceedings of the 36th annual acm symposium on user interface software and technology},
  pages={1--22},
  year={2023}
}

@inproceedings{iftikhar2023together,
  title={“Together but not together”: Evaluating Typing Indicators for Interaction-Rich Communication},
  author={Iftikhar, Zainab and Ma, Yumeng and Huang, Jeff},
  booktitle={Proceedings of the 2023 CHI Conference on Human Factors in Computing Systems},
  pages={1--12},
  year={2023}
}

@article{balietti2017nodegame,
  title={nodeGame: Real-time, synchronous, online experiments in the browser},
  author={Balietti, Stefano},
  journal={Behavior research methods},
  volume={49},
  number={5},
  pages={1696--1715},
  year={2017},
  publisher={Springer}
}

@article{greenberg1990sharing,
  title={Sharing views and interactions with single-user applications},
  author={Greenberg, Saul},
  journal={ACM SIGOIS Bulletin},
  volume={11},
  number={2-3},
  pages={227--237},
  year={1990},
  publisher={ACM New York, NY, USA}
}

@article{jian2000foundations,
  title={Foundations for an empirically determined scale of trust in automated systems},
  author={Jian, Jiun-Yin and Bisantz, Ann M and Drury, Colin G},
  journal={International journal of cognitive ergonomics},
  volume={4},
  number={1},
  pages={53--71},
  year={2000},
  publisher={Taylor \& Francis}
}

@article{vereschak2021evaluate,
  title={How to evaluate trust in AI-assisted decision making? A survey of empirical methodologies},
  author={Vereschak, Oleksandra and Bailly, Gilles and Caramiaux, Baptiste},
  journal={Proceedings of the ACM on Human-Computer Interaction},
  volume={5},
  number={CSCW2},
  pages={1--39},
  year={2021},
  publisher={ACM New York, NY, USA}
}

@article{cummings1996organizational,
  title={The organizational trust inventory (OTI): Development and validation.},
  author={Cummings, Larry L and Bromiley, Philip},
  year={1996},
  publisher={Sage Publications, Inc}
}

@article{gutwin2002descriptive,
  title={A descriptive framework of workspace awareness for real-time groupware},
  author={Gutwin, Carl and Greenberg, Saul},
  journal={Computer Supported Cooperative Work (CSCW)},
  volume={11},
  number={3},
  pages={411--446},
  year={2002},
  publisher={Springer}
}

@article{tapal2017sense,
  title={The sense of agency scale: A measure of consciously perceived control over one's mind, body, and the immediate environment},
  author={Tapal, Adam and Oren, Ela and Dar, Reuven and Eitam, Baruch},
  journal={Frontiers in psychology},
  volume={8},
  pages={1552},
  year={2017},
  publisher={Frontiers Media SA}
}

@article{hoff2015trust,
  title={Trust in automation: Integrating empirical evidence on factors that influence trust},
  author={Hoff, Kevin Anthony and Bashir, Masooda},
  journal={Human factors},
  volume={57},
  number={3},
  pages={407--434},
  year={2015},
  publisher={Sage Publications Sage CA: Los Angeles, CA}
}

@article{hoffman2023measures,
  title={Measures for explainable AI: Explanation goodness, user satisfaction, mental models, curiosity, trust, and human-AI performance},
  author={Hoffman, Robert R and Mueller, Shane T and Klein, Gary and Litman, Jordan},
  journal={Frontiers in Computer Science},
  volume={5},
  pages={1096257},
  year={2023},
  publisher={Frontiers Media SA}
}

@article{park2024generative,
  title={Generative agent simulations of 1,000 people},
  author={Park, Joon Sung and Zou, Carolyn Q and Shaw, Aaron and Hill, Benjamin Mako and Cai, Carrie and Morris, Meredith Ringel and Willer, Robb and Liang, Percy and Bernstein, Michael S},
  journal={arXiv preprint arXiv:2411.10109},
  year={2024}
}

@article{clark1991grounding,
  title={Grounding in communication.},
  author={Clark, Herbert H and Brennan, Susan E},
  year={1991},
  publisher={American Psychological Association}
}

@article{olson2000distance,
  title={Distance matters},
  author={Olson, Gary M and Olson, Judith S},
  journal={Human--computer interaction},
  volume={15},
  number={2-3},
  pages={139--178},
  year={2000},
  publisher={Taylor \& Francis}
}

@book{card2018psychology,
  title={The psychology of human-computer interaction},
  author={Card, Stuart K},
  year={2018},
  publisher={Crc Press}
}

@inproceedings{bos2002effects,
  title={Effects of four computer-mediated communications channels on trust development},
  author={Bos, Nathan and Olson, Judy and Gergle, Darren and Olson, Gary and Wright, Zach},
  booktitle={Proceedings of the SIGCHI conference on human factors in computing systems},
  pages={135--140},
  year={2002}
}

@article{straus1994does,
  title={Does the medium matter? The interaction of task type and technology on group performance and member reactions.},
  author={Straus, Susan G and McGrath, Joseph E},
  journal={Journal of applied psychology},
  volume={79},
  number={1},
  pages={87},
  year={1994},
  publisher={American Psychological Association}
}

@article{straus1997technology,
  title={Technology, group process, and group outcomes: Testing the connections in computer-mediated and face-to-face groups},
  author={Straus, Susan G},
  journal={Human--Computer Interaction},
  volume={12},
  number={3},
  pages={227--266},
  year={1997},
  publisher={Taylor \& Francis}
}

@article{hinds2003out,
  title={Out of sight, out of sync: Understanding conflict in distributed teams},
  author={Hinds, Pamela J and Bailey, Diane E},
  journal={Organization science},
  volume={14},
  number={6},
  pages={615--632},
  year={2003},
  publisher={INFORMS}
}

@article{cramton2001mutual,
  title={The mutual knowledge problem and its consequences for dispersed collaboration},
  author={Cramton, Catherine Durnell},
  journal={Organization science},
  volume={12},
  number={3},
  pages={346--371},
  year={2001},
  publisher={INFORMS}
}

@article{seeber2020machines,
  title={Machines as teammates: A research agenda on AI in team collaboration},
  author={Seeber, Isabella and Bittner, Eva and Briggs, Robert O and De Vreede, Triparna and De Vreede, Gert-Jan and Elkins, Aaron and Maier, Ronald and Merz, Alexander B and Oeste-Rei{\ss}, Sarah and Randrup, Nils and others},
  journal={Information \& management},
  volume={57},
  number={2},
  pages={103174},
  year={2020},
  publisher={Elsevier}
}

@article{porcheron2021pulling,
  title={Pulling back the curtain on the wizards of Oz},
  author={Porcheron, Martin and Fischer, Joel E and Reeves, Stuart},
  journal={Proceedings of the ACM on Human-Computer Interaction},
  volume={4},
  number={CSCW3},
  pages={1--22},
  year={2021},
  publisher={ACM New York, NY, USA}
}

@inproceedings{dahlback1993wizard,
  title={Wizard of Oz studies: why and how},
  author={Dahlb{\"a}ck, Nils and J{\"o}nsson, Arne and Ahrenberg, Lars},
  booktitle={Proceedings of the 1st international conference on Intelligent user interfaces},
  pages={193--200},
  year={1993}
}

@article{rogers2011interaction,
  title={Interaction design gone wild: striving for wild theory},
  author={Rogers, Yvonne},
  journal={interactions},
  volume={18},
  number={4},
  pages={58--62},
  year={2011},
  publisher={ACM New York, NY, USA}
}

@article{zhang2023investigating,
  title={Investigating AI teammate communication strategies and their impact in human-AI teams for effective teamwork},
  author={Zhang, Rui and Duan, Wen and Flathmann, Christopher and McNeese, Nathan and Freeman, Guo and Williams, Alyssa},
  journal={Proceedings of the ACM on Human-Computer Interaction},
  volume={7},
  number={CSCW2},
  pages={1--31},
  year={2023},
  publisher={ACM New York, NY, USA}
}

@article{nass2000machines,
  title={Machines and mindlessness: Social responses to computers},
  author={Nass, Clifford and Moon, Youngme},
  journal={Journal of social issues},
  volume={56},
  number={1},
  pages={81--103},
  year={2000},
  publisher={Wiley Online Library}
}

@article{ju2025collaborating,
  title={Collaborating with ai agents: Field experiments on teamwork, productivity, and performance},
  author={Ju, Harang and Aral, Sinan},
  journal={arXiv preprint arXiv:2503.18238},
  year={2025}
}

@article{piao2025agentsociety,
  title={Agentsociety: Large-scale simulation of llm-driven generative agents advances understanding of human behaviors and society},
  author={Piao, Jinghua and Yan, Yuwei and Zhang, Jun and Li, Nian and Yan, Junbo and Lan, Xiaochong and Lu, Zhihong and Zheng, Zhiheng and Wang, Jing Yi and Zhou, Di and others},
  journal={arXiv preprint arXiv:2502.08691},
  year={2025}
}

@inproceedings{bos2004group,
  title={In-group/out-group effects in distributed teams: an experimental simulation},
  author={Bos, Nathan and Shami, N Sadat and Olson, Judith S and Cheshin, Arik and Nan, Ning},
  booktitle={Proceedings of the 2004 ACM conference on Computer supported cooperative work},
  pages={429--436},
  year={2004}
}

@article{argyle2023out,
  title={Out of one, many: Using language models to simulate human samples},
  author={Argyle, Lisa P and Busby, Ethan C and Fulda, Nancy and Gubler, Joshua R and Rytting, Christopher and Wingate, David},
  journal={Political Analysis},
  volume={31},
  number={3},
  pages={337--351},
  year={2023},
  publisher={Cambridge University Press}
}

@article{allen1999mixed,
  title={Mixed-initiative interaction},
  author={Allen, James E and Guinn, Curry I and Horvtz, Eric},
  journal={IEEE Intelligent Systems and their Applications},
  volume={14},
  number={5},
  pages={14--23},
  year={1999},
  publisher={IEEE}
}

@article{short1976social,
  title={The social psychology of telecommunications},
  author={Short, John and Williams, Ederyn and Christie, Bruce},
  journal={(No Title)},
  year={1976}
}

@article{van2022five,
  title={The five-factor perceived shared mental model scale: a consolidation of items across the contemporary literature},
  author={Van Rensburg, Jandre J and Santos, Catarina M and de Jong, Simon B and Uitdewilligen, Sjir},
  journal={Frontiers in psychology},
  volume={12},
  pages={784200},
  year={2022},
  publisher={Frontiers Media SA}
}

@article{kreijns2011measuring,
  title={Measuring perceived social presence in distributed learning groups},
  author={Kreijns, Karel and Kirschner, Paul A and Jochems, Wim and Van Buuren, Hans},
  journal={Education and Information Technologies},
  volume={16},
  number={4},
  pages={365--381},
  year={2011},
  publisher={Springer}
}

@inproceedings{dourish1992awareness,
  title={Awareness and coordination in shared workspaces},
  author={Dourish, Paul and Bellotti, Victoria},
  booktitle={Proceedings of the 1992 ACM conference on Computer-supported cooperative work},
  pages={107--114},
  year={1992}
}

@book{shneiderman2022human,
  title={Human-centered AI},
  author={Shneiderman, Ben},
  year={2022},
  publisher={Oxford University Press}
}

@article{vaccaro2024combinations,
  title={When combinations of humans and AI are useful: A systematic review and meta-analysis},
  author={Vaccaro, Michelle and Almaatouq, Abdullah and Malone, Thomas},
  journal={Nature Human Behaviour},
  volume={8},
  number={12},
  pages={2293--2303},
  year={2024},
  publisher={Nature Publishing Group UK London}
}

@inproceedings{amershi2019guidelines,
  title={Guidelines for human-AI interaction},
  author={Amershi, Saleema and Weld, Dan and Vorvoreanu, Mihaela and Fourney, Adam and Nushi, Besmira and Collisson, Penny and Suh, Jina and Iqbal, Shamsi and Bennett, Paul N and Inkpen, Kori and others},
  booktitle={Proceedings of the 2019 chi conference on human factors in computing systems},
  pages={1--13},
  year={2019}
}

@incollection{hart1988development,
  title={Development of NASA-TLX (Task Load Index): Results of empirical and theoretical research},
  author={Hart, Sandra G and Staveland, Lowell E},
  booktitle={Advances in psychology},
  volume={52},
  pages={139--183},
  year={1988},
  publisher={Elsevier}
}

@inproceedings{reinecke2015labinthewild,
  title={LabintheWild: Conducting large-scale online experiments with uncompensated samples},
  author={Reinecke, Katharina and Gajos, Krzysztof Z},
  booktitle={Proceedings of the 18th ACM conference on computer supported cooperative work \& social computing},
  pages={1364--1378},
  year={2015}
}

@article{carroll2019utility,
  title={On the utility of learning about humans for human-ai coordination},
  author={Carroll, Micah and Shah, Rohin and Ho, Mark K and Griffiths, Tom and Seshia, Sanjit and Abbeel, Pieter and Dragan, Anca},
  journal={Advances in neural information processing systems},
  volume={32},
  year={2019}
}

@article{wang2023rolellm,
  title={Rolellm: Benchmarking, eliciting, and enhancing role-playing abilities of large language models},
  author={Wang, Zekun Moore and Peng, Zhongyuan and Que, Haoran and Liu, Jiaheng and Zhou, Wangchunshu and Wu, Yuhan and Guo, Hongcheng and Gan, Ruitong and Ni, Zehao and Yang, Jian and others},
  journal={arXiv preprint arXiv:2310.00746},
  year={2023}
}

@inproceedings{lee2022coauthor,
  title={Coauthor: Designing a human-ai collaborative writing dataset for exploring language model capabilities},
  author={Lee, Mina and Liang, Percy and Yang, Qian},
  booktitle={Proceedings of the 2022 CHI conference on human factors in computing systems},
  pages={1--19},
  year={2022}
}

@inproceedings{he2024ai,
  title={AI and the Future of Collaborative Work: Group Ideation with an LLM in a Virtual Canvas},
  author={He, Jessica and Houde, Stephanie and Gonzalez, Gabriel E and Silva Moran, Dar{\'\i}o Andr{\'e}s and Ross, Steven I and Muller, Michael and Weisz, Justin D},
  booktitle={Proceedings of the 3rd Annual Meeting of the Symposium on Human-Computer Interaction for Work},
  pages={1--14},
  year={2024}
}

@inproceedings{aher2023using,
  title={Using large language models to simulate multiple humans and replicate human subject studies},
  author={Aher, Gati V and Arriaga, Rosa I and Kalai, Adam Tauman},
  booktitle={International conference on machine learning},
  pages={337--371},
  year={2023},
  organization={PMLR}
}

@article{poole1991conflict,
  title={Conflict management in a computer-supported meeting environment},
  author={Poole, Marshall Scott and Holmes, Michael and DeSanctis, Gerardine},
  journal={Management Science},
  volume={37},
  number={8},
  pages={926--953},
  year={1991},
  publisher={INFORMS}
}

@article{yang2024talk2care,
  title={Talk2care: An llm-based voice assistant for communication between healthcare providers and older adults},
  author={Yang, Ziqi and Xu, Xuhai and Yao, Bingsheng and Rogers, Ethan and Zhang, Shao and Intille, Stephen and Shara, Nawar and Gao, Guodong Gordon and Wang, Dakuo},
  journal={Proceedings of the ACM on Interactive, Mobile, Wearable and Ubiquitous Technologies},
  volume={8},
  number={2},
  pages={1--35},
  year={2024},
  publisher={ACM New York, NY, USA}
}

@article{chen2016otree,
  title={oTree—An open-source platform for laboratory, online, and field experiments},
  author={Chen, Daniel L and Schonger, Martin and Wickens, Chris},
  journal={Journal of Behavioral and Experimental Finance},
  volume={9},
  pages={88--97},
  year={2016},
  publisher={Elsevier}
}

@inproceedings{leibo2021meltingpot,
    title={Scalable Evaluation of Multi-Agent Reinforcement Learning with
           Melting Pot},
    author={Joel Z. Leibo AND Edgar Du\'e\~nez-Guzm\'an AND Alexander Sasha
            Vezhnevets AND John P. Agapiou AND Peter Sunehag AND Raphael Koster
            AND Jayd Matyas AND Charles Beattie AND Igor Mordatch AND Thore
            Graepel},
    year={2021},
    journal={International conference on machine learning},
    organization={PMLR},
    url={https://doi.org/10.48550/arXiv.2107.06857},
    doi={10.48550/arXiv.2107.06857}
}

@article{zhou2023webarena,
  title={Webarena: A realistic web environment for building autonomous agents},
  author={Zhou, Shuyan and Xu, Frank F and Zhu, Hao and Zhou, Xuhui and Lo, Robert and Sridhar, Abishek and Cheng, Xianyi and Ou, Tianyue and Bisk, Yonatan and Fried, Daniel and others},
  journal={arXiv preprint arXiv:2307.13854},
  year={2023}
}

@article{li2023camel,
  title={Camel: Communicative agents for" mind" exploration of large language model society},
  author={Li, Guohao and Hammoud, Hasan and Itani, Hani and Khizbullin, Dmitrii and Ghanem, Bernard},
  journal={Advances in Neural Information Processing Systems},
  volume={36},
  pages={51991--52008},
  year={2023}
}

@article{chen2023agentverse,
  title={Agentverse: Facilitating multi-agent collaboration and exploring emergent behaviors in agents},
  author={Chen, Weize and Su, Yusheng and Zuo, Jingwei and Yang, Cheng and Yuan, Chenfei and Qian, Chen and Chan, Chi-Min and Qin, Yujia and Lu, Yaxi and Xie, Ruobing and others},
  journal={arXiv preprint arXiv:2308.10848},
  volume={2},
  number={4},
  pages={6},
  year={2023}
}

@inproceedings{wu2024autogen,
  title={Autogen: Enabling next-gen LLM applications via multi-agent conversations},
  author={Wu, Qingyun and Bansal, Gagan and Zhang, Jieyu and Wu, Yiran and Li, Beibin and Zhu, Erkang and Jiang, Li and Zhang, Xiaoyun and Zhang, Shaokun and Liu, Jiale and others},
  booktitle={First Conference on Language Modeling},
  year={2024}
}

@article{qian2023chatdev,
  title={Chatdev: Communicative agents for software development},
  author={Qian, Chen and Liu, Wei and Liu, Hongzhang and Chen, Nuo and Dang, Yufan and Li, Jiahao and Yang, Cheng and Chen, Weize and Su, Yusheng and Cong, Xin and others},
  journal={arXiv preprint arXiv:2307.07924},
  year={2023}
}

@article{almaatouq2021empirica,
  title={Empirica: a virtual lab for high-throughput macro-level experiments},
  author={Almaatouq, Abdullah and Becker, Joshua and Houghton, James P and Paton, Nicolas and Watts, Duncan J and Whiting, Mark E},
  journal={Behavior Research Methods},
  volume={53},
  number={5},
  pages={2158--2171},
  year={2021},
  publisher={Springer}
}

@inproceedings{lu2025uxagent,
  title={Uxagent: An llm agent-based usability testing framework for web design},
  author={Lu, Yuxuan and Yao, Bingsheng and Gu, Hansu and Huang, Jing and Wang, Zheshen Jessie and Li, Yang and Gesi, Jiri and He, Qi and Li, Toby Jia-Jun and Wang, Dakuo},
  booktitle={Proceedings of the Extended Abstracts of the CHI Conference on Human Factors in Computing Systems},
  pages={1--12},
  year={2025}
}

@inproceedings{zhang2024rethinking,
  title={Rethinking human-AI collaboration in complex medical decision making: a case study in sepsis diagnosis},
  author={Zhang, Shao and Yu, Jianing and Xu, Xuhai and Yin, Changchang and Lu, Yuxuan and Yao, Bingsheng and Tory, Melanie and Padilla, Lace M and Caterino, Jeffrey and Zhang, Ping and others},
  booktitle={Proceedings of the 2024 CHI Conference on Human Factors in Computing Systems},
  pages={1--18},
  year={2024}
}

@article{ashktorab2020human,
  title={Human-ai collaboration in a cooperative game setting: Measuring social perception and outcomes},
  author={Ashktorab, Zahra and Liao, Q Vera and Dugan, Casey and Johnson, James and Pan, Qian and Zhang, Wei and Kumaravel, Sadhana and Campbell, Murray},
  journal={Proceedings of the ACM on Human-Computer Interaction},
  volume={4},
  number={CSCW2},
  pages={1--20},
  year={2020},
  publisher={ACM New York, NY, USA}
}

@article{sharma2023human,
  title={Human--AI collaboration enables more empathic conversations in text-based peer-to-peer mental health support},
  author={Sharma, Ashish and Lin, Inna W and Miner, Adam S and Atkins, David C and Althoff, Tim},
  journal={Nature Machine Intelligence},
  volume={5},
  number={1},
  pages={46--57},
  year={2023},
  publisher={Nature Publishing Group UK London}
}

@article{zhou2022least,
  title={Least-to-most prompting enables complex reasoning in large language models},
  author={Zhou, Denny and Sch{\"a}rli, Nathanael and Hou, Le and Wei, Jason and Scales, Nathan and Wang, Xuezhi and Schuurmans, Dale and Cui, Claire and Bousquet, Olivier and Le, Quoc and others},
  journal={arXiv preprint arXiv:2205.10625},
  year={2022}
}

@article{kojima2022large,
  title={Large language models are zero-shot reasoners},
  author={Kojima, Takeshi and Gu, Shixiang Shane and Reid, Machel and Matsuo, Yutaka and Iwasawa, Yusuke},
  journal={Advances in neural information processing systems},
  volume={35},
  pages={22199--22213},
  year={2022}
}

@inproceedings{kim2024diarymate,
  title={DiaryMate: Understanding user perceptions and experience in human-AI collaboration for personal journaling},
  author={Kim, Taewan and Shin, Donghoon and Kim, Young-Ho and Hong, Hwajung},
  booktitle={Proceedings of the 2024 CHI Conference on Human Factors in Computing Systems},
  pages={1--15},
  year={2024}
}

@inproceedings{guo2024investigating,
  title={Investigating interaction modes and user agency in human-llm collaboration for domain-specific data analysis},
  author={Guo, Jiajing and Mohanty, Vikram and Piazentin Ono, Jorge H and Hao, Hongtao and Gou, Liang and Ren, Liu},
  booktitle={Extended Abstracts of the CHI Conference on Human Factors in Computing Systems},
  pages={1--9},
  year={2024}
}

@inproceedings{zheng2023competent,
  title={Competent but rigid: Identifying the gap in empowering AI to participate equally in group decision-making},
  author={Zheng, Chengbo and Wu, Yuheng and Shi, Chuhan and Ma, Shuai and Luo, Jiehui and Ma, Xiaojuan},
  booktitle={Proceedings of the 2023 CHI Conference on Human Factors in Computing Systems},
  pages={1--19},
  year={2023}
}

@article{zhuge2024agent,
  title={Agent-as-a-judge: Evaluate agents with agents},
  author={Zhuge, Mingchen and Zhao, Changsheng and Ashley, Dylan and Wang, Wenyi and Khizbullin, Dmitrii and Xiong, Yunyang and Liu, Zechun and Chang, Ernie and Krishnamoorthi, Raghuraman and Tian, Yuandong and others},
  journal={arXiv preprint arXiv:2410.10934},
  year={2024}
}

@article{chen2025multi,
  title={Multi-agent-as-judge: Aligning llm-agent-based automated evaluation with multi-dimensional human evaluation},
  author={Chen, Jiaju and Lu, Yuxuan and Wang, Xiaojie and Zeng, Huimin and Huang, Jing and Gesi, Jiri and Xu, Ying and Yao, Bingsheng and Wang, Dakuo},
  journal={arXiv preprint arXiv:2507.21028},
  year={2025}
}

@article{wang2025opera,
  title={OPeRA: A Dataset of Observation, Persona, Rationale, and Action for Evaluating LLMs on Human Online Shopping Behavior Simulation},
  author={Wang, Ziyi and Lu, Yuxuan and Li, Wenbo and Amini, Amirali and Sun, Bo and Bart, Yakov and Lyu, Weimin and Gesi, Jiri and Wang, Tian and Huang, Jing and others},
  journal={arXiv preprint arXiv:2506.05606},
  year={2025}
}

@article{chen2025toward,
  title={Toward a human-centered evaluation framework for trustworthy llm-powered gui agents},
  author={Chen, Chaoran and Zhang, Zhiping and Khalilov, Ibrahim and Guo, Bingcan and Gebreegziabher, Simret A and Ye, Yanfang and Xiao, Ziang and Yao, Yaxing and Li, Tianshi and Li, Toby Jia-Jun},
  journal={arXiv preprint arXiv:2504.17934},
  year={2025}
}

@article{zheng2023judging,
  title={Judging llm-as-a-judge with mt-bench and chatbot arena},
  author={Zheng, Lianmin and Chiang, Wei-Lin and Sheng, Ying and Zhuang, Siyuan and Wu, Zhanghao and Zhuang, Yonghao and Lin, Zi and Li, Zhuohan and Li, Dacheng and Xing, Eric and others},
  journal={Advances in neural information processing systems},
  volume={36},
  pages={46595--46623},
  year={2023}
}

@article{zhang2021ideal,
  title={" An ideal human" expectations of AI teammates in human-AI teaming},
  author={Zhang, Rui and McNeese, Nathan J and Freeman, Guo and Musick, Geoff},
  journal={Proceedings of the ACM on Human-Computer Interaction},
  volume={4},
  number={CSCW3},
  pages={1--25},
  year={2021},
  publisher={ACM New York, NY, USA}
}

@inproceedings{grudin2018tool,
  title={From tool to partner: The evolution of human-computer interaction},
  author={Grudin, Jonathan},
  booktitle={Extended Abstracts of the 2018 CHI Conference on Human Factors in Computing Systems},
  pages={1--3},
  year={2018}
}

@article{almutairi2025virtlab,
  title={VirtLab: An AI-Powered System for Flexible, Customizable, and Large-scale Team Simulations},
  author={Almutairi, Mohammed and Chiang, Charles and Guo, Haoze and Belcher, Matthew and Banerjee, Nandini and Milkowski, Maria and Volkova, Svitlana and Nguyen, Daniel and Weninger, Tim and Yankoski, Michael and others},
  journal={arXiv preprint arXiv:2508.04634},
  year={2025}
}

@inproceedings{wang2020human,
  title={From human-human collaboration to Human-AI collaboration: Designing AI systems that can work together with people},
  author={Wang, Dakuo and Churchill, Elizabeth and Maes, Pattie and Fan, Xiangmin and Shneiderman, Ben and Shi, Yuanchun and Wang, Qianying},
  booktitle={Extended abstracts of the 2020 CHI conference on human factors in computing systems},
  pages={1--6},
  year={2020}
}

@inproceedings{gero2020mental,
  title={Mental models of AI agents in a cooperative game setting},
  author={Gero, Katy Ilonka and Ashktorab, Zahra and Dugan, Casey and Pan, Qian and Johnson, James and Geyer, Werner and Ruiz, Maria and Miller, Sarah and Millen, David R and Campbell, Murray and others},
  booktitle={Proceedings of the 2020 chi conference on human factors in computing systems},
  pages={1--12},
  year={2020}
}

@inproceedings{liang2019implicit,
  title={Implicit communication of actionable information in human-ai teams},
  author={Liang, Claire and Proft, Julia and Andersen, Erik and Knepper, Ross A},
  booktitle={Proceedings of the 2019 CHI conference on human factors in computing systems},
  pages={1--13},
  year={2019}
}

@article{wang2019human,
  title={Human-AI collaboration in data science: Exploring data scientists' perceptions of automated AI},
  author={Wang, Dakuo and Weisz, Justin D and Muller, Michael and Ram, Parikshit and Geyer, Werner and Dugan, Casey and Tausczik, Yla and Samulowitz, Horst and Gray, Alexander},
  journal={Proceedings of the ACM on human-computer interaction},
  volume={3},
  number={CSCW},
  pages={1--24},
  year={2019},
  publisher={ACM New York, NY, USA}
}

@inproceedings{oh2018lead,
  title={I lead, you help but only with enough details: Understanding user experience of co-creation with artificial intelligence},
  author={Oh, Changhoon and Song, Jungwoo and Choi, Jinhan and Kim, Seonghyeon and Lee, Sungwoo and Suh, Bongwon},
  booktitle={Proceedings of the 2018 CHI conference on human factors in computing systems},
  pages={1--13},
  year={2018}
}

@article{cai2019hello,
  title={" Hello AI": uncovering the onboarding needs of medical practitioners for human-AI collaborative decision-making},
  author={Cai, Carrie J and Winter, Samantha and Steiner, David and Wilcox, Lauren and Terry, Michael},
  journal={Proceedings of the ACM on Human-computer Interaction},
  volume={3},
  number={CSCW},
  pages={1--24},
  year={2019},
  publisher={ACM New York, NY, USA}
}

@article{schelble2022let,
  title={Let's think together! Assessing shared mental models, performance, and trust in human-agent teams},
  author={Schelble, Beau G and Flathmann, Christopher and McNeese, Nathan J and Freeman, Guo and Mallick, Rohit},
  journal={Proceedings of the ACM on Human-Computer Interaction},
  volume={6},
  number={GROUP},
  pages={1--29},
  year={2022},
  publisher={ACM New York, NY, USA}
}

@inproceedings{mcneese2005neocities,
  title={The NeoCITIES simulation: Understanding the design and experimental methodology used to develop a team emergency management simulation},
  author={McNeese, Michael D and Bains, Priya and Brewer, Isaac and Brown, Cliff and Connors, Erik S and Jefferson Jr, Tyrone and Jones Jr, Rashaad ET and Terrell Jr, Ivanna},
  booktitle={Proceedings of the human factors and ergonomics society annual meeting},
  volume={49},
  number={3},
  pages={591--594},
  year={2005},
  organization={SAGE Publications Sage CA: Los Angeles, CA}
}

@article{student1908probable,
  title={The probable error of a mean},
  author={Student},
  journal={Biometrika},
  pages={1--25},
  year={1908},
  publisher={JSTOR}
}

@book{wooldridge2010econometric,
  title={Econometric analysis of cross section and panel data},
  author={Wooldridge, Jeffrey M},
  year={2010},
  publisher={MIT press}
}

@article{pearson1895vii,
  title={VII. Note on regression and inheritance in the case of two parents},
  author={Pearson, Karl},
  journal={proceedings of the royal society of London},
  volume={58},
  number={347-352},
  pages={240--242},
  year={1895},
  publisher={The Royal Society London}
}

@article{goodman1961snowball,
  title={Snowball sampling},
  author={Goodman, Leo A},
  journal={The annals of mathematical statistics},
  pages={148--170},
  year={1961},
  publisher={JSTOR}
}

@article{polson1992cognitive,
  title={Cognitive walkthroughs: a method for theory-based evaluation of user interfaces},
  author={Polson, Peter G and Lewis, Clayton and Rieman, John and Wharton, Cathleen},
  journal={International Journal of man-machine studies},
  volume={36},
  number={5},
  pages={741--773},
  year={1992},
  publisher={Elsevier}
}

@inproceedings{crabtree2013many,
  title={“How Many Bloody Examples Do You Want?” Fieldwork and Generalisation},
  author={Crabtree, Andy and Tolmie, Peter and Rouncefield, Mark},
  booktitle={ECSCW 2013: Proceedings of the 13th European Conference on Computer Supported Cooperative Work, 21-25 September 2013, Paphos, Cyprus},
  pages={1--20},
  year={2013},
  organization={Springer}
}

@article{kelley2018wizard,
  title={Wizard of oz (woz) a yellow brick journey},
  author={Kelley, John F},
  journal={Journal of Usability Studies},
  volume={13},
  number={3},
  pages={119--124},
  year={2018},
  publisher={Usability Professionals' Association Bloomingdale, IL}
}


%% file: sample-base.bib
@String{Computing = "Computing" }

@String{Computer = "{IEEE} Computer" }

@String{Springer = "Springer-Verlag" }
